\documentclass[12pt,preprint]{aastex}

\slugcomment{Submitted to The Astrophysical Journal - revised version}

\shortauthors{Limongi, Chieffi} \shorttitle{The Al26 and Fe60 gamma ray emitters}

\begin{document}

\newcommand{\msun}{$\rm M_\odot~$}
\newcommand{\rsun}{$\rm R_\odot~$}
\newcommand\nuk[2]{$\rm ^{\rm #2} #1$}

\title{The nucleosynthesis of \nuk{Al}{26} and \nuk{Fe}{60} in solar metallicity stars
extending in mass from 11 to 120 \msun: the hydrostatic and explosive contributions}

\author{Marco Limongi\altaffilmark{1,3} and Alessandro Chieffi\altaffilmark{2,3}}

\altaffiltext{1}{Istituto Nazionale di Astrofisica - Osservatorio Astronomico di Roma, Via Frascati 33, I-00040, Monteporzio Catone, Italy;
marco@oa-roma.inaf.it}

\altaffiltext{2}{Istituto Nazionale di Astrofisica - Istituto di Astrofisica Spaziale e Fisica Cosmica, Via Fosso del Cavaliere, I-00133, Roma, Italy;
alessandro.chieffi@iasf-roma.inaf.it}

\altaffiltext{3}{Centre for Stellar \& Planetary Astrophysics, 
School of Mathematical Sciences, P.O. Box, 28M, Monash University, Victoria 3800, Australia}

\begin{abstract} 

We present the \nuk{Al}{26} and \nuk{Fe}{60} yields produced by a generation of solar metallicity stars ranging in mass between 11 
and 120\msun. We discuss the production sites of these $\gamma$ ray emitters and quantify the relative contributions of the various 
components. More specifically we provide the separate contribution of the wind, the C convective shell and the explosive Ne/C 
burning to the total \nuk{Al}{26} yield per each stellar model in our grid. We provide the contributions of the He convective shell, 
the C convective shell and the explosive Ne/C burning to the \nuk{Fe}{60} yield as well. From these computations we conclude that, 
at variance with current beliefs, \nuk{Al}{26} is mainly produced by the explosive C/Ne burning over most of the mass interval 
presently analyzed while \nuk{Fe}{60} is mainly produced by the C convective shell and the He convective shell.

By means of these yields we try to reproduce two quite strong observational constraints related to the abundances of these nuclei in 
the interstellar medium, i.e. the number of $\gamma_{1.8}$ photons per Lyman continuum photon, $\rm R_{GxL}$, and the 
\nuk{Fe}{60}$/$\nuk{Al}{26} $\gamma$-ray line flux ratio. $\rm R_{GxL}$ is found to be roughly constant along the galactic plane 
\citep{k99b} (and of the order of $1.25\times10^{- 11}$), while the \nuk{Fe}{60}$/$\nuk{Al}{26} ratio has beed recently measured by 
both RHESSI $(0.17\pm 0.05)$ and SPI(INTEGRAL) $(0.11\pm 0.03)$. We can quite successfully fit simultaneously both ratios for a 
quite large range of exponents of the power law initial mass function.

We also address the fit to $\gamma^2$ Velorum and we find that a quite large range of initial masses, at least from 40 to 60\msun, 
do eject an amount of \nuk{Al}{26} (through the wind) compatible with the current upper limit quoted for this WR star: such a result 
removes a longstanding discrepancy between the models and the observational data.

\end{abstract}


\section{Introduction}

The quest for the main \nuk{Al}{26} source(s) started as soon as the first observational evidence of the presence of live 
\nuk{Al}{26} in the inner Galaxy was demonstrated by the HEAO 3 experiment \citep[1984]{ma82} and it still continues at the present 
days. Its presence in the interstellar medium was not a surprise because it was theoretically predicted since the middle of the 
seventies that explosive Ne/C burning in core collapse supernovae may produce such a nucleus \citep{a77,rl77,tc78,aw78,ww80}. Later 
on \citet{db85} pointed out that \nuk{Al}{26} may by injected into the interstellar medium also by Wolf-Rayet stars through stellar 
wind and since then a number of works have been devoted to the prediction of the amount \nuk{Al}{26} ejected by these stars 
\citep[e.g.][]{pc86,wm89,me97,vu04,pa05}. \citet{ar80} and \citet{cla84}, on the basis of the theoretical models available at that 
time, proposed also the novae as main \nuk{Al}{26} producers. All these candidates, with the further addition of the AGB stars 
\citep{mm00}, could in principle account for a significant fraction of the observed flux due to the uncertainties in both the 
frequencies of the various candidates and the yields itself. \citet{lc85} were the first to point out that the lack of information 
about the distribution of the \nuk{Al}{26} in the Galaxy severely limits the possibility to identify the main \nuk{Al}{26} donor(s). 
A clear snapshot of the situation just before the launch of the Compton Gamma Ray Observatory (CGRO) is provided by the 
\citet{pra91} review paper. 

The launch of the CGRO in 1991, and its continuous operation up to 2001, allowed the first mapping of the 1809 KeV line all over the 
Galaxy and put firm constraints on the possible main sources of \nuk{Al}{26}. By the way, the latest all-sky map has been published 
by \citet{plu01}. This observational campaign showed beyond any doubt that the \nuk{Al}{26} distribution was strongly confined 
towards the galactic disk and quite clumpy and asymmetric, strongly pushing towards a massive star parent for the bulk of the 
\nuk{Al}{26}. We refer the reader to the reviews by \citet{pd96} and \citet{dt98} for an excellent clear picture of the scenarios 
that were under debate as soon as the first results from COMPTEL aboard the CGRO became available. A further milestone progress in 
the identification of the main \nuk{Al}{26} producers was obtained when the \nuk{Al}{26} map was compared to a series of all sky 
maps obtained at various wavelengths. \citet{k99a} and \citet{k99b} found that the 53 GHz microwave free- free emission all-sky map 
was the only one to almost perfectly overlap the \nuk{Al}{26} all-sky map. The 53 GHz free-free emission line maps the regions of 
ionized gas and such a sustained ionization may be powered only by the Lyman continuum photons provided by the very massive stars 
and disappears on a timescale of 1 Myr after the switch off of the Lyman continuum "lamp". Hence the existence of such a good 
overlap between the two all-sky maps demonstrates beyond any reasonable doubt that the bulk of the \nuk{Al}{26} is produced in the 
same regions where massive stars are present. Such a finding did not close at all the quest of the \nuk{Al}{26} progenitors but it 
simply shifted such a quest towards which subclass of massive stars do mainly contribute to the \nuk{Al}{26} budget. The two 
classical competing candidates were identified as the Type II supernovae and the Wolf Rayet (WR) stars (through the stellar wind). A 
further question concerned the possibility, sometimes suggested, that a consistent fraction of the 1.8 MeV $\gamma$-flux 
($\gamma_{1.8}$) were due to close localized sources and not to Galaxy wide sources. Such a doubt was mainly due to the fact that 
the massive star census in the Galaxy is well known only out to distances of a few kiloparsecs and many regions of the Galaxy are 
obscured for direct measurements. This problem was recently definitely solved by SPI(INTEGRAL) \citep{diehl06} that measured for the 
first time with great accuracy the doppler shift of the \nuk{Al}{26} line along the inner Galaxy clearly demonstrating that the 
\nuk{Al}{26} corotates with the Galaxy, hence supporting a Galaxy-wide origin. The total amount of \nuk{Al}{26} present in the disk 
of our Galaxy has been estimated by assuming a variety of distributions of the \nuk{Al}{26} sources and it ranges between 1 and 3 
$\rm M_\odot$.

Together with the search for a signal from the decay of \nuk{Al}{26}, also a signal from another gamma ray emitter, namely 
\nuk{Fe}{60}, has been searched for thoroughly. The production of this nucleus has been historically attributed to the Ne explosive 
burning in core collapse supernovae \citep{c82}, hence it has always been considered as the litmus paper to discriminate among the 
various candidates as major \nuk{Al}{26} contributors. Unfortunately, all the experiments up to 2002 could only provide upper limits 
on the abundance of this nucleus in the interstellar medium. With the launch of the Reuven Ramaty High Energy Solar Spectroscopic 
Imager (RHESSI) and of the International Gamma-Ray Astrophysics Laboratory (INTEGRAL) the situation changed drastically since both 
RHESSI and SPI aboard INTEGRAL were able to measure an \nuk{Fe}{60} signal from the inner radiant of our Galaxy. The latest 
simultaneous measurements of \nuk{Al}{26} and \nuk{Fe}{60} performed by RHESSI and SPI at the time of writing give a 
\nuk{Fe}{60}$/$\nuk{Al}{26} $\gamma$-ray line flux ratio equal to $(0.17\pm 0.05)$ \citep[RHESSI,][]{smi05} and $(0.11\pm 0.03)$ 
\citep[SPI,][]{har05}, respectively. 

The theoretical models that constitute the base of the various interpretations of these data may be divided in two broad categories: 
the first one includes stars less massive than, say, 35-40\msun and a second one that includes the more massive stars. This limiting 
mass roughly marks the minimum mass that becomes a WR star. The evolution of the stars in the first group has been extensively 
explored through all the hydrostatic burnings up to the core collapse together with the following explosive burnings. Yields for 
both \nuk{Al}{26} and \nuk{Fe}{60} have been published, for this mass range, over the years by several groups 
\citep{ww95,tnh96,rhhw02,lc03,cl04}. The evolution of the stars in the upper mass range, which spend part of their lifetime as WR 
stars, have never been followed all along their full evolution (and final explosion) but only through the H and He burning (plus, 
quite recently, the central C burning) phases; obviously also the explosive burnings of these very massive stars have never been 
addressed. As a consequence of this very biased knowledge of the evolution of these very massive stars, only the fraction of 
\nuk{Al}{26} yield ejected by the stellar wind has been predicted up to now \citep[e.g.][]{pra86a,pra86b,wm89,me97,pa05}, and 
unfortunately quite often this partial prediction has been interpreted as the total \nuk{Al}{26} yield produced by these very 
massive stars. In other words, it has become a deep-seated notion that the stars that do not become WR contribute to both the 
\nuk{Al}{26} and \nuk{Fe}{60} yields while those that become WR do contribute only to the synthesis of the \nuk{Al}{26}.  Within 
this accepted framework, an "observed" \nuk{Fe}{60}$/$\nuk{Al}{26} flux ratio of 0.11-0.17 would favor the masses that do not become 
WR as the main producers of both \nuk{Al}{26} and \nuk{Fe}{60} \citep{tim95}. Indeed, the WW95 models do predict a 
\nuk{Fe}{60}$/$\nuk{Al}{26} flux ratio of 0.18 (averaged over a power law IMF - $\rm dn/dm=M^{\rm (-2.35)}$ - ranging between 11 and 
40\msun). However, more recent models \citep{rhhw02,lc03,cl04} do predict much larger \nuk{Fe}{60}$/$\nuk{Al}{26} flux ratios (0.5-
1.0), leaving therefore plenty of room for an important contribution of the WR stars to the synthesis of the \nuk{Al}{26}. For sake 
of completeness, it must be mentioned that \citet{wlw95} published the presupernova evolution and explosive yields of massive pure 
He cores that should be associated to WR stars. However, unfortunately, these bare cores can hardly be reconciled with a population 
of (single) WR stars because all of them correspond to models that lose the H-rich mantle before the central He ignition while 
consistent models of massive stars with mass loss (see below) show that this happens only in stars more massive than 80 $\rm 
M_\odot$.

In this paper we completely reanalyze and rediscuss the interpretation of the observed fluxes of both \nuk{Al}{26} and \nuk{Fe}{60} 
in the light of a new large grid of massive star models of solar metallicity with mass loss. These homogeneous sets of models extend 
in mass between 11 and 120\msun and cover the full hydrostatic evolution from the Main Sequence up to the core collapse together 
with the following explosive burnings. These models will be published shortly. Given the importance of the problems connected with 
the $\gamma$ ray emitters \nuk{Al}{26} and \nuk{Fe}{60}, we think that it is worth dedicating a full paper to them. We are aware 
that these models are just solar metallicity models and that it would be important to have supersolar models too, but we think that 
the availability of a full set of models from 11 to 120\msun covering all the hydrostatic and explosive burnings is already a big 
step forward in the understanding of these nuclei. We also address the fit to $\gamma^2$ Velorum, the closest WR star.

The paper is organized as follows: first, the new set of models is briefly presented, then two distinct sections discuss separately 
the theoretical yields of these two nuclei as a function of the initial mass. Their dependence on the mass loss rate adopted in the 
WNE+WCO phase is addressed in Section 5. A discussion of the scenarios that comes out from the adoption of these new yields and a 
final brief conclusion follow.

\section{The stellar models}

The \nuk{Al}{26} and \nuk{Fe}{60} yields used in the present paper have been extracted from a new generation of stellar models that 
will be available soon. These new models have been computed with the latest release (5.050218) of the FRANEC (Frascati RAphson 
Newton Evolutionary Code). By the way, from now on the release number will follow the following convention: the integer part of the 
number refers to the major release (the fifth at present) while its decimal part refers to the date in which it has been released, 
in the form YYMMDD. The main features of this new release, with respect to that described in \citet{lc03}, are the following. First 
of all the convective mixing and the nuclear burning are coupled together and solved simultaneously. More specifically, the set of 
equations describing the chemical evolution of the matter are now:

$$
{dY_{i} \over dt} = \Bigl( {{\partial Y_{i}}\over{\partial t}} \Bigr)_{\rm nuc} + {{\partial}\over{\partial m}} \Bigl[ (4\pi r^{2} \rho)^{2} D {{\partial Y_{i}}\over{\partial m}} \Bigr]
$$

where the diffusion coefficient $D$ is given by $D={{1/3}} v_{c} l$, the convective velocity $v_{c}$ is computed by means of the 
mixing length theory and $l$ is the mixing lenght parameter. This set of equations is linearized and solved by means of a Raphson 
Newton technique. Let us also remind that, as in all our computations, the nuclear energy generation rate is always computed with 
the same, full, network adopted to describe the chemical evolution of the matter, without any kind of approximation. The network 
adopted in the present computations is the same as in \citet{lc03}. $\rm 0.2~H_{\rm p}$ of overshooting has been assumed at the top 
of the convective core in central H burning. The nuclear cross sections have been updated with respect to those adopted in 
\citet{lc03} whenever possible. Table \ref{tabele} shows the full reference matrix of all the processes taken into account in the 
network, together with its proper legend. Note that for the weak interaction rates, $\beta^{+}$ and $\beta^{-}$ mean the sum of both 
the electron capture and the $\beta^{+}$ decay and the positron capture and the $\beta^{-}$ decay, respectively. The dependence on 
the temperature of the weak interaction rates has been considered following \citet{odaetal94}, \citet{FFN82}, \citet{lp00} and 
\citet{ty87}, see Table \ref{tabele}. The $\rm ^{26}Al$ requires a special treatment since the ground (Alg6 entry in Table 
\ref{tabele}) $\rm ^{26}Al^{g}$ ($J^{\pi}=5^{+}$, $\tau=1.03\times 10^{6}~yr$) and the isomeric states (Alm6 entry in Table 
\ref{tabele}) $\rm ^{26}Al^{m}$ ($J^{\pi}=0^{+}$, $\tau=9.15~s$) do not come into statistical equilibrium for temperatures $\rm T 
\leq 10^{9}~K$ \citep{wf80,gm01}. For this reason we treat the ground and isomeric states of $\rm ^{26}Al$ as separate species for 
$\rm T\leq 10^{9}~K$  while we assume the two states to be in statistical equilibrium (and hence consider just one isotope) above 
this temperature.  By the way, the Al26 entry in Table \ref{tabele} refers to the total amount of $\rm ^{26}Al$.

Mass loss has been included following the prescriptions of \citet[2001]{val00} for the blue supergiant phase ($\rm 
T_{eff}>12000~K$), \citet{dejag88} for the red supergiant phase ($\rm T_{eff}<12000~K$) and \citet{nl00} for the Wolf-Rayet phase. 
Additional models have been computed with the \citet{la89} mass loss rate in the WNE+WCO phases.

The explosion of the mantle of the star is started by imparting istantaneously an initial velocity $v_{0}$ to a mass coordinate of 
$\sim 1~M_\odot$ of the presupernova model, i.e., well within the iron core. The propagation of the shock wave is computed by means 
of an explosive simulation code, developed by us, that solves the fully compressible reactive hydrodynamic equations using the 
Piecewise Parabolic Method (PPM) of \citep{cw84} in the lagrangean form. $v_{0}$ is properly tuned in order to eject all the mass 
above the Fe core. The chemical evolution of the matter is computed by coupling the same nuclear network adopted in the hydrostatic 
calculations to the system of hydrodynamic equations. The nuclear energy generation is not taken into account in these simulations 
since we assume that it is always negligible with respect to both the kinetic and the internal energies. By taking advantage of the 
fact that the final yields mainly depend on the mass cut location (the mass coordinate that separates the ejecta from the compact 
remnant) and only mildly on the properties of the parametrized explosion, yields corresponding to different amounts of $\rm ^{\rm 
56}Ni$ ejected can be easily obtained by fixing the mass cut by hand a posteriori \citep{lc03}. Here, for each star, we chose the 
mass cut in order to obtain $\rm 0.1~M_\odot$ of $\rm ^{\rm 56}Ni$ ejected. By the way, let us note that since both $\rm ^{\rm 26} 
Al$ and $\rm ^{\rm 60}Fe$ are synthesized in quite external regions, their final yields do not depend on the particular choice of 
the mass cut since we assume in any case that at least a minimum amount of \nuk{Ni}{56} is ejected.

All these solar models has been computed by adopting an initial He abundance equal to 0.285 (by mass fraction), a global metallicity 
Z$=$0.02 (by mass fraction) - the relative abundances among the various nuclear species are taken from the \citet{ag89} solar 
distribution. They range in mass between 11\msun and 120\msun, covering therefore the full range of masses that are expected to give 
rise to the Type II/Ib/Ic Supernovae as well as to the WR populations. These models and the corresponding full set of yields will be 
published, and discussed, shortly; for the moment we anticipate a few basic properties in Table \ref{taba}. The following data are 
reported: the initial and final mass in solar masses (cols. 1 and 2), the central H and He burning lifetimes in Myr (cols. 3 and 4), 
the lifetime after the central He exhaustion in yr (col. 5), the maximum sizes of the He and CO core masses in solar masses (cols. 6 
and 7), the bottom and top mass locations of the last C convective shell in solar masses (col. 8), the central C mass fraction at 
the central He exhaustion (col. 9), the total amount of time spent as O star (col. 10) and WR star (col. 11) in yr and eventually 
the lifetimes as WNL (col. 12), WNE (col. 13) and WCO (col. 14) all in yr. We assume that the star enters the WR phase when $\rm 
Log(T_{\rm eff})>4$ and $\rm H_{\rm surf}<0.4$ and we adopt the following usual definitions for the various WR phases: WNL ($\rm 
10^{-5}<H_{\rm surf}<0.4$), WNE ($\rm H_{\rm surf}<10^{-5}$ and $\rm (C/N)_{\rm surf}<0.1$), WNC ($\rm 0.1<(C/N)_{\rm surf}<10$) and 
WCO ($\rm (C/N)_{\rm surf}>10$). The main data concerning the \nuk{Al}{26} and \nuk{Fe}{60} yields are summarized in Table 
\ref{tabb}. The first column refers to the initial mass in solar masses. Columns 2 to 5 refer to the \nuk{Al}{26} yield (all in 
solar masses): total (col. 2), the wind contribution (col. 3), the C\&Ne shell contribution (col. 4) and the explosive component 
(col. 5). Columns 6 to 10 refer to the \nuk{Fe}{60} yield (all in solar masses): total (col. 6), the contributions of the radiative 
and convective He burning shell (cols. 7 to 9), the convective C shell contribution (col. 10) and the explosive component (col. 11).

\section{\nuk{Al}{26}}

\nuk{Al}{26} is an unstable nucleus (its terrestrial half life is $\rm \tau_{1/2}\simeq 7.17\times 10^5 y$) produced almost 
exclusively by proton capture on \nuk{Mg}{25}. Its synthesis occurs, in massive stars, essentially in three different specific 
environments, i.e., the core H burning, the C and Ne convective shells and the explosive Ne burning. Let us discuss each of these 
environments separately.

All over the mass range presently analyzed, \nuk{Al}{26} is produced in Main Sequence by the \nuk{Mg}{25}(p,$\gamma$)\nuk{Al}{26} 
reaction (hereinafter MG25PG) and mainly destroyed by the $\rm \beta^+$ decay into \nuk{Mg}{26} since the competing destruction 
process, i.e. the \nuk{Al}{26}(p,$\gamma$)\nuk{Si}{27} process, becomes efficient for $\rm T > 50\times10^6$ K and only our most 
massive model barely reaches such a temperature towards the end of the central H burning. 

\begin{figure} 
\rotate
\begin{center}
\epsscale{1}
\plotone{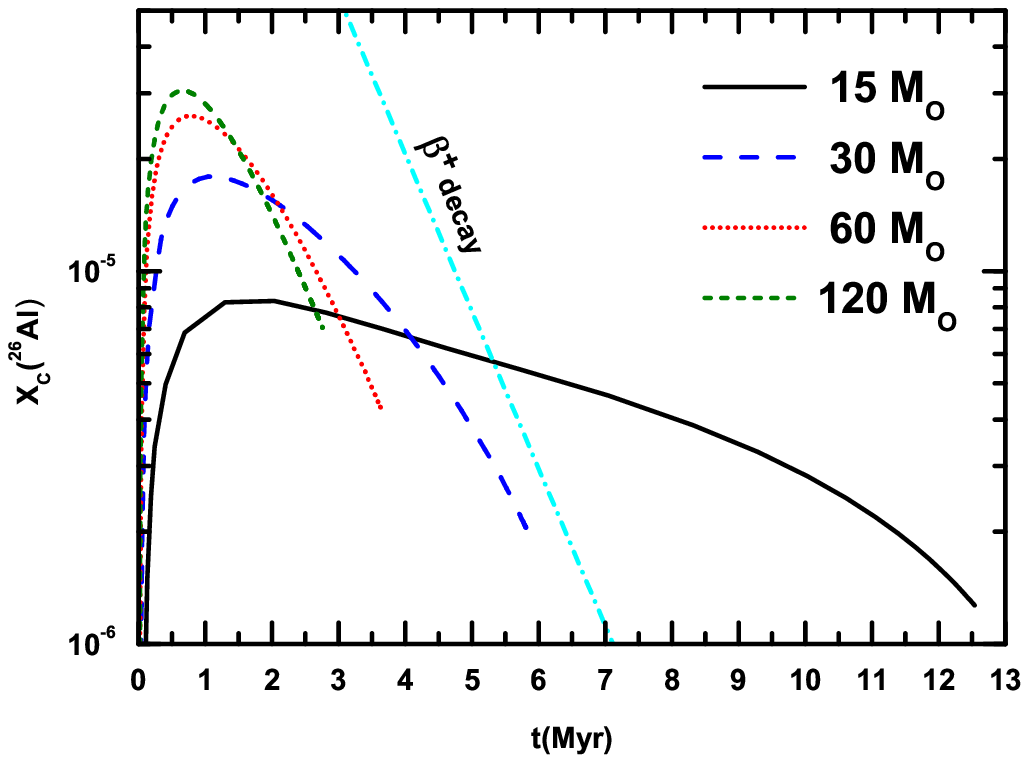}
\end{center}

\caption{Temporal evolution of the central \nuk{Al}{26} mass fraction during core H burning. The {\rm 
black solid}, {\rm blue long dashed}, {\rm red dotted} and {\rm green short dashed} lines refer to the 15, 
30, 60 and 120\msun models, respectively. The cyan dot-dashed line refers to a pure \nuk{Al}{26} decay.}

\label{al26temporal}
\end{figure}

The conversion of the initial abundance of \nuk{Mg}{25} into \nuk{Al}{26} starts as soon as a star settles on the Main Sequence 
(since $\rm T_{\rm central} \geq 30\times10^6$ K) and comes to completion on a timescale of few million years, i.e. well within the 
central H burning lifetime. Figure \ref{al26temporal} shows the temporal evolution of the \nuk{Al}{26} mass fraction in the H 
convective core: a maximum is reached shortly after the beginning of the central H burning, followed by a progressive decrease 
mainly due to the $\rm \beta^+$ decay. Since \nuk{Al}{26} continues to be produced also after the maximum, its decline is shallower 
than that of a pure decay (shown in Figure \ref{al26temporal} as the cyan dot-dashed line). By the way, it is such a prolonged 
production that preserves some \nuk{Al}{26} in the H convective core up to the end of the central H burning. In fact, if the slope 
of the decline would follow a pure decay after the maximum, essentially all the \nuk{Al}{26} would decay by the time the star 
exhausts the H in the center. The {\rm solid}, {\rm long dashed}, {\rm dotted} and {\rm short dashed} lines refer to the 15, 30, 60 
and 120 \msun, respectively. 

Since the efficiency of the MG25PG reaction rate scales directly with the initial mass, the larger the initial mass the higher the 
maximum \nuk{Al}{26} mass fraction abundance, the faster the \nuk{Mg}{25} consumption and hence the faster the \nuk{Al}{26} decline. 
Hence, if the efficiency of the MG25PG were the only parameter to control the scaling of the \nuk{Al}{26} mass fraction abundance at 
the central H exhaustion with the initial mass of the star, one should obtain an inverse scaling of the central \nuk{Al}{26} mass 
fraction at the central H exhaustion with the mass. Vice versa the models show a strong direct scaling. Such a result is the obvious 
consequence of the strong shortening of the central H burning lifetime with the initial mass that freezes the central \nuk{Al}{26} 
abundance progressively at earlier times (and hence at higher abundances). Of course, a direct scaling between the \nuk{Al}{26} mass 
fraction and the initial mass of a star does not imply, by itself, a direct scaling between the total amount of \nuk{Al}{26} (in 
solar masses) produced by the H burning and the initial mass. In fact, it is the size of the convective core that really determines 
how much \nuk{Al}{26} is produced in total, and since it scales directly with the initial mass, the direct scaling between 
\nuk{Al}{26} produced and initial mass is secured.

\begin{figure} 
\begin{center}
\epsscale{1}
\plotone{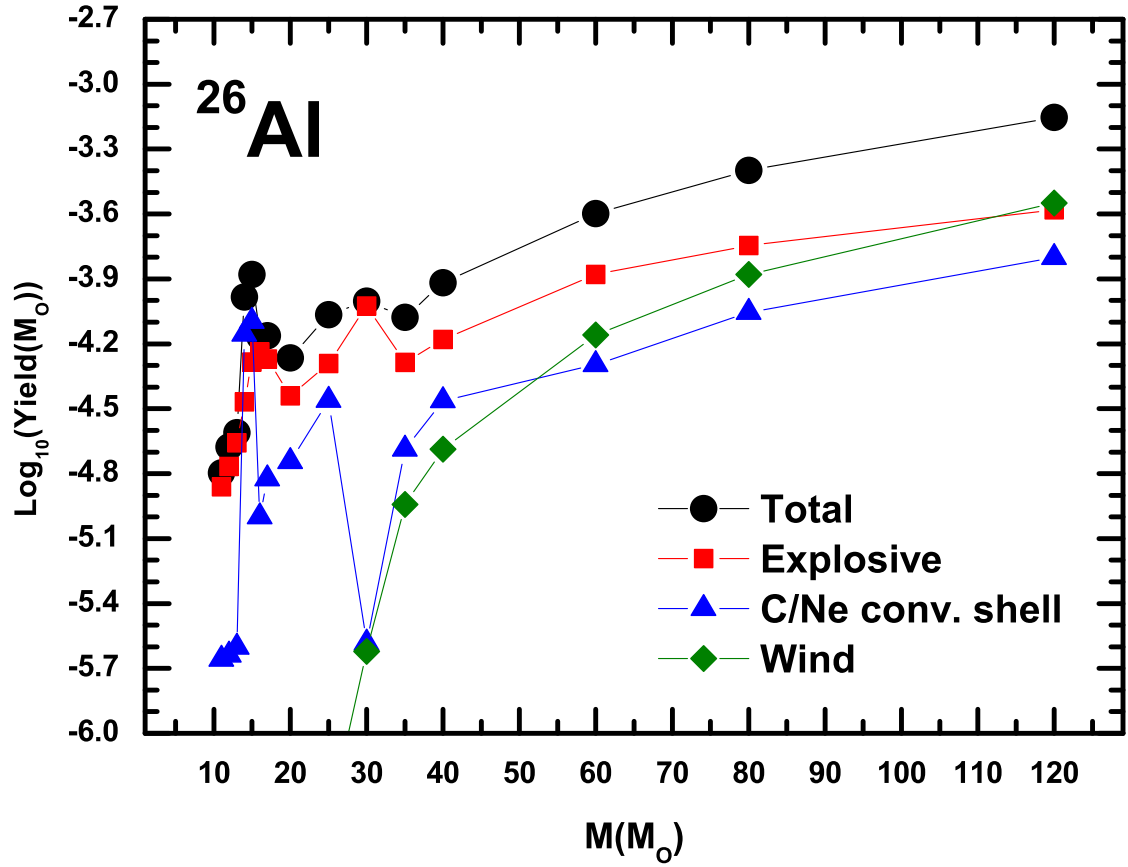}
\end{center}

\caption{\nuk{Al}{26} yields as a function of the initial mass. The symbol key is found in the figure.}

\label{al26}
\end{figure}

At central H exhaustion, the \nuk{Al}{26} is located both in the He core and in the region of variable H left behind by the receding 
convective core. Since the He burning easily and quickly destroys the \nuk{Al}{26} (via the $(n,\alpha)$ and $(n,p)$ reactions - the 
neutrons being released by the $\rm ^{13}C(\alpha,n)^{16}O$ process) the amount of \nuk{Al}{26} synthesized by central H burning and 
really preserved up to the explosion is just the one located in the H rich layers plus the one locked in the fraction of the He core 
that will not be affected by the He burning. If the dredge up and the mass loss were not effective, this \nuk{Al}{26} would be 
ejected all together by the explosion, while the occurrence of these two phenomena may anticipate such an ejection at earlier times 
and also increase the amount of \nuk{Al}{26} ejected into the interstellar medium. In our models three mass intervals may be 
identified: 1) stars between 11 \msun and 30 \msun undergo a dredge up episode that does not enter into the He core and the mass 
loss is weak enough that all these stars end their life with an H-rich envelope; since the He convective shell extends almost up to 
the base of the H burning shell, only the tiny amount of \nuk{Al}{26} present in the region of variable H left by the receding H 
convective core and engulfed in the convective envelope is ejected in the interstellar medium. 2) stars between 35\msun and 40\msun 
experience a dredge up episode that does not enter the He core as well, but the mass loss is strong enough that also a fraction of 
the He core is ejected outward; since the mass loss erodes the He core before the formation of the He convective shell, the 
\nuk{Al}{26} present in these layers is preserved by the destruction and goes to increase the amount of \nuk{Al}{26} ejected in the 
interstellar medium. 3) stars more massive than 40\msun do not show any dredge up episode and the mass loss is so strong that a 
substantial fraction of the He core is ejected through the stellar wind; as a consequence a larger amount of \nuk{Al}{26} present in 
the He core is preserved from the destruction and ejected. As for the first two mass intervals, the wind starts expelling 
\nuk{Al}{26} as soon as the star becomes a red (super)giant. Vice versa, stars in the upper mass interval begin to eject 
\nuk{Al}{26} only when the total mass of the star reduces enough that layers processed by the H burning are exposed to the surface 
(i.e. when the star becomes a WR). Let us eventually note that only a very tiny fraction of the original \nuk{Mg}{25} is ejected in 
the interstellar medium as \nuk{Al}{26}(mainly because of the substantial \nuk{Al}{26} decay during the core H burning) and also 
that this \nuk{Al}{26} is of semi-secondary origin since only the \nuk{Mg}{25} depends on the initial metallicity Z and obviously 
not the protons (the true secondary nuclei are the ones formed by reactions in which both the target and the projectile depend on 
the initial metallicity, for example, as in the s process production). Figure \ref{al26} shows, as green diamonds, the amount of 
\nuk{Al}{26} expelled through the stellar wind as a function of the initial mass; the observed trend is the obvious consequence of 
what has been discussed above, i.e., it is the result of the direct scaling of (1) the size of the H convective core, (2) the 
\nuk{Al}{26} mass fraction left by the central H burning and (3) the mass loss efficiency with the initial mass.

Once the star moves beyond the central H burning, the "habitat" becomes very hostile to the survival of \nuk{Al}{26}. First of all, 
how already mentioned above, the He burning simply destroys \nuk{Al}{26} through neutron captures. Second, the half life of 
\nuk{Al}{26} has a tremendous dependence on the temperature: it reduces to $\rm \sim 0.19$ yr already at $\rm Log(T)=8.4$, to $\rm 
\sim 13$ h at $\rm Log(T)=8.6$, and to $\rm 13\div2$ m in the range $\rm 9<Log(T)<9.6$. Third, efficient photodisintegration quickly 
destroy \nuk{Al}{26} when the temperature raises above, say, $\rm Log(T)=9.3$. Such a temperature limit obviously constrains a 
possible \nuk{Al}{26} production to the C and/or Ne burning shells. 

As for the C shell, its typical temperature ($\rm Log(T)<9.08$) does not allow in general a substantial production of \nuk{Al}{26}. 
However, after the central Si burning, the strong contraction (and heating) of the inner core that precedes the final gravitational 
collapse, induces a strong temperature increase of the C burning shell (up to, roughly, $\rm Log(T)\sim9.255$). If, at this stage, 
such a burning occurs in an efficient convective shell, a substantial amount of \nuk{Al}{26} is produced. Let us discuss such a 
production in a more detailed way. The processes that produce and destroy \nuk{Al}{26} in these conditions are once again the MG25PG 
and the beta decay. The MG25PG reaction rate depends on the MG25PG cross section but also on the abundances of \nuk{Mg}{25} and 
protons. The \nuk{Mg}{25} that now enters in the \nuk{Al}{26} production comes directly from the initial CNO abundance, via the 
sequence $\rm (CNO)_{ini}\to$\nuk{N}{14}$\to$\nuk{Ne}{22}$\to$\nuk{Mg}{25} while the protons are mainly produced by the 
\nuk{C}{12}(\nuk{C}{12},p)\nuk{Na}{23} and the \nuk{Na}{23}($\alpha$,p)\nuk{Mg}{26} processes. By the way, note that the main proton 
poison is not the \nuk{Mg}{25} but, instead, the \nuk{Na}{23} through the (p,$\alpha$) channel. The high temperature influences the 
\nuk{Al}{26} production either because it raises the MG25PG cross section but also because it strongly increases the proton density 
by increasing the cross sections of both the processes that provide the bulk of protons. On the other hand, the presence of a 
convective environment is also crucial because it continuously brings fresh \nuk{C}{12} and \nuk{Mg}{25} where the \nuk{Al}{26} 
production occurs and simultaneously brings the freshly made \nuk{Al}{26} to much lower temperatures where its lifetime increases 
enormously. 

At variance with the C burning, \nuk{Al}{26} is always produced by the Ne burning (both in the center and in the shell). A 
substantial amount of \nuk{Mg}{25} is left unburned by the C burning and the protons needed to activate the $\rm 
^{25}Mg(p,\gamma)^{26}Al$ reaction mainly come from the $\rm ^{23}Na(\alpha,p)^{26}Mg$ reaction ($\rm ^{23}Na$ being one of the 
products of C burning), the $\alpha's$ being provided by the $\rm ^{20}Ne(\gamma,\alpha)^{16}O$ photodisintegration. Again, the most 
effective protons poison is the $\rm ^{23}Na(p,\alpha)^{20}Ne$. In a radiative environment the \nuk{Al}{26} "equilibrium" abundance 
depends on the local balance between production and destruction so that, as soon as the \nuk{Ne}{20} photodisintegration weakens for 
lack of fuel, the abundance of the $\alpha's$ drops as well and the proton production channels dry up: no more \nuk{Al}{26} may be 
produced and its short lifetime at these high temperatures leads quickly to its total destruction. Once again only the presence of 
an efficient convective shell could act as a preservation buffer. Since most of the stars do not reach the onset of the explosion 
with an efficient Ne convective shell, in general only the small equilibrium abundance of \nuk{Al}{26} located within the radiative 
Ne burning shell is present at the beginning of the core collapse. There are two exceptions: the 14 and the 15 $\rm M_{\odot}$ 
models. Stars within this mass interval are characterized by the lack of a C convective shell, at the end of the central Si burning, 
and by the formation of a Ne convective shell very close to the region where it was previously efficient a carbon convective shell. 
During the last strong contraction of the core that precedes the final collapse, such a Ne convective shell penetrates the C rich 
region with the consequence of producing a huge amount of \nuk{Al}{26}. 

The \nuk{Al}{26} synthesized in the C (or Ne/C) convective shell is located close enough to the iron core that is partially 
destroyed by the passage of the shock wave. In general, the larger the mass the larger the amount of \nuk{Al}{26} that survives the 
explosion (Table \ref{tabb}). The blue triangles in Figure \ref{al26} show the \nuk{Al}{26} yield produced in this convective shell 
that survived to the explosion as a function of the initial mass. The general increase of this yield with the mass depends on the 
fact that both the mass size and the peak temperature of the C (or Ne/C) convective shell scale directly with the mass. The peak 
around the 15\msun model, as already discussed above, is due to the presence of an extended Ne convective shell that engulfs part of 
the C rich region. The minimum yield around the 30\msun is the natural consequence of the lack of an active C (and Ne) convective 
shell at the moment of the explosion. The \nuk{Al}{26} produced by the C and/or Ne shell is of semi-secondary origin (as the one 
produced by the H burning) because the \nuk{Mg}{25} is secondary while protons are primary.

\begin{figure} 
\begin{center}
\epsscale{1}
\plotone{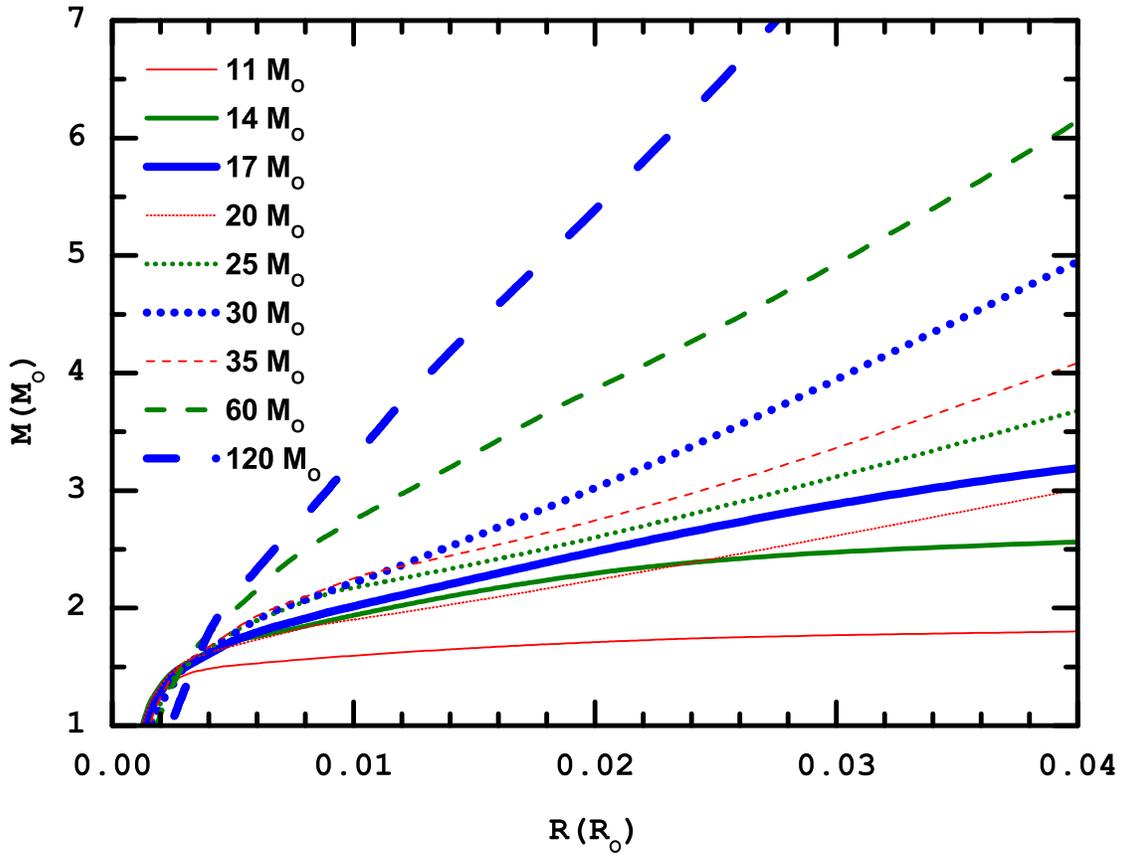}
\end{center}

\caption{Final Mass-Radius relation at the beginning of the core collapse for a subset of the grid models. 
The line key is found in the figure.}

\label{mara}
\end{figure}

\nuk{Al}{26} is also produced during the explosion at a typical temperature of the order of $\rm \sim 2.3$ billion degrees. Such a 
condition occurs within the C convective shell and the main process that controls its production is once again the MG25PG process 
while its destruction is now controlled, roughly parithetically, by the $\rm (n,p)$ and $\rm (n,\alpha)$ processes. The \nuk{Mg}{25} 
now comes mainly from the $\rm (n,\gamma)$ capture on \nuk{Mg}{24}, this isotope being a primary outcome of the C and Ne burning. 
The neutron density that enters in both the production and destruction of \nuk{Al}{26} is determined by the competition among 
several processes: the main neutron producers are $\rm (\alpha,n)$ captures on \nuk{Mg}{26}, \nuk{Mg}{25}, \nuk{Ne}{21}, 
\nuk{Si}{29} plus the $\rm (p,n)$ capture on \nuk{Al}{28} while the main neutron poisons are the $\rm (n,\gamma)$ captures on 
\nuk{Mg}{24}, \nuk{O}{16} and \nuk{Ne}{20}. The proton density that enters the MG25PG rate is mainly determined by the competition 
between the production that occurs via the $\rm (\alpha,p)$ captures on \nuk{Ne}{20}, \nuk{Mg}{24}, \nuk{Al}{27} and \nuk{Na}{23} 
and the destruction that occurs via the $\rm (p,\gamma)$ capture on \nuk{Mg}{26}, \nuk{Ne}{20}, \nuk{Mg}{24}, \nuk{Al}{27}, 
\nuk{Mg}{25}, \nuk{Si}{30} plus the $\rm (p,n)$ reaction on \nuk{Al}{28}. Figure \ref{al26} shows the contribution of the explosive 
burning to the synthesis of \nuk{Al}{26} as filled red squares. To understand the trend of the explosive yield with the mass it is 
necessary to remind that the final Mass Radius (MR) relation at the moment of the core collapse plays a fundamental role for the 
synthesis of many nuclei, because it mainly determines the amount of matter exposed to the various explosive burnings 
\citep{cls00,cl02}. Such a relation strongly depends on the behavior of the C, Ne and O shell burnings all along the evolution of a 
star. The general rule is that the stronger is a shell burning, the slower is the contraction and hence the shallower is the final 
MR relation; it is also worth reminding that a stronger shell burning usually implies a wider (in mass) convective shell. Figure 
\ref{mara} shows the final MR relation for a subset of stars in the present grid. By the way, since the blast wave is essentially 
radiation dominated, the spatial location where the peak temperature drops to, say, 2.3 billion degrees occurs around 0.01-0.02 $\rm 
R_{\odot}$, the exact value depending (quite mildly) on the explosion energy. The Figure clearly shows that there is a non monotonic 
dependence on the initial mass: though a general trend exists, in the sense that the larger the mass the steeper this relation, two 
inversions exist: the first one between 17\msun and 20\msun and the second one between 30\msun and 35\msun. The first inversion 
marks the first mass (20\msun) forming a very extended last C convective shell that slows down the gravitational contraction of the 
core and hence halts the direct scaling between the compactness of the inner layers and the initial mass. The second inversion, 
occurring around the 35\msun model, marks the minimum mass that forms a single well settled C convective shell that lasts up to the 
explosion. Also in this case the formation of a single very active C convective shell counterbalances the increase in the 
gravitational contraction induced by the increase of the total mass. The explosive yields shown in Figure \ref{al26} follow almost 
exactly the general trend shown by the MR relation: there is a general direct scaling between the explosive \nuk{Al}{26} and the 
initial mass, but with two minima corresponding to the two masses quoted above.

The black dots in Figure \ref{al26} show the total yield of \nuk{Al}{26} as a function of the initial mass. It is quite evident that 
for the largest majority of the stars in the present mass range $\rm 11<M/M_{\odot}<120$ the explosive component is the major 
contributor to the total yield. Only around 15\msun the \nuk{Al}{26} produced in the advanced hydrostatic burnings dominates the 
final yield. The wind component to the total is always negligible up to a mass of, say, 60\msun. Above this mass, it contributes to 
roughly 30\% of the global outcome. Note, however, that though the explosive burning is the overall major producer of \nuk{Al}{26} 
within the full initial mass interval presently analyzed, the \nuk{Al}{26} produced by both the H burning and the C/Ne shell cannot 
be ignored in the computation of the final budget of \nuk{Al}{26} produced by a generation of massive stars.

There are many quite uncertain basic stellar parameters that may significantly alter the final \nuk{Al}{26} yields shown in Figure 
\ref{al26}. We have explored a few of them:

1) the adopted initial \nuk{Mg}{25} abundance (usually scaled solar) constitutes the main buffer from which the \nuk{Al}{26} is 
produced in MS and a change of its initial (assumed) abundance reflects linearly on the final amount of \nuk{Al}{26} ejected by the 
wind. In other words, a doubling of the initial \nuk{Mg}{25} leads to a doubling of the \nuk{Al}{26} ejected through the wind.

2) the size of the H convective core certainly plays a pivotal role since it affects both directly and indirectly the final 
\nuk{Al}{26} yield. The direct influence occurs in MS because it contributes to determine its abundance, by mass fraction, as well 
as its distribution within the star at the end of the core H burning; the indirect influence occurs through its major role in 
determining the He core mass, which is the main parameter driving all the further evolution of a star. In order to study the 
dependence of the \nuk{Al}{26} yield on the size of the H convective core we recomputed the full presupernova evolution plus the 
explosive burning of a 60 and a 120\msun by adopting 0.5 Hp of overshooting (instead of the basic 0.2 Hp) in core H burning. The 
main results are shown in the last two rows of Table \ref{taba} and Table \ref{tabb}. The first thing worth noting is that the 
increase of the size of the H convective core has two opposite effects on the two masses: the total \nuk{Al}{26} yield increases by 
18\% in the 60\msun while it remains roughly constant in the 120\msun. A closer look to the various components shows that the wind 
component reduces by 16\% in the 60\msun while it increases by 24\% in the 120\msun. The sum of the other two components shows, on 
the other hand, the opposite trend, i.e. it increases in the smaller mass and reduces in the larger one. Such results are the 
consequence of a complex interplay among several phenomena. First of all an increase of the convective core has two opposite 
effects: it increases the region where the initial \nuk{Mg}{25} is converted in \nuk{Al}{26} but it also leads to a lower 
\nuk{Al}{26} abundance by mass fraction at the core H exhaustion (because the H-burning lifetime increases). The final balance 
between these two opposite phenomena is not predictable a priori, but it happens that the first effect prevails on the second one so 
that at last more \nuk{Al}{26} (in solar masses) is produced overall. However, such an occurrence is not sufficient to determine how 
the \nuk{Al}{26} wind component depends on the size of the convective core because another crucial role is played by mass loss: a 
change of the size of the convective core strongly affects the path followed by a star in the HR diagram and hence the total amount 
of mass it loses. In the 60\msun case, the standard and test models lose similar amounts of mass in MS; however, while the standard 
model spends a consistent fraction of the central He burning phase as a Red Supergiant where it loses a very large amount of mass, 
the model with the larger convective core moves much earlier to the blue where the average mass loss rates are significantly lower. 
The final result is that the test model loses less mass, reaching the final collapse with a total mass of 19.93\msun, i.e. 2.8\msun 
more than the standard case. This explains why the test 60\msun ejects less \nuk{Al}{26} through the wind than its standard 
counterpart. In the 120\msun case the opposite happens: the test model loses much more mass than the standard model: in fact its 
final mass is 26.83\msun, i.e. 4\msun less than the standard model. In this specific case the luminosity increase in central H 
burning (due to the increase of the convective core) allows the model to enter (for a consistent amount of time) a region of the HR 
diagram where mass loss is very efficient \citep[see][]{val00,val01}. This explains why the amount of \nuk{Al}{26} present in the 
wind is larger in the test than in the standard model. As for the \nuk{Al}{26} produced by the advanced (C/Ne) burnings plus the 
explosion, we found that these components scale directly with the size of the convective core in the 60\msun and inversely in the 
120\msun. The reason is simply that the amount of \nuk{Al}{26} produced both in the advanced and in the Ne explosive burnings scale 
directly with the size of the He core mass, and Table \ref{taba} shows that this parameter increases with the size of the convective 
core in the 60\msun, while it decreases in the 120\msun case: this last (apparently) unexpected result is due to the strong increase 
of the mass loss rate in the 120 \msun stellar model computed with 0.5 Hp of overshooting.

\begin{figure} 
\begin{center}
\epsscale{1}
\plotone{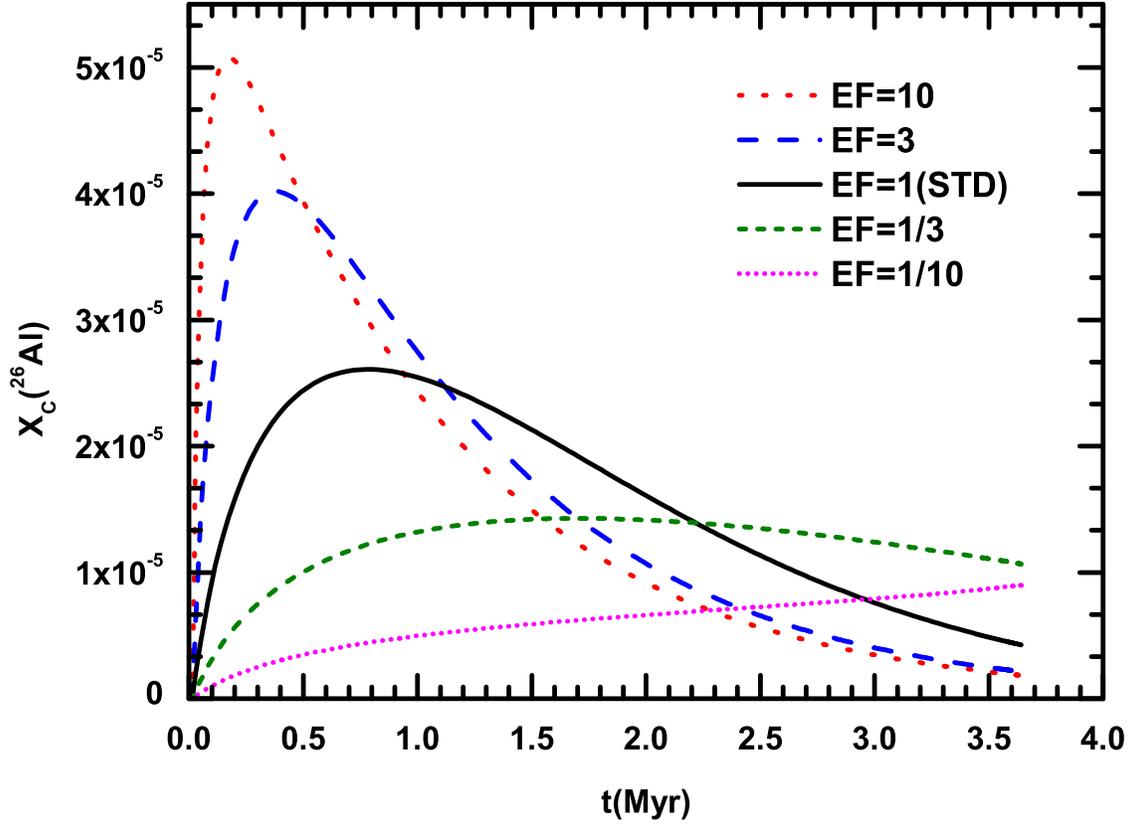}
\end{center}

\caption{Temporal variation of the central \nuk{Al}{26} mass fraction during core H burning in a 
60\msun. The {\rm black solid} line refers to the standard evolution while the other four lines show the cases 
in which the cross section of the \nuk{Mg}{25}$\rm (p,\gamma)$ reaction is multiplied by a factor EF. The 
line key is found in the Figure.}

\label{al26time}
\end{figure}

3) We have also explored the influence of the MG25PG cross section on the \nuk{Al}{26} produced by the H burning and expelled 
through the wind. Though the uncertainty quoted for this process \citep[e.g.][]{ili01} is of the order of a factor of 2 in the 
temperature range $0.02 \le T9 \le 0.2$, we preferred to explore the general theoretical dependence of the \nuk{Al}{26} (ejected by 
the wind) on the MG25PG cross section in a range of values wider than (presently) expected. Hence we have performed a few tests on 
both the 60\msun and the 120\msun by multiplying and dividing this cross section by a factor of 3 and 10. Table \ref{tabc} shows the 
results of these tests in the first four rows: the first column shows the cross section adopted in the test while the forth and 
fifth columns show the amount of \nuk{Al}{26} (in solar masses) present in the wind  for the 60\msun and the 120\msun star models, 
respectively. These tests show that a) the amount of \nuk{Al}{26} ejected by the wind does not scale monotonically with the 
efficiency of this cross section but, instead, that it firstly increases  as the cross section decreases and then it begins to 
decrease when the cross section drops below $1/3$ of the reference one, and b) the amount of \nuk{Al}{26} ejected by the wind 
depends only mildly on a systematic variation of the cross section, since it varies by less than a factor of two over a huge 
variation of the MG25PG cross section. To understand these results it is necessary to analyze in more details the synthesis of 
\nuk{Al}{26} in H burning. In the following we will discuss the behavior of the 60\msun but the same analysis does hold over the 
full mass range presently analyzed. Figure \ref{al26time} shows the variation of the abundance of the \nuk{Al}{26} in the convective 
core as a function of time for the five values of the cross section: the red sparse dotted line refers to the case in which the 
enhancement factor (EF) of the standard cross section is 10, the blue long dashed one to an EF=3, the black solid line to the 
standard case, the green short dashed line to an EF=1/3 while the magenta short dotted line refers to an EF=0.1. The test with the 
highest enhancement factor simulates the extreme case in which the \nuk{Mg}{25} is completely converted in \nuk{Al}{26} just at the 
beginning of the H burning. In this case the maximum \nuk{Al}{26} abundance is the highest but also the following decline is the 
steepest (it practically follows a pure decay) because no more \nuk{Mg}{25} is left to feed the \nuk{Al}{26} abundance at late 
times. As the enhancement factor reduces, the maximum equilibrium abundance of the \nuk{Al}{26} drops accordingly while the final 
\nuk{Al}{26} abundance progressively increases due to the larger availability of \nuk{Mg}{25} at late time. This trend, however, 
naturally comes to an end when the cross section becomes so slow that the timescale over which \nuk{Mg}{25} converts in \nuk{Al}{26} 
becomes longer than the H burning timescale. At this point the final abundance of \nuk{Al}{26} is mainly controlled by the total 
amount of \nuk{Mg}{25} destroyed: the lower the rate, the smaller the amount of \nuk{Mg}{25} converted in \nuk{Al}{26} and hence the 
smaller its final abundance. A cross section 10 times lower than our standard case is low enough to enter such a regime. By the way, 
it goes without saying that the physical evolution of the star does not depend at all on the rate of this process, so that all the 
timescales, the amount of mass lost, the penetration of the dredge up and so on remain rigorously identical in all these tests. 

4) Since most of the \nuk{Al}{26} ejected in the interstellar medium is anyway synthesized by the passage of the shock wave, we have 
also explored its sensitivity to the rates of the processes involved in its production in Ne explosive burning. Of course the number 
of factors that may influence the explosive production of \nuk{Al}{26} is very large and includes also all the factors that may 
affect the final mass-radius relation (as, e.g., the $\rm ^{12}C(\alpha,\gamma)^{16}O$); such a deep analysis goes well beyond the 
purposes of this paper. We have explored the influence of a few rates, namely the two main processes involved in the synthesis of 
\nuk{Al}{26}, i.e. the $\rm ^{24}Mg(n,\gamma)^{25}Mg$ and the $\rm ^{25}Mg(p,\gamma)^{26}Al$, as well as the two processes that 
control its destruction, i.e. the $\rm ^{26}Al(n,p)^{26}Mg$ and the $\rm ^{26}Al(n,\alpha)^{23}Na$. In particular we have recomputed 
three explosions for the two masses 25 and 60\msun, the first one doubling the cross sections of both the neutron captures on 
\nuk{Al}{26}, a second one doubling the n capture on \nuk{Mg}{24} and a last one by doubling the MG25PG. The results are shown in 
rows 8 to 11 in Table \ref{tabc}. Since the efficiency of any process depends on its rate (i.e. the cross section times the 
abundances of the involved nuclei) rather than just on its cross section, the tests shown in Table \ref{tabc} may also be 
reinterpreted as due to a change by a factor of two of any of the two nuclear species involved in the given process. This means that 
the tests concerning in particular the influence of the $\rm ^{24}Mg(n,\gamma)^{25}Mg$, the $\rm ^{26}Al(n,p)^{26}Mg$ and the $\rm 
^{26}Al(n,\alpha)^{23}Na$ on the explosive synthesys of the \nuk{Al}{26}, may also be seen as tests in which it is the neutron 
density to have been increased by a factor of two. Since a doubling of the $\rm ^{24}Mg(n,\gamma)^{25}Mg$ rate increase the 
explosive \nuk{Al}{26} by roughly 60\% while the simultaneous doubling of both the $\rm ^{26}Al(n,p)^{26}Mg$ and the $\rm 
^{26}Al(n,\alpha)^{23}Na$ rates reduces the \nuk{Al}{26} by roughly 40\%, it is clear that the two effects almost cancel out, so 
that we feel confident to conclude that a changing of the neutron density plays a minor role on the explosive synthesys of the 
\nuk{Al}{26}. Let us stress here, as a general warning, that the neutron and proton equilibrium abundances that enter in the various 
processes depend on the balance among a variety of similarly efficient processes, so that all of them should be carefully checked in 
order to understand if these equilibrium abundances are reliable or not.

Let us eventually explicitly mention other important sources of uncertainty that could affect the \nuk{Al}{26} yield:

a) the C abundance left by the He burning provides the main fuel that feeds both the C and the Ne burning (whose abundance derives 
directly from the C one) and hence it regulates the birth and the development of all the convective shells and hence, in turn, the 
final MR relation. Unfortunately, as it is well known, this value is not firmly established because it depends on the still very 
uncertain cross section of the \nuk{C}{12}($\alpha$,$\gamma$)\nuk{O}{16} process together with the adopted mixing scheme in central 
He burning. 

b) we have shown how critical the presence of a convective shell and the efficiency of mixing is to preserve the \nuk{Al}{26} 
produced in the C (Ne/C) burning shell; any change in their mass size and/or in the efficiency of mixing could significantly alter 
their contribution to the total yield (see also Weaver \& Woosley 1993).

It goes without saying that only the computation of a series of evolutionary tracks may really provide a quantitative estimate of 
these additional important uncertainties.

\begin{figure} 
\begin{center}
\epsscale{1}
\plotone{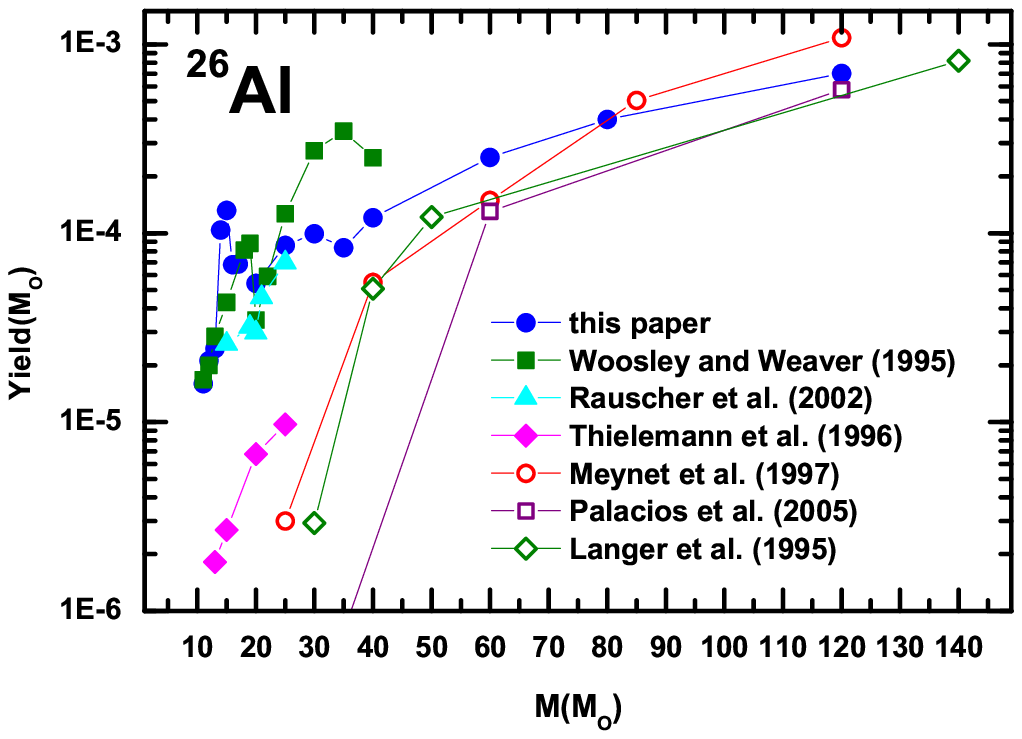}
\end{center}

\caption{Comparison among the \nuk{Al}{26} yields provided by various authors. The symbol key is found in the 
Figure.}

\label{conf1al26}
\end{figure}

Before closing this section let us compare our final yields with the other ones available in the literature. We already published 
yields for \nuk{Al}{26} in 2003 and 2004 \citep{lc03,cl04}; those yields (the \nuk{Al}{26} ones only), unfortunately, must be 
totally disregarded since, when we updated the full set of cross sections adopted in the evolutionary code, we found that the $\rm 
^{26}Al(p,\gamma)^{27}Si$ cross section present in the REACLIB database, and based on the \citet{ca88} formula, was wrong. Figure 
\ref{conf1al26} shows the \nuk{Al}{26} yields by WW95 (green filled squares), \citet{rhhw02} (cyan filled triangles), \citet{tnh96} 
(magenta filled rhombs), \citet{me97} (red open dots), \citet{pa05} (magenta open squares) and \citet{lbf95} (green open rhombs) 
together with the present ones (marked as "this paper" - blue filled dots). The first thing worth noting is that our set of models 
is the only one that fully extends over the full range of the massive stars. How we have already said above, for historical reasons 
two parallel fields of research may be identified: the first one (mainly involving Woosley and coauthors), addressed the full 
evolution of stellar models (including the computation of the passage of the shock wave) but did not extend the explored mass 
interval above 35-40\msun and a second one (that includes mainly European groups) involved in the study of the central H and He 
burning phases in a wide mass interval (up to 120\msun) by including the mass loss and hence fully exploring the properties of the 
WR stars. The WW95 \nuk{Al}{26} yields show remarkable similarities with ours: both sets of yields show two maxima, a first one 
around 15-20\msun and a second one around 30-35\msun, the WW95 ones being slightly shifted towards more massive stars. We already 
discussed above how these two peaks are related to the temporal evolution of the C convective shell, and hence these strong 
similarities clearly point towards a quite similar presupernova evolution. The quite larger WW95 yields between 25 and 40\msun are 
probably at least partially due to the contribution of the $\nu$ process that we neglect. WW95 quote the \nuk{Al}{26} yield for the 
25\msun without the $\nu$ contribution and the value closely matches our yield. The \citet{rhhw02} yields are an upgrade of the WW95 
for the solar metallicity; note that they do not show any more a strong peak around a 20\msun, which means that something in the 
properties of the C convective shell has changed with respect to the older WW95 stellar models. Anyway, we feel confident to say 
that these three sets of yields lead to quite similar predictions. An additional set of yields for this mass interval was provided 
by \citet{tnh96}: these yields are much lower than all the other ones probably because they come from the evolution of just pure He 
cores computed by adopting a quite poor network in the hydrostatic burnings. Historically, stars more massive than, say, 35\msun 
have been studied almost exclusively up to the end of the central He burning, so that only the \nuk{Al}{26} present in the wind has 
been (widely) discussed in the literature. The latest \nuk{Al}{26} yields in this mass interval are those provided by \citet{lbf95}, 
\citet{me97} and \citet{pa05}. Note that, for homogeneity reason, only the non rotating models are shown. Figure \ref{conf1al26} 
shows that the differences among these theoretical yields are confined within a factor of two.

\section{\nuk{Fe}{60}}

\nuk{Fe}{60} is an unstable nucleus (its terrestrial half life is $\rm \tau_{1/2}\simeq 1.5\times 10^6 y$) that lies slightly out of 
the stability valley, its closest stable neighbor being \nuk{Fe}{58}. It is mainly produced by neutron capture on the unstable 
nucleus \nuk{Fe}{59} and destroyed by the $\rm (n,\gamma)$ process. Since its closest parent, \nuk{Fe}{59}, is unstable, the 
\nuk{Fe}{59}$\rm (n,\gamma)$ process must compete with the \nuk{Fe}{59}$(\beta^{-})$ decay to produce an appreciable amount of 
\nuk{Fe}{60}. An order of magnitude estimate of the neutron densities needed to cross the \nuk{Fe}{59} bottleneck may be derived by 
equating the (n,$\gamma$) and $\beta^{-}$ decay rates: a value of the order of $\rm 3\times10^{\rm 10}~n/cm^3$ is obtained at 
temperatures lower than Log(T)=8.7. Above this temperature the half life of \nuk{Fe}{59} (whose terrestrial value is $\rm \simeq 44$ 
d) reduces systematically, dropping to $\rm \sim 6$ d at Log(T)=9.0 and to $\sim 1$ hour at Log(T)=9.6, while the \nuk{Fe}{59}$\rm 
(n,\gamma)$ cross section varies very mildly with the temperature. As a consequence, the neutron density needed to cross the 
\nuk{Fe}{59} bottleneck steeply increases with the temperature: it raises to $\rm \simeq3\times10^{\rm 11}~n/cm^3$ at Log(T)=9 and 
to $\rm \simeq6\times10^{\rm 12}~n/cm^3$ at Log(T)=9.3. Also the half life of \nuk{Fe}{60} has a steep dependence on the 
temperature: it remains of the order of the terrestrial value up to Log(T)=8.6 and then drops to half a year at Log(T)=9, to $\sim 
1.5$ d at Log(T)=9.2 and to $\sim 14$ m at Log(T)=9.4. In spite of this strong dependence of its half life on the temperature, and 
contrarily to what happens to \nuk{Al}{26}, \nuk{Fe}{60} is mainly destroyed by the (n,$\gamma$) reaction because the 
\nuk{Fe}{60}(n,$\gamma$) rate always overcomes the decay rate at the extremely high neutron densities required to produce it. A 
temperature of the order of 2 billion degrees represents an upper temperature limit for the synthesis of \nuk{Fe}{60} because above 
this temperature the $\rm (\gamma,n)$ and the $\rm (\gamma,p)$ photodisintegrations of both \nuk{Fe}{60} and \nuk{Fe}{59} become 
tremendously efficient. Such an occurrence limits a possible \nuk{Fe}{60} production to the He, C and Ne burning phases. 

\begin{figure} 
\begin{center}
\epsscale{1}
\plotone{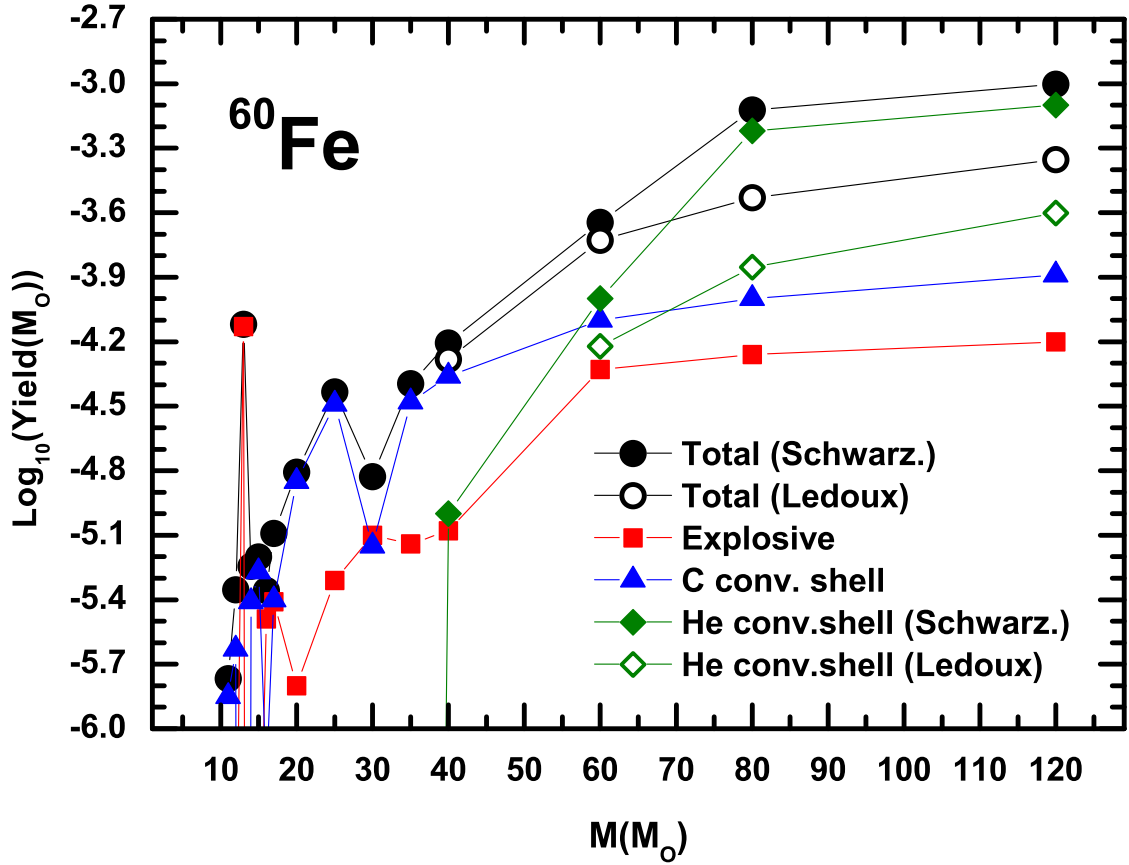}
\end{center}

\caption{\nuk{Fe}{60} yields as a function of the initial mass. The black filled and open dots refer to the 
total amount of \nuk{Fe}{60} ejected as a function of the initial mass, for two different choices of the 
stability criterion in the He convective shell, i.e. the Schwarzschild and the Ledoux criterions, 
respectively. The other lines show the various contributions to the total yield and the symbol key is found in the 
Figure.}

\label{fe60}
\end{figure}

In He burning, where the main neutron donor is the \nuk{Ne}{22}($\alpha$,n)\nuk{Mg}{25} process, a temperature of the order of 
$4\times10^8$ K would be required to reach the threshold neutron density of $\rm 3\times10^{\rm 10}~n/cm^3$. In central He burning 
the temperature remains well below $3\times10^8$ K so that the neutron density never exceeds $\rm 10^{\rm 7}~n/cm^3$ and no 
appreciable production of \nuk{Fe}{60} occurs. In shell He burning, vice versa, the temperature may raise up to and above 
$4\times10^8$ K so that even a large amount of \nuk{Fe}{60} may be synthesized. To understand its synthesis in this phase it is 
necessary to briefly remind a few firmly established characteristics of the shell He burning. At variance with the central H 
burning, where the mass size of the convective core shrinks together with the H content, central He burning is characterized by a 
convective core that advances progressively in mass (or remains stable at most) as the He is depleted. As a consequence, the He 
profile left by the central He burning shows a sharp discontinuity in correspondence to the location of the maximum size of the 
convective core and the He convective shell forms outside this discontinuity in a region where the He abundance is flat and equal to 
the one left by the H burning; since all the advanced burning phases are very rapid with respect to the shell He burning timescale, 
only a modest amount of He is burnt in the shell before the core collapse so that its final abundance barely drops below 0.9 by mass 
fraction. The temperature at the base of this convective shell never raises enough to make the \nuk{Fe}{59}$\rm (n,\gamma)$ process 
competitive with respect to the \nuk{Fe}{59}$(\beta^{-})$ decay so that no \nuk{Fe}{60} is synthesized by these stars. Such a 
feature holds up to the first mass that becomes a WR star of the WNE/WCO kind (40\msun). In fact, these stars experience such a 
strong mass loss that they firstly lose all their H rich envelope and then they continue eroding the He core up to the moment of the 
core collapse. Since the properties of the He burning depend on the He core mass size, the core "feels" a progressively smaller mass 
as the He core reduces and hence the mass size of the convective core reduces accordingly: a He profile is left in these stars at 
the central He exhaustion and its shape will depend on the balance between the speed of the central burning and the mass loss rate. 
The He convective shell forms, in these stars, within the region of variable He abundance and hence the well known problem arises if 
the Schwarzschild criterion or the Ledoux one must be used to determine if (and on which timescale) a convective region forms. We do 
not want to address such a question here, mainly because a full non local treatment of the mixing is still missing and only semi 
phenomenological (and parametric) solutions have been proposed over the years to treat these layers. We decided, vice versa, to 
explore the two limiting cases: the first one in which the Schwarzschild criterion is adopted  and a second one in which the Ledoux 
criterion is used. Columns 8 and 9 in Table \ref{tabb} show the amount of \nuk{Fe}{60} that is produced in the two limiting cases: 
it is evident that the presence of a fully convective region largely increases the amount of \nuk{Fe}{60} synthesized by the shell 
He burning. Typical neutron densities reached in the He shell in these very massive stars range between $\rm 6\times10^{\rm 
10}~n/cm^3$ and $\rm 1\times10^{\rm 12}~n/cm^3$. A rather modest, but worth being mentioned, additional amount of \nuk{Fe}{60} is 
also produced in the radiative tail of the He burning shell of the less massive stars. These stars experience a significant 
contraction of the core (including the radiative tail of the He shell) after the end of the central Oxygen burning phase that raises 
the temperature (and hence the neutron density) enough to allow the synthesis of some \nuk{Fe}{60} (column 7 in Table \ref{tabb}). 
As the initial mass increases, this phenomenon disappears because the region external to the CO core becomes progressively more 
insensitive to the evolution of the deep interior of the star (mainly because of the presence of a very active and stable C 
convective shell). The contribution of the He convective shell to the synthesis of \nuk{Fe}{60} is shown in Figure \ref{fe60} as 
green filled rhombs when the Schwarzschild criterion is adopted and as green empty rhombs when the Ledoux criterion is adopted.

C burning behaves similarly to the He burning. In central C burning the neutron density does not exceed a few times $\rm 10^{\rm 
7}~n/cm^3$, so that also in this case no \nuk{Fe}{60} may be produced. The main neutron donor is, once again, the 
\nuk{Ne}{22}($\alpha$,n)\nuk{Mg}{25} process and the reason why it can't provide a high neutron rate is, in this case, the 
relatively low concentration of $\alpha$ particles provided by the \nuk{C}{12}(\nuk{C}{12},$\alpha$)\nuk{Ne}{20} reaction. In 
analogy with the He burning, the larger burning temperature ($\rm T \ge 10^9~K$) at which the shell C burning occurs, allows a much 
larger production of $\alpha$ particles that translates into a much higher neutron density and hence in a conspicuous \nuk{Fe}{60} 
production. The typical neutron density we find ranges between $\rm 6\times10^{\rm 11}~n/cm^3$ and $\rm 2\times10^{\rm 12}~n/cm^3$ 
in the mass range 20 to 120 \msun while it drops significantly at lower masses. The mild dependence of the neutron density at the 
base of the C-convective shell on the initial mass (for M$>$17\msun) explains why the \nuk{Fe}{60} produced by the C convective 
shell (blue filled triangles in Figure \ref{fe60} and col. 10 in Table \ref{tabb}) increases only modestly with the initial mass. It 
must be noted, at this point, that the presence of a convective shell plays a crucial role also for the synthesis of \nuk{Fe}{60}. 
In fact it has the double responsibility of bringing new fuel ($\alpha$ particles and \nuk{Ne}{22}) in the region where the active 
burning occurs and simultaneously of bringing the freshly made \nuk{Fe}{60} outward in mass, i.e. at lower temperatures where the 
neutron density becomes negligible and its half life increases substantially. The 30\msun model is the best example of the 
importance of the convective burning: in fact, the minimum in the yield of the \nuk{Fe}{60} produced by the C shell (see Figure 
\ref{fe60}) corresponding to the 30\msun model is the direct consequence of the lack of an efficient C convective shell lasting all 
along the advanced evolutionary phases (see the previous section). 

Ne burning would produce \nuk{Fe}{60} too, because of the large abundance of the $\alpha$ particles and neutrons provided in this 
case by the \nuk{Na}{23}($\alpha$,n)\nuk{Al}{26} reaction, but the lack of an extended and stable convective shell lasting up to the 
moment of the explosion prevents the build up of a significant amount of \nuk{Fe}{60}. In fact, the \nuk{Fe}{60} produced by the 
advancing Ne burning shell drops quickly to zero together with the local Ne abundance because either the production channel dries up 
and the $\beta^-$ decay speeds up consistently. By the way, the Ne convective shell that forms in the 14 and 15\msun after the end 
of the central Si burning, and that is responsible of the strong peak in the \nuk{Al}{26} yield produced by these masses (see Figure 
\ref{al26}), does not lead to a similar effect for the \nuk{Fe}{60} because the neutron density remains too low to cross the 
\nuk{Fe}{59} bottleneck.

The last episode of synthesis of \nuk{Fe}{60} occurs when the blast wave crosses the mantle of the star on its way out, in the 
region where the peak temperature is of the order of $\rm 2.2\times10^9$ K and hence roughly in the same region where the 
(explosive) synthesis of \nuk{Al}{26} occurs. In most cases such a temperature is reached either at the base or within the C 
convective shell so that the amount of explosive \nuk{Fe}{60} produced will depend on the local abundances of \nuk{Ne}{20}, 
\nuk{C}{12}, \nuk{Na}{23}, \nuk{Ne}{22} left by the last C convective shell episode and on the final MR relation at the moment of 
the core collapse. The red filled squares in Figure \ref{fe60} and col. 11 in Table \ref{tabb} show the contribution of the 
explosive burning to the synthesis of \nuk{Fe}{60} as a function of the initial mass. Two local minima corresponding to the two 
inversions in the MR relation are quite evident also in this case (see the previous section). In the mass interval 11 to 15\msun, 
the explosive contribution to the synthesis of \nuk{Fe}{60} is almost negligible because of the low concentration of the nuclei that 
are necessary for its synthesis. The 13\msun constitutes, however, a striking exception because a large amount of \nuk{Fe}{60} is 
synthesized by the blast wave in this case. The reason is that the peak temperature of $\rm 2.2\times10^9$ K occurs beyond the outer 
border of the C convective shell where the abundances of \nuk{C}{12} and \nuk{Ne}{22}, in particular, are much higher than in the C 
convective shell. Since such a phenomenon occurs for just a single mass, one could question its reality. A close look at the models 
shows that, while the radius at which the blast wave synthesizes \nuk{Fe}{60} in these low mass stars remains roughly constant 
($\sim 0.013$\rsun), the radius of the outer border of the last C convective shell at the moment of the core collapse shows a non 
monotonic trend with the initial mass and, in particular, a minimum value of 0.012\rsun for the 13\msun occurs. Such a minimum 
corresponds to a local minimum in the mass size of the last C convective shell as well. It is not easy to understand why the last C 
convective shell has a minimum mass size for this mass because the temporal evolution and mass (and radial) extension of the various 
convective episodes that follow each other depend on a non linear interplay among various factors (the main ones being the mass 
sizes of both the He and the CO core and the amount of C left by the central He burning). To better assess the range of masses for 
which this peculiar explosive production occurs, the computation of a finer mass grid would be required: at present we can just say 
that it is confined somewhere between 12 and 14\msun. Let us eventually note that no \nuk{Fe}{60} is lost by the stellar wind 
because, in the present computations, no fraction of the He convective shell is lost through the wind during the advanced WR phase.

The black filled dots in Figure \ref{fe60} and col. 6 in Table \ref{tabb} show the total yield of \nuk{Fe}{60} as a function of the 
initial mass in the case in which the Schwarzschild criterion is assumed in the He convective shell. Its trend obviously reflects 
the trends of the main contributors at the various masses. There is a global direct scaling with the initial mass and a quite 
monotonic behavior: the two exceptions are the 13\msun and the 30\msun. Below the 40\msun the total yield is dominated by the 
contribution of the C convective shell  while above this mass it is the He convective shell to play the major role. The explosive 
burning almost always plays a minor role. The black open dots refer to the case in which the Ledoux criterion is assumed in the He 
convective shell: its contribution to the global synthesis of \nuk{Fe}{60} is strongly reduced in this case.

Similarly to the \nuk{Al}{26} case, the treatment of the convective layers and the cross section of the 
\nuk{C}{12}($\alpha$,$\gamma$)\nuk{O}{16} are strong sources of uncertainty in the prediction of the \nuk{Fe}{60} yield 
(see the previous section). The situation is even worse in this case because the cross sections of all the key processes involved in 
the synthesis of \nuk{Fe}{60} are purely theoretical: no experimental data exist for the \nuk{Fe}{59}(n,$\gamma$) and the 
\nuk{Fe}{60}(n,$\gamma$) rates as well as the dependence of their $\beta^-$ decay on the temperature. Let us eventually note that, 
since \nuk{Fe}{60} is mainly synthesized by the C and He shells, it may well be considered a pure secondary element because both the 
target (Fe) and the bullet (neutrons) depend on the initial metallicity (the neutrons, in fact, mainly come from the 
\nuk{Ne}{22}($\alpha$,n) process). Therefore also the rate of this process as well as the adopted initial [CNO/Fe] ratio (scaled 
solar or not) will significantly influence the final \nuk{Fe}{60} outcome .

\begin{figure} 
\begin{center}
\epsscale{1}
\plotone{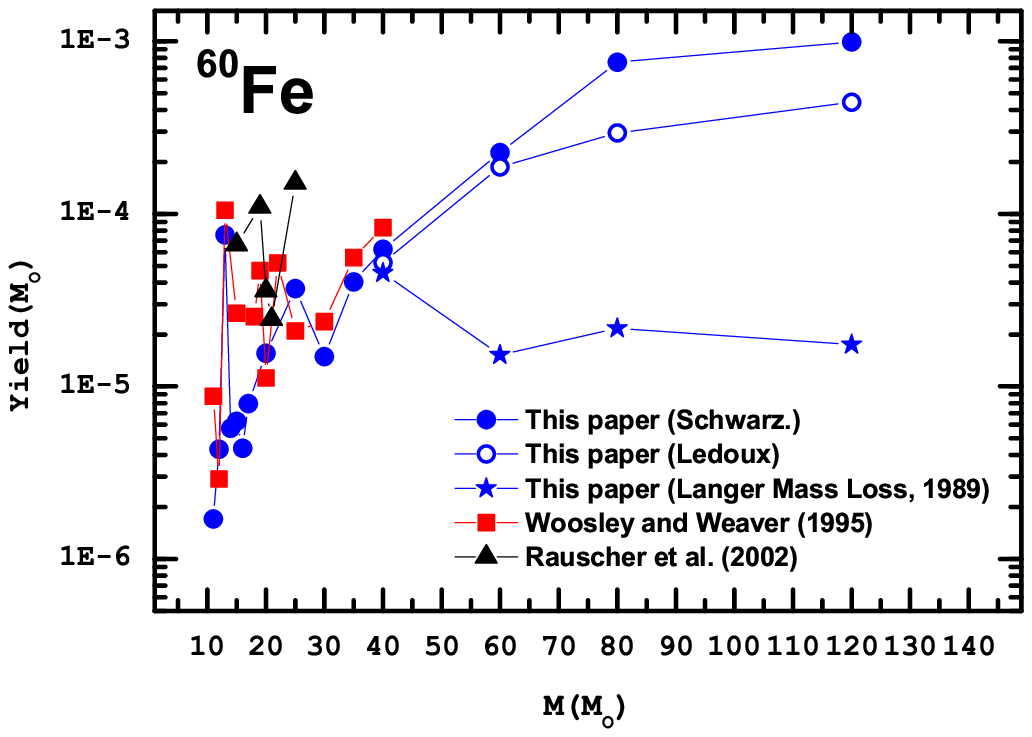}
\end{center}

\caption{Comparison among the \nuk{Fe}{60} yields provided by various authors. The symbol key is found in the 
Figure.}

\label{conf1fe60}
\end{figure}

\nuk{Fe}{60} yields have been published, as far as we know, by \citet{tim95}, (they are part of the WW95 large database) and by 
\citet{rhhw02} and are shown in Figure \ref{conf1fe60} as red filled squares and black filled triangles, respectively. Our yields 
obtained for the two extreme cases are shown as filled and open blue dots. For sake of completeness also the yields obtained by 
adopting the \citet{la89} mass loss rate are shown (see next section). The first thing worth noting is that our yields are the 
only ones available above 40\msun. Second, a good overall agreement exists between the WW95 and our yields: both sets show a peak 
towards the lower end (13\msun in both cases) and a minimum (20\msun in WW95 and 30\msun in our models) that clearly indicate how 
the two sets of models behave similarly (at least qualitatively). As for \citet{rhhw02}, their yields are systematically and 
significantly larger than both the WW95 and our yields. We do not have an definite explanation for such a difference since our 
models, computed with the latest input physics available today, show a remarkable agreement with the WW95 ones that have been 
computed with the input physics available at that time and not with the \citet{rhhw02} one that has probably been computed with a 
much closer input physics. Let us eventually remark that our \nuk{Fe}{60} yields already published in 2003 and 2004 
\citep{lc03,cl04} are in very close agreement with the present ones.

\section{Dependence of the yields on the Mass Loss rate adopted beyond the WNL phase}

\begin{figure} 
\begin{center}
\epsscale{1}
\plotone{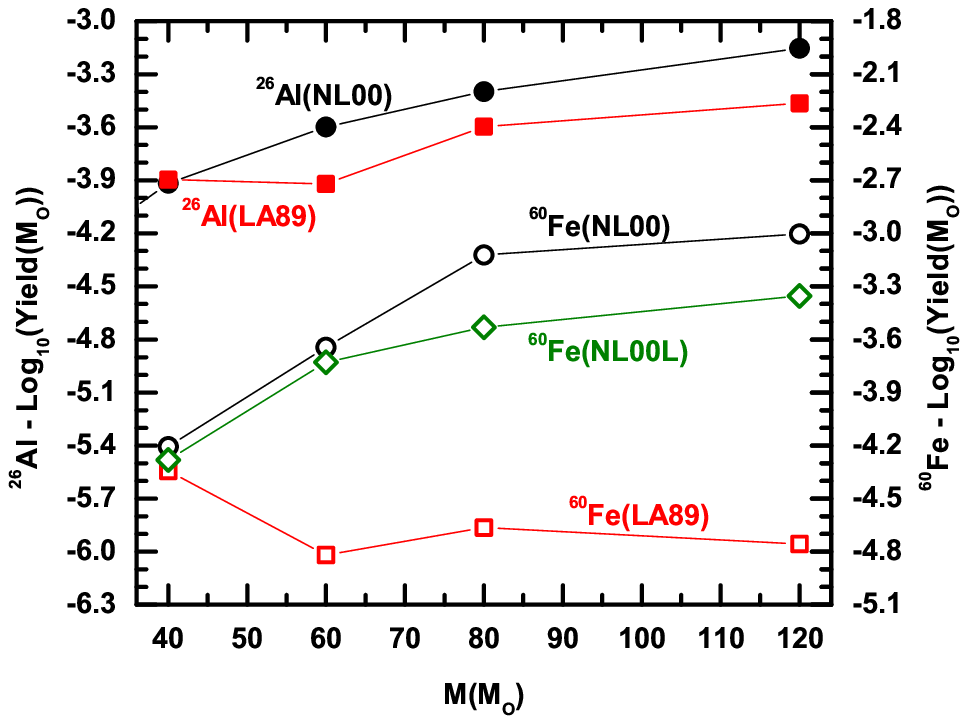}
\end{center}

\caption{\nuk{Al}{26} yields (left axis) as a function of the initial mass: the red filled squares refer to the models computed with 
the \citet{la89} mass loss rate (LA89) while the black filled dots refer to the models computed with the \citet{nl00} mass loss rate 
(NL00). \nuk{Fe}{60} yields (right axis) as a function of the initial mass: the red open squares refer to the models computed with 
the \citet{la89} mass loss rate (LA89), the black open dots and the green open rhombs refer to the models computed with the 
\citet{nl00} mass loss rate adopting the Schwarzschild (NL00) and the Ledoux (NL00L) criterions in the He convective shell 
respectively.}

\label{alfelanger}
\end{figure}

In the previous sections we have discussed at some extent the yields of both the \nuk{Al}{26} and \nuk{Fe}{60} as a function of the 
initial mass as well as their dependence on some nuclear processes and some physical phenomena (like the efficiency of the 
convective mixing). In this section we show how the yields of these two nuclei depend on the mass loss rate beyond the WNL phase. As 
it is well known, mass loss is still one of the main uncertain physical phenomenon that influence the evolution of a star. To 
explore the influence of the mass loss in the WNE+WCO phases we have recomputed the evolution of the stellar models that enter the 
WNE (i.e. stars in the range 40 to 120\msun) by adopting the mass loss rate proposed by \citet{la89} and widely adopted in the past. 
This older mass loss rate is much stronger than the \citet{nl00} one and it scales as $\rm M^{2.5}$, so that all the very massive 
models tend now to converge towards a similar quite small final mass (Column 2 in Table \ref{langer}). Moreover, since such a strong 
reduction of the total mass occurs during the central He burning, all these models tend to have a quite similar CO core mass (column 
3 in Table \ref{langer}) and hence quite similar advanced burning phases. This is the reason why the \nuk{Al}{26} and \nuk{Fe}{60} 
produced by the advanced burnings (hydrostatic plus explosive) tend to become quite insensitive to the initial mass (Columns 5 and 7 
in Table \ref{langer}). Of course the \nuk{Al}{26} lost by the wind (Column 4 in Table \ref{langer}) preserves, vice versa, a strong 
dependence on the initial mass because it is ejected in the earliest phases of the evolution of these stars. Note that the 
\nuk{Al}{26} ejected by the wind is slightly larger in the Langer case because a larger fraction of the He core is ejected before 
being reached by the He burning. The total amount of \nuk{Al}{26} obtained by adopting the \citet{la89} mass loss rate (Column 4 in 
Table \ref{langer}) is compared to the one obtained by adopting the \citet{nl00} mass loss rate in Figure \ref{alfelanger} (left 
axis). A similar comparison for the \nuk{Fe}{60} is shown in the same Figure (right axis). It is clear from these comparisons that 
while the \nuk{Al}{26} yield is reduced by roughly a factor of two, the \nuk{Fe}{60} one shows a dramatic dependence on the adopted 
mass loss rate. The reason is that the \nuk{Fe}{60} yield is dominated, in these very massive stellar models, by the He convective 
shell contribution (see the previous section) and the very strong mass loss implied by the Langer prescription completely kills such 
a contribution because it reduces enormously the final CO core mass. Hence, also the stability criterion adopted in the He 
convective shell does not play any significant role in this case.

\section{Comparison with the observations}

The clumpy and patchy distribution of the \nuk{Al}{26} in the galactic plane shown by COMPTEL, coupled to the strong correlation 
between its all-sky map and the 53 GHz free-free emission map, point towards a massive star parent for the \nuk{Al}{26} and, more 
specifically, towards the subset of massive stars that are also responsible for the Lyman continuum photons that power the 
ionization of the interstellar medium. By comparing the 53 GHz free-free emission to the \nuk{Al}{26} decay map, \citet{k99b} 
determined also that the scaling factor between the 1.8 MeV gamma ray line ($\gamma_{1.8}$) flux and the Lyman continuum photons 
(LCP) flux remains roughly constant all over the galactic plane. He quantified this occurrence by identifying the average amount of 
\nuk{Al}{26} ($\rm Y_{26}^{\rm O7 V}=10^{-4}$\msun) that must be associated to the LCP flux ($\rm Q_0$) of an equivalent O7 V star 
($\rm Log(Q_0)=49.05$) in order to obtain the observed correlation between the two aforementioned all-sky maps; see \citet{k99b} for 
details. For reasons of convenience we prefer to reinterpret these data in terms of number of $\gamma_{1.8}$ photons per ionizing 
photon (GPL); this number, that we name for convenience $\rm R_{\rm GxL}$, results to be equal to $1.25\times10^{- 11}$ GPL. Let us 
remind once again that the determination of $\rm R_{\rm GxL}$ does not require any explicit assumption about the sources of the 
ionizing photons and of the \nuk{Al}{26} but only that the \nuk{Al}{26} nuclei and the LCP share the same spatial distribution 
\citep{k99b}.

We have discussed in the previous sections essentially three sets of models that differ because of the mass loss rate adopted in the 
WNE+WCO phases and because of the stability criterion adopted in the He convective shell. For sake of clarity, the set of models 
computed by adopting the \citet{nl00} mass loss rate in the WNE+WCO phases and the Schwarzschild criterion in the He convective 
shell will be referred to as the NL00 models, the one computed with the same mass loss rate but the Ledoux criterion in the He 
convective shell will be referred to as the NL00L models while the set computed by adopting the \citet{la89} mass loss rate will be 
referred to as the LA89 models. Since all these different choices concern only stars more massive than 35\msun, all these sets share 
the same stellar models between 11 and 35\msun.

From the theoretical side we can easily determine the $\rm R_{\rm GxL}$ factor(s) for our grid of models. Let us remind that at the 
base of this derivation there is the usual assumption of a steady state condition, i.e. of the equivalence between destruction and 
production rates: this implies that the observed decay rate of the \nuk{Al}{26} is equal to the average ejection rate of this 
nucleus from the presently evolving stars. We have estimated the number of Lyman continuum photons produced by each of our models by 
means of table 3 of \citet{sk97}; the two following relations provide both the average $\rm Log(Q_0)$ and the lifetime of a star as 
a function of the initial mass (for our grid of models):

$$
Log_{10}(Q_0)=34.4906+14.90772 \times Log_{10}(M)-3.64592 \times Log_{10}(M)^2 ~~~~~ (LCP)~s^{-1}
$$
$$
Log_{10}(t)=9.598-2.879 \times Log_{10}(M)+0.6679 \times Log_{10}(M)^2~~~~~~~(yr)
$$

\begin{figure} 
\begin{center}
\epsscale{1}
\plotone{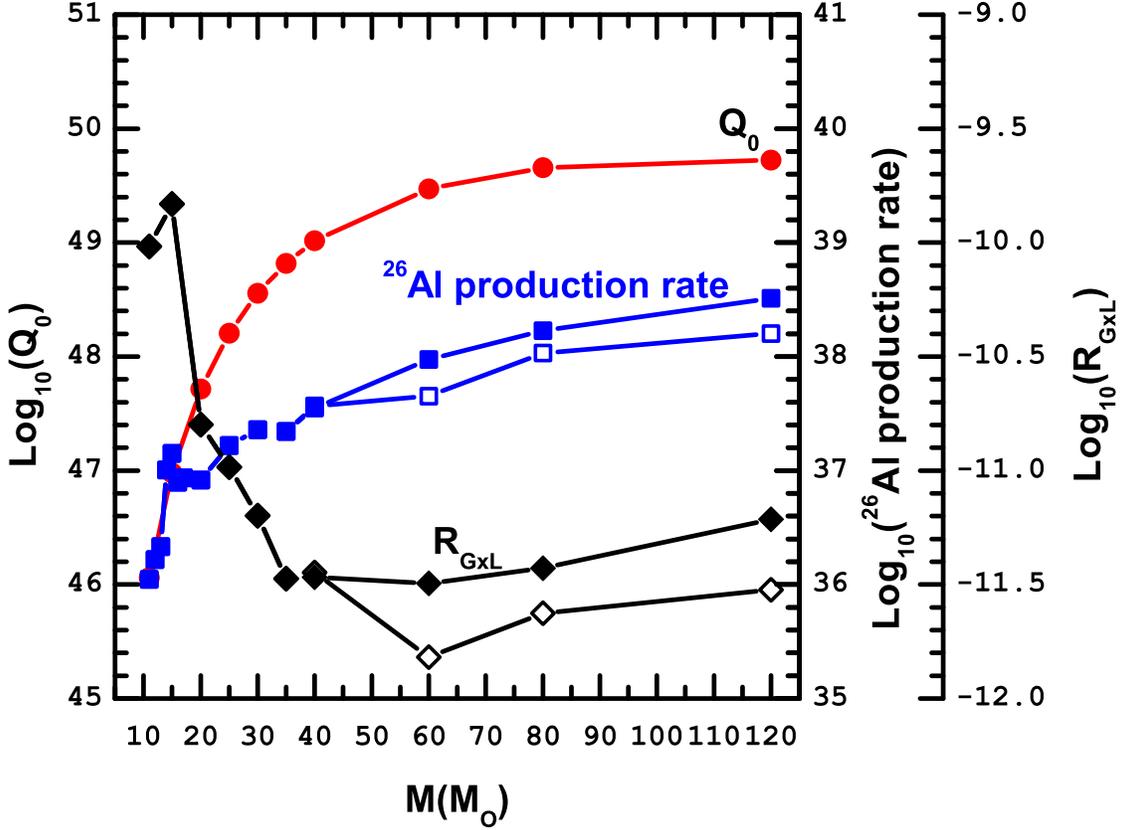}
\end{center}

\caption{Lyman continuum photons per second ($\rm Q_{0}$) (red filled dots, left axis) and average \nuk{Al}{26} production rate 
(blue filled and open squares for the NL00 and LA89 models respectively, first right axis) as a function of the initial mass. The 
rhombs refer to the number of $\gamma_{\rm 1.8}$ photons per ionizing photon ($\rm R_{\rm GxL}$, second right axis). In particular 
the black filled and open rhombs refer to the NL00 and LA89 models, respectively.}

\label{gplxm}
\end{figure}

Note that these relations are obviously identical for the three sets of models. Figure \ref{gplxm} shows the trend of both $\rm 
Log_{10}(Q_0)$ (red filled dots, left axis) and the average \nuk{Al}{26} production rate (first right axis) with the initial mass. 
The blue filled and open squares refer to the NL00 and the LA89 models, respectively. By the way, the average \nuk{Al}{26} 
production rate is simply the total amount of \nuk{Al}{26} produced by a star divided by its lifetime. The $\rm R_{\rm GxL}$ factor 
per each stellar mass is also shown in the same Figure and its scale is on the second right axis. The black filled and open rhombs 
refer, respectively, to the NL00 and LA89 cases. $\rm R_{\rm GxL}$ remains roughly constant above 30\msun because both $\rm 
Log_{10}(Q_0)$ and the \nuk{Al}{26} production rate run almost parallel; below this mass, on the contrary, the Lyman continuum 
photons drop down faster than the \nuk{Al}{26} production rate, so that $\rm R_{\rm GxL}$ increases significantly. Note that $\rm 
R_{\rm GxL}$ varies by roughly a factor of two between the NL00 and the LA89 stellar models. 

\begin{figure} 
\begin{center}
\epsscale{1}
\plotone{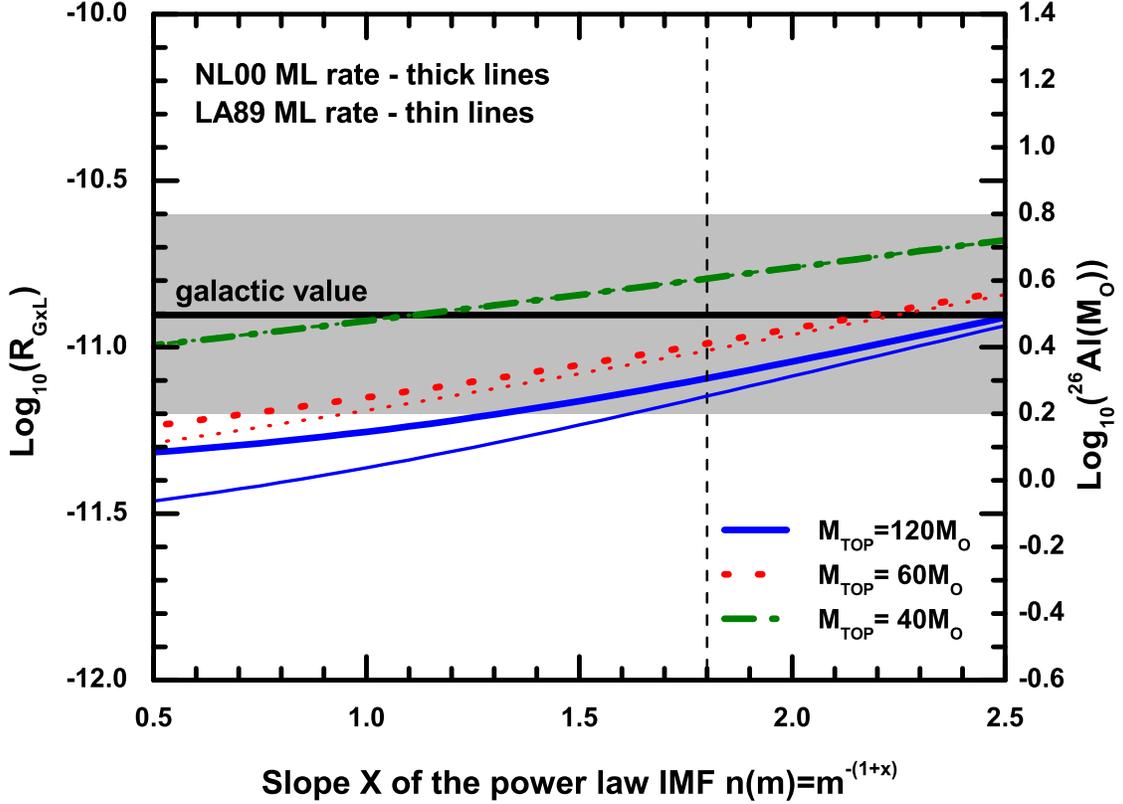}
\end{center}

\caption{$\rm R_{\rm GxL}$ as a function of the IMF slope x for three different values of the IMF upper mass limit $\rm M_{\rm 
TOP}$, namely 40 (green dashed lines), 60 (red dotted lines) and 120\msun (blue solid lines). The thick and thin lines refer to the 
NL00 and the LA89 models respectively. The corresponding total amount of $\rm ^{26}Al$ present in the Galaxy obtained by adopting 
$\rm Q_{\rm MW}=3.5 \times 10^{53} \rm ~ photons ~ s^{-1}$ (see text) is shown on the right axis. The horizontal thick solid black 
line marks the observed galactic value derived by \citet{k99b}, but expressed in terms of $\rm R_{\rm GxL}$, while the shaded area 
reflects the range of values that correspond to an uncertainty of a factor of two in the observed value. The vertical black dashed 
line marks a conservative representation of the actual IMF slope of massive stars, x=1.8, as reported by \citet{kw03} and 
\citet{k04}.}

\label{gplximf}
\end{figure}

The data shown in Figure \ref{gplxm} cannot be directly used to interpret the observed galactic value of $\rm R_{\rm GxL}$ because 
stars of different masses contribute simultaneously to both the LCP and the $\gamma_{1.8}$ fluxes, and hence a proper initial mass 
function (IMF) must be taken into account. The IMF is usually expressed as a power law, namely:
 
$$
n(m)~=~K~\times~m^{-(1+x)}
$$

where n(m) is the number of stars of mass m. The constant K is fixed by the condition:

$$
N_{TOT}~=~\int_{M_{BOT}}^{M_{TOP}}~K~\times m^{-(1+x)} dm = 1
$$

where $\rm M_{\rm BOT}$ and $\rm M_{\rm TOP}$ represent the limiting masses that bracket the range of the stars that explode as core 
collapse supernovae. As for the lowest mass that explodes as a core collapse supernova, we have adopted the smallest star in our 
grid, i.e. the 11\msun. One could question that even less massive stars do explode as core collapse supernovae, but this is 
practically uninfluent with respect to the computation of the $\rm R_{\rm GxL}$ parameter because below 11\msun both the LCP and the 
\nuk{Al}{26} drop practically to zero. On the other hand, the choice of the upper limit may play a role. Figure \ref{gplximf} shows 
the theoretical $\rm R_{\rm GxL}$ for a range of possible values of the power law slope, x, and three different values of the upper 
mass limit $\rm M_{\rm TOP}$. The blue solid, red dotted and green dashed lines refer to three different choices of the upper mass 
limit: 120, 60 and 40\msun respectively. The thick and thin lines refer to the NL00 and LA89 models respectively. It is quite 
evident that, fortunately, $\rm R_{\rm GxL}$ does not depend significantly on the adopted mass loss rate and also the dependence on 
$\rm M_{\rm TOP}$ is quite modest; in fact, even a drastic change of $\rm M_{\rm TOP}$ from, e.g., 120 to 40\msun affects $\rm 
R_{\rm GxL}$ by no more than a factor of two. The reason for such a low dependence of $\rm R_{\rm GxL}$ on $\rm M_{\rm TOP}$ is that 
above 30\msun the ratio between the \nuk{Al}{26} production rate and Lyman continuum luminosity remains roughly constant (see Figure 
\ref{gplxm}) for all sets of models.

The horizontal thick solid black line marks the observed galactic value derived by \citet{k99b}, but expressed in terms of $\rm 
R_{\rm GxL}$, while the shaded area reflects the range of values that correspond to an uncertainty of a factor of two in the 
observed value. \citet{kw03} and \citet{k04} have recently rediscussed the IMF slope for massive stars and came to the conclusion 
that it must be larger than, at least, x=1.8. If we take this specific value (marked by the thin vertical black dashed line in 
Figure \ref{gplximf}) as a conservative representation of the actual IMF slope of these stars, it turns out that a good fit to the 
galactic $\rm R_{\rm GxL}$ exists for any $\rm M_{\rm TOP}$ mass ranging between, at least, 120 and 40\msun and both sets of models. 
The fit improves for steeper IMF slopes. We therefore conclude that either the NL00 and the LA89 sets of models can naturally 
explain and reproduce the constant galactic $\rm R_{\rm GxL}$.

The good fit to this parameter, however, does not carry any information, by itself, about the total amount of \nuk{Al}{26} present 
in our Galaxy. In order to determine such an amount by means of $\rm R_{\rm GxL}$, it is necessary to know the total Lyman continuum 
luminosity provided by our Galaxy ($\rm Q_{\rm MW}$). Once this value is known, the global \nuk{Al}{26} production rate, $\rm P_{\rm 
MW}^{\rm ^{26}Al}$, is given by $\rm P_{\rm MW}^{\rm ^{26}Al}=Q_{\rm MW}\times R_{\rm GxL}$ and the present steady state abundance 
of \nuk{Al}{26} in our Galaxy would be given by $\rm P_{\rm MW}^{\rm ^{26}Al}\times \tau_{dec}^{^{26}Al}$.

\begin{figure} 
\begin{center}
\epsscale{1}
\plotone{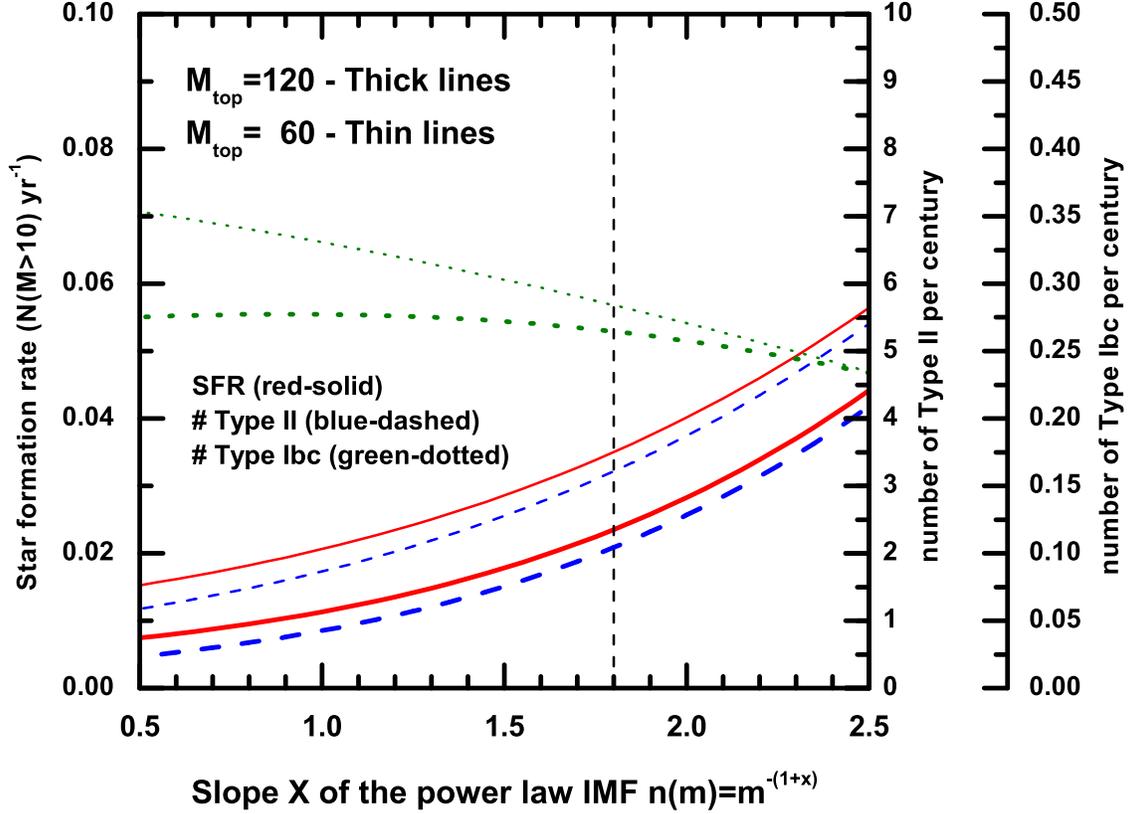}
\end{center}

\caption{Star formation rate (number of stars greater than 10\msun per year, left axis, red solid lines), 
number of Type II (first right axis, blue dashed lines) and Type Ibc (second right axis, green dotted lines) 
supernovae per century as a function of the IMF slope x for two choices of the upper mass limit $\rm M_{\rm 
TOP}$ of the IMF, namely 60\msun (thin lines) and 120\msun (thick lines). A total galactic Lyman continuum 
luminosity $\rm Q_{\rm MW}=3.5 \times 10^{53} \rm ~ photons ~ s^{-1}$ has been assumed (see text). The 
vertical black dashed line has the same meaning as in Figure \ref{gplximf}.}

\label{sfr}
\end{figure}

\begin{figure} 
\begin{center}
\epsscale{1}
\plotone{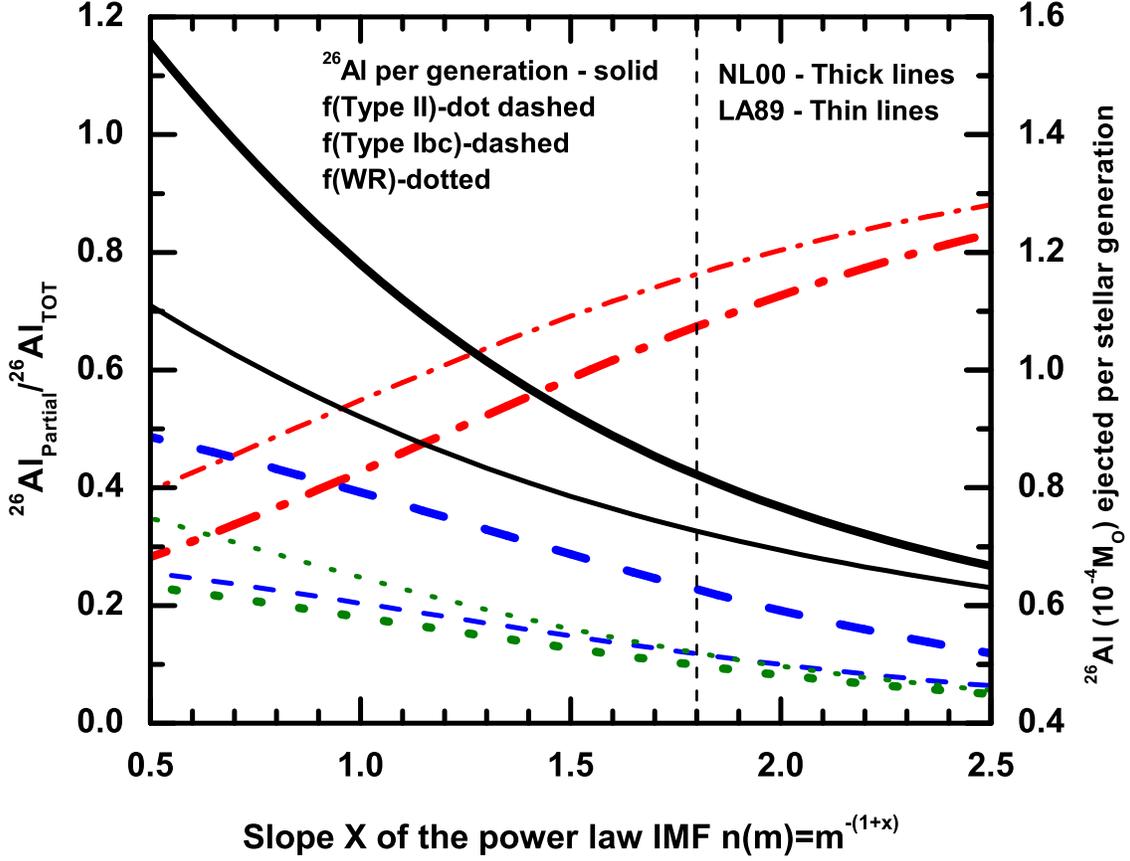}
\end{center}

\caption{Total amount of \nuk{Al}{26} ejected by a generation of massive stars in the range 11-120\msun (black solid lines, right 
axis) as a function of the slope x of the IMF. Percentage contribution of the Type II supernovae (red dot dashed lines), the Type 
Ibc supernovae (blue dashed lines) and the Wolf-Rayet stars (green dotted lines) to the synthesis of $\rm ^{26}Al$ as a function of 
the IMF slope x. The thick and thin lines refer to the NL00 and LA89 models respectively. The vertical black dashed line has the 
same meaning as in Figure \ref{gplximf}.}

\label{fracal26}
\end{figure}

\begin{figure} 
\begin{center}
\epsscale{1}
\plotone{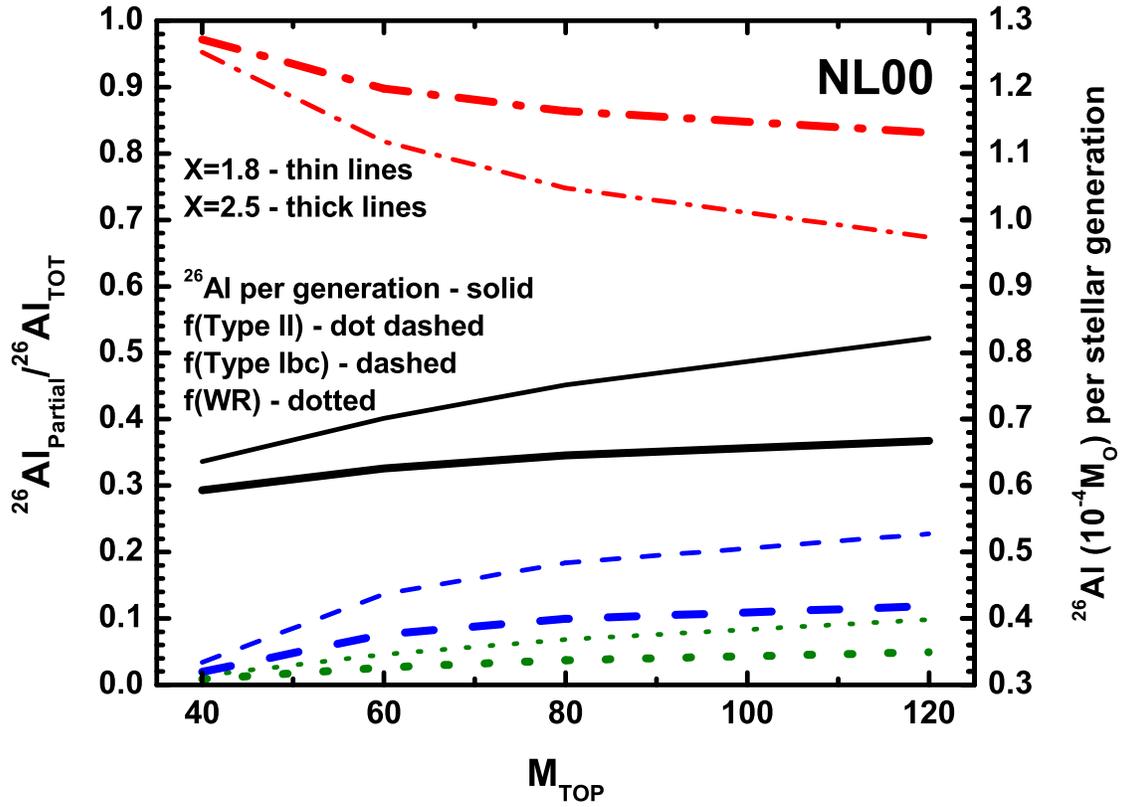}
\end{center}

\caption{Same quantities shown in Figure 12 as a function of $\rm M_{\rm TOP}$ for the NL00 models and two specific slopes of the 
IMF, namely x=1.8 and x=2.5.}

\label{al26mtopnl}
\end{figure}

\begin{figure} 
\begin{center}
\epsscale{1}
\plotone{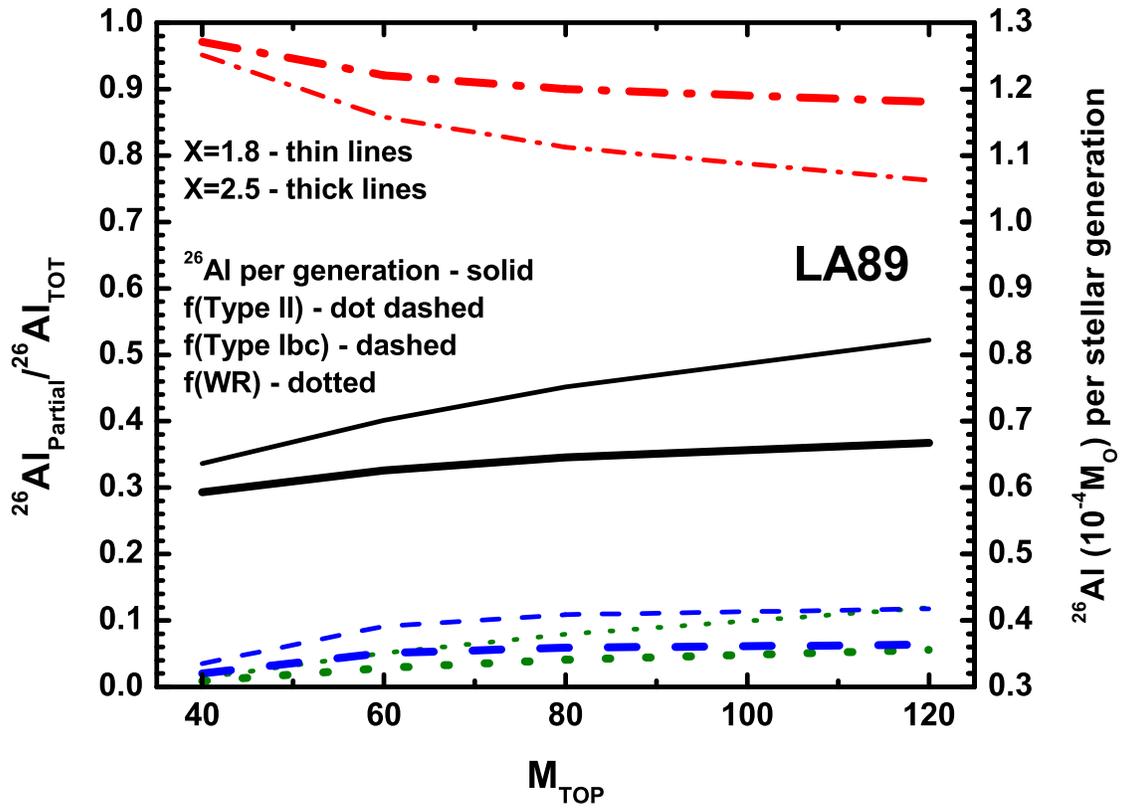}
\end{center}

\caption{Same as Figure 13 but for the LA89 models.}

\label{al26mtopla}
\end{figure}

Recent determinations of the galactic Lyman continuum luminosity range between $\rm Q_{\rm MW}=3.5 \times 10^{53} \rm ~ photons ~ 
s^{-1}$ \citep{ben94} and $\rm Q_{\rm MW}=2.6 \times 10^{53} \rm ~ photons ~ s^{-1}$ \citep{mw97} with about 50\% uncertainty in 
both cases (see the cited papers for details). If one would adopt the highest value \citep{ben94}, an IMF slope x=1.8 and $\rm 
M_{\rm TOP}=120$\msun, the present steady state abundance of \nuk{Al}{26} in our Galaxy would amount to 1.97\msun for the NL00 
models and to 1.70\msun for the LA89 case. Figures \ref{gplximf} (right scale) and \ref{sfr} show the trends of the \nuk{Al}{26} 
abundance, of the SFR, of the number of Type II supernovae and of the Type Ibc supernovae as a function of the slope x for a $\rm 
Q_{\rm MW}=3.5 \times 10^{53} \rm ~ photons ~ s^{- 1}$. In Figure \ref{sfr}, the thick lines refer to an $\rm M_{\rm TOP}=120$\msun 
while the thin lines refer to an $\rm M_{\rm TOP}=60$\msun. Since all these relations scale linearly with $\rm Q_{\rm MW}$, the 
adoption of the value provided by \citep{mw97} would simply imply a downward scaling by a factor equal to $2.6/3.5$. The quantities 
shown in Figure \ref{sfr} obviously do not depend at all neither on the adopted mass loss beyond the WNL phase nor on the stability 
criterion adopted in the He convective shell. Figure \ref{fracal26} shows the total amount of \nuk{Al}{26} ejected by a stellar 
generation (solid lines - right axis) together with the percentage contributions of the Type II supernovae (dot dashed lines), the 
Type Ibc supernovae (dashed lines) and the WR stars (dotted lines), all as a function of the IMF slope x: the thick and thin lines 
refer to the NL00  and the LA89 cases, respectively. The observed trends may be readily understood by reminding that a) the LA89 
models predict sistematically less \nuk{Al}{26} than the NL00 ones (see Table \ref{langer}) because of the reduced production both 
in the C convective shell and Ne explosive burnings (see the previous section) and b) the total amount of \nuk{Al}{26} ejected by 
the Type II supernovae is identical in both sets because they share the same stellar models below 40\msun. Since the integration 
over the IMF may depend on the adopted upper mass limit ($\rm M_{\rm TOP}$), we show in Figures \ref{al26mtopnl} (for the NL00 case) 
and \ref{al26mtopla} (for the LA89 case) the same quantities shown in Figure \ref{fracal26} as a function of $\rm M_{\rm TOP}$ and 
two specific IMF slopes, i.e. x=1.8 and x=2.5. Both these Figures clearly show that, indipendently on $\rm M_{\rm TOP}$ and on the 
mass loss rate adopted beyond the WNL stage, the largest contribution  to the total \nuk{Al}{26} produced by a stellar generation 
comes from the Type II supernovae, while both the WR and the Type Ibc contribute at most (complessively) for a 30\% in the most 
favoreable case. The present result, i.e. that most of the \nuk{Al}{26} comes from the Type II supernovae must be interpreted as a 
property of our sets of models. We have already shown in Figure \ref{conf1al26} that the yields produced by different authors differ 
even significantly in the mass range they share; it is therefore very probable that they could have been different also in the mass 
interval not explored by other groups so that even the relative contributions of the various mass intervals could change 
significantly from one author to another. In other words, it would be a mistake to use a set of models that extend only up to 30-
40\msun to infer the \nuk{Al}{26} produced by a generation of massive stars basing such a choice on the properties of the present 
models. It is not known a priori the role that would play the more massive stars within other sets of models. In this respect it is 
of overwhelming importance that also other groups compute grid of models extended in mass at least as much as in the present ones, 
because only a comparison between indipendent computations would allow a better understanding of the role really played by the 
various mass intervals to the global budget of the \nuk{Al}{26}. It goes without saying that the same holds for the \nuk{Fe}{60}.

\begin{figure} 
\begin{center}
\epsscale{1}
\plotone{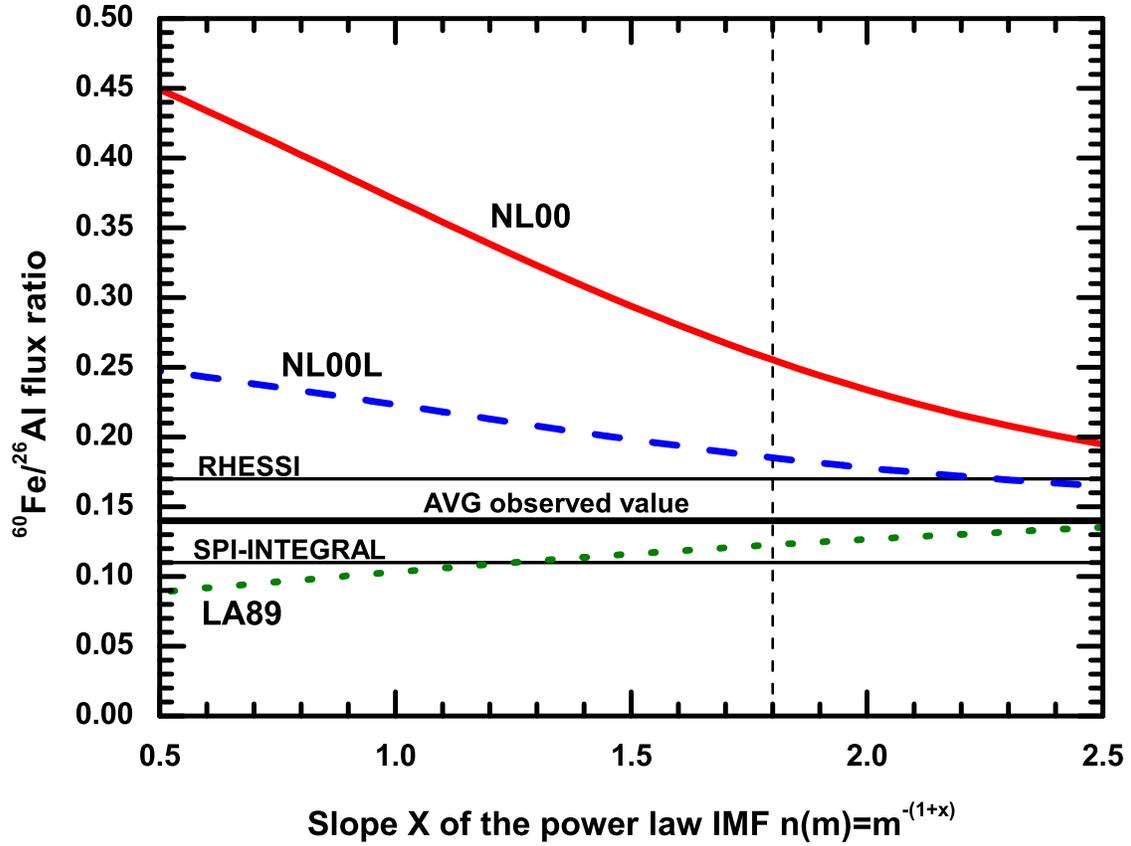}
\end{center}

\caption{Theoretical $\rm ^{60}Fe/^{26}Al$ $\gamma$-ray line flux ratio as a function of the IMF slope x, for the three sets of 
models, namely NL00 (red solid line), NL00L (blue dashed line) and LA89 (green dotted line). The thick horizontal black solid line 
shows the value averaged between the detections reported by RHESSI and INTEGRAL/SPI (thin black solid lines) and taken as a 
representative value for both the experiments.}

\label{fealsux}
\end{figure}

\begin{figure} 
\begin{center}
\epsscale{1}
\plotone{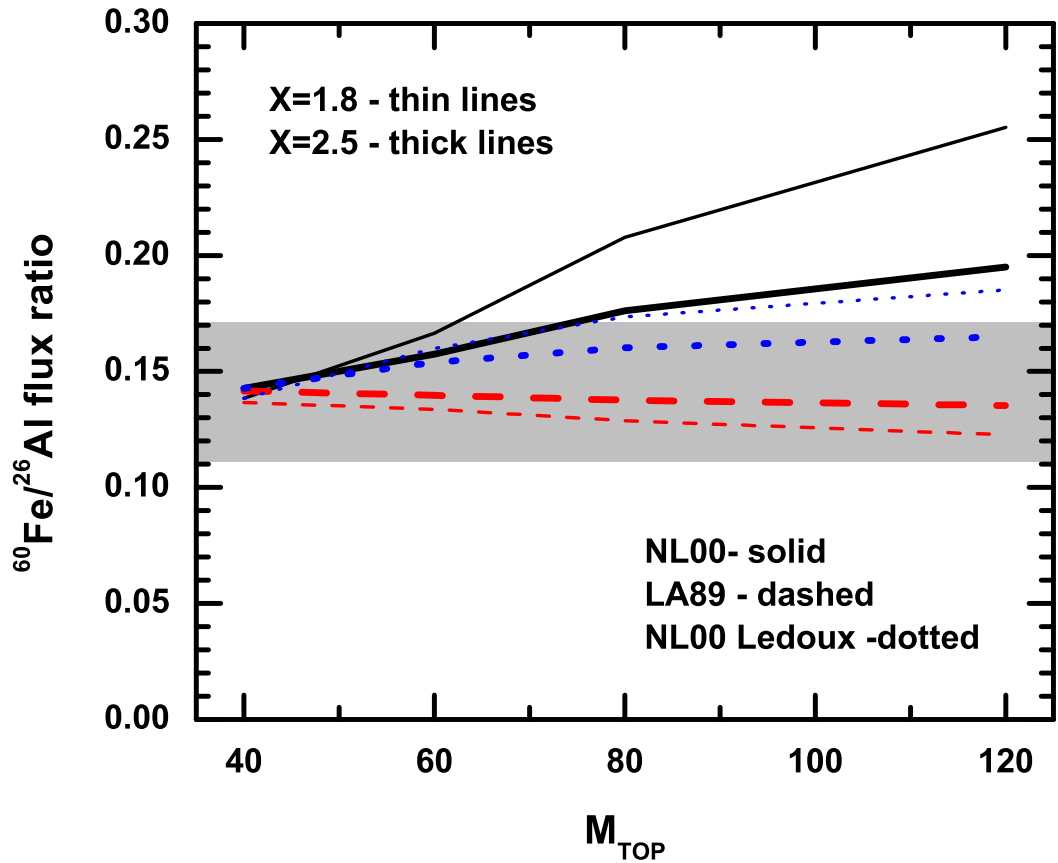}
\end{center}

\caption{Theoretical $\rm ^{60}Fe/^{26}Al$ $\gamma$-ray line flux ratio as a function of $\rm M_{\rm TOP}$ for two selected IMF 
slopes, namely x=1.8 (thin lines) and x=2.5 (thick lines). The black solid, the blue dotted and the red dashed lines refer to the 
NL00, NL00L and LA89 models, respectively.}

\label{fealsumtop}
\end{figure}

Both RHESSI and SPI (INTEGRAL) have reported a measurement of the \nuk{Fe}{60}$/$\nuk{Al}{26} flux ratio towards the central radiant 
of our Galaxy (see the introduction). The latest values, at the time of writing, are $(0.17\pm 0.05)$ \citep[RHESSI,][]{smi05} and 
$(0.11\pm 0.03)$ \citep[SPI,][]{har05}. Since the two experimental data show a compatibility range around 
\nuk{Fe}{60}$/$\nuk{Al}{26}=0.14, we decided to tentatively (and arbitrarily) adopt this value as representative of both the 
experiments and to consider as a typical error the semi difference between the quoted values. This representative value is shown as 
a horizontal black thick line in Figure \ref{fealsux}, while the two experimental values are shown as thin black solid lines in the 
same Figure. Figure \ref{fealsux} also shows our theoretical predictions for the three sets of models, namely the NL00 (red solid 
line), the NL00L (blue dashed line) and the LA89 (green dotted line) and $\rm M_{\rm TOP}$=120\msun. This Figure shows that the NL00 
models predict a too large \nuk{Fe}{60}$/$\nuk{Al}{26} flux ratio for any slope in the explored range. Such a large flux ratio is 
due to the very large amount of \nuk{Fe}{60} produced in the He convective shell in stars more massive than 35\msun. The adoption of 
the Ledoux criterion in the He convective shell significantly dumps out the \nuk{Fe}{60} production in this region so that the 
theoretical flux ratio reduces somewhat though it remains larger than the observed value. The fit improves as the IMF slope 
increases because the contribution of the more massive stars reduces correspondingly. The LA89 models, vice versa, predict an 
\nuk{Fe}{60}$/$\nuk{Al}{26} flux ratio in excellent agreement with the observed value over a very large range of IMF slopes, and 
this is the consequence of the very strong mass loss that prevents either the production of \nuk{Fe}{60} in the He convective shell 
and reduces also its production in the C convective shell and in the Ne explosive burning. Figure \ref{fealsumtop} shows, for two 
selected IMF slopes, the dependence of this flux ratio on $\rm M_{\rm TOP}$. Only the NL00 models show a strong dependence of the 
flux ratio on $\rm M_{\rm TOP}$ and this is due to the fact that this is the only set for which a substantial contribution to the 
\nuk{Fe}{60} comes from the stars more massive than 40\msun. 

\begin{figure} 
\begin{center}
\epsscale{1}
\plotone{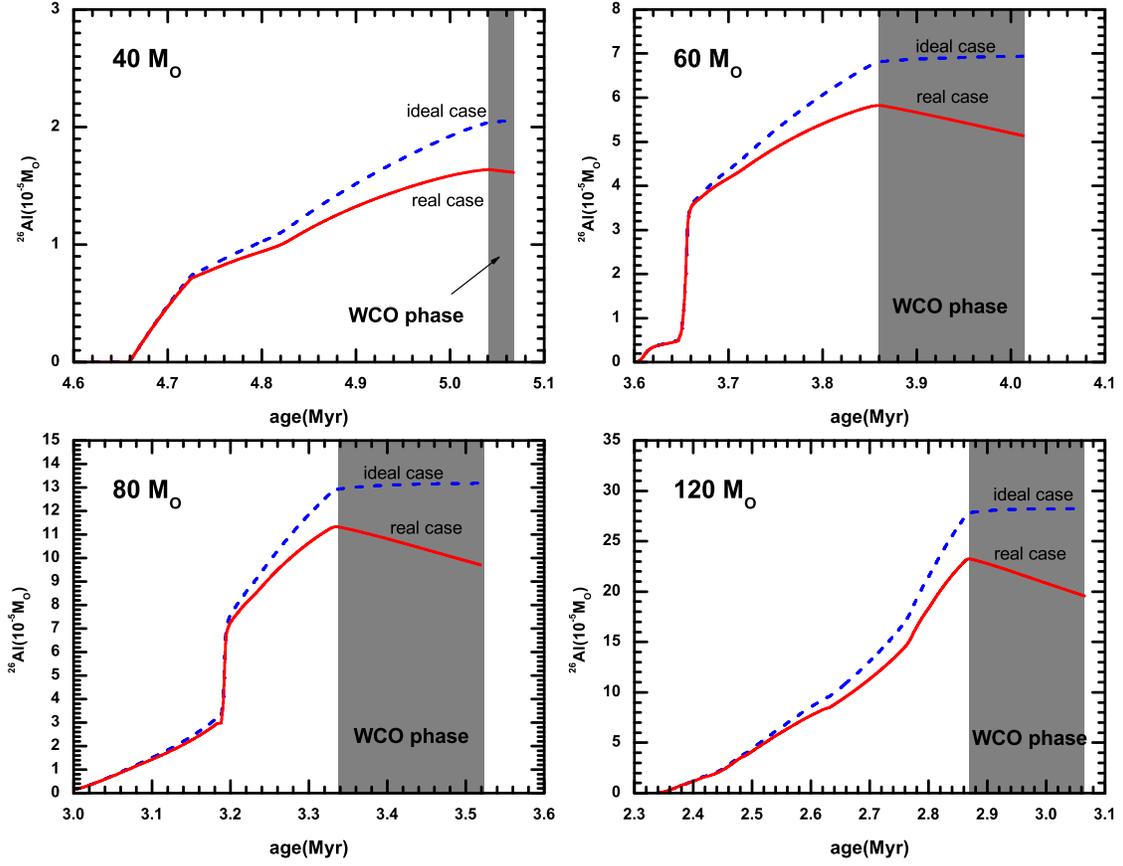}
\end{center}

\caption{Cumulative abundance of \nuk{Al}{26} present in the ejecta as a function of the age of the star. 
The four panels refer to the 40 (upper left), 60 (upper right), 80 (lower left) and 120\msun (lower right). 
The blue dashed thick lines show the case in which the \nuk{Al}{26} is assumed to accumulate in the 
interstellar medium without decaying; the red solid thick lines represent the real total amount of 
\nuk{Al}{26} one would find in the interstellar medium as a function of time, taking into account that the 
\nuk{Al}{26} starts to decay as soon as it is ejected into the interstellar medium. The gray area mark the 
temporal phase in which each star appears as a WCO star.}

\label{al26vt}
\end{figure}

A paper devoted to the presentation of a new generation of \nuk{Al}{26} yields cannot escape the temptation to reanalyze the fit to 
$\gamma^2$ Velorum, the closest WR star (WR11) presently known. As far as we know, the latest very detailed and comprehensive 
analysis of this WR star has been presented by \citet{ober00} and we refer the reader to this paper and to the references therein 
for a detailed discussion of both the observational aspects and the theoretical problems involved in the interpretation of this star 
(either in the single and binary stars scenarios). Here it suffices to say that the typical initial mass quoted for this star is 
$60\pm15$\msun, that its initial metallicity was very probably solar and the binary system is wide enough that probably the 
evolution of $\gamma^2$ Velorum was not affected much by the binarity environment. No \nuk{Al}{26} has been detected around this 
star but an upper limit of $6.3^{+2.1}_{- 1.4}\times10^{- 5}$\msun has been derived by \citet{ober00}. This value represents a 
problem because the theoretical models predict a much larger amount of \nuk{Al}{26}. \citet{me97} find that a 60\msun solar 
metallicity star ejects $1.49\times10^{-4}$\msun of \nuk{Al}{26} through the wind while \citet{pa05} quote a value of 
$1.3\times10^{- 4}$\msun for the same initial mass; also the older models by \citet{lbf95} predict an amount of \nuk{Al}{26} equal 
to $1.2\times10^{-4}$\msun for a 50\msun star. Also for a mass as low as 40\msun the models predict rather large values that are 
only barely compatible with the quoted upper limit: $5.5\times10^{-5}$\msun \citep{me97} and $5.09\times10^{-5}$\msun \citep{lbf95}. 
Such a failure in reproducing the closest WR star is very embarrassing because this is a member of the class of stars that are very 
probably responsible for the production of the bulk of the \nuk{Al}{26} presently in the disk of our Galaxy. An inspection to Table 
\ref{tabb} shows that the wind component of the \nuk{Al}{26} yield in our models is roughly a factor of two lower than that obtained 
by the other authors, so that a large part of the discrepancy is removed. In addition to this, there is another important point that 
has been neglected up to now: the timescale over which the \nuk{Al}{26} is injected into the interstellar medium. Figure 
\ref{al26vt} shows the cumulative abundance of \nuk{Al}{26} present in the ejecta as a function of the age of the star and the four 
panels refer to the 40, 60, 80 and 120\msun: the blue dashed thick lines show the case in which the \nuk{Al}{26} is assumed to 
accumulate in the interstellar medium without decaying (unrealistic, but usually adopted, ideal case in which the \nuk{Al}{26} is 
assumed to start decaying after it has been completely ejected) while the red solid thick lines represent the real total amount of 
\nuk{Al}{26} one would find in the interstellar medium as a function of time (taking into account the fact that the \nuk{Al}{26} 
starts to decay as soon as it is ejected in the interstellar medium). The gray area mark the temporal phase in which each star 
appears as a WCO star. If $\gamma^2$ Velorum is now just at the beginning of its WCO phase (the worst case because the {\it real} 
\nuk{Al}{26} abundance is at its maximum) the predicted amount of \nuk{Al}{26} would be 15 to 20\% smaller than the value shown in 
Table \ref{tabb} for masses in the range 40 to 120\msun. In particular, an abundance of $5.82\times10^{-5}$\msun is predicted in the 
interstellar medium in the case of a 60\msun model, a value well below the upper limit quoted at present. This means that at least 
the full range of masses between 40\msun and 60\msun is now compatible with the observed upper limit. There is even room for an 
initial mass larger than 60\msun. We can't be more precise at the moment because our nearest grid mass is the 80\msun. The adoption 
of the LA89 models would not change these results appreciably. Hence we conclude that a proper treatment of the injection of the 
\nuk{Al}{26} ejected into the interstellar medium coupled to our new yields removes the longstanding discrepancy between the upper 
limit quoted for $\gamma^2$ Velorum and the theoretical predictions. 

\section{Conclusions}

We have extensively discussed the production sites of the two gamma ray emitters \nuk{Al}{26} and \nuk{Fe}{60} over the range of 
massive stars that may contribute significantly to either the production of these nuclei and to the Lyman continuum luminosity, i.e. 
the range 11 to 120\msun. These theoretical predictions fully cover all the evolutionary phases from the pre main sequence to the 
explosive burnings. At variance with current ideas, \nuk{Al}{26} is mainly produced by the Ne/C explosive burning over most of the 
mass interval analyzed. The main production site of the \nuk{Fe}{60}, vice versa, strongly depends on the adopted mass loss rate. In 
the LA89 case the main \nuk{Fe}{60} producer is always the C convective shell while in the NL00 case the main producer is still the 
C convective shell for masses lower than 40\msun while above this mass a strong contribution comes from the He convective shell. 
Since the He convective shell forms, in these stars, in a region where a gradient in the He abundance is present, its contribution 
to the synthesys of \nuk{Fe}{60} depends significantly on the adopted stability criterion (Schwarzschild or Ledoux). We have used 
these yields to address the problem of the diffuse abundances of \nuk{Al}{26} and \nuk{Fe}{60} in the Galaxy.

The discovery of an almost constant flux ratio all over the galactic disk between the Lyman continuum photons (derived from the 53 
GHz free-free emission all-sky map) and the $\gamma_{1.8}$ MeV line (that we express as $\rm R_{\rm GxL}$, number of $\gamma_{1.8}$ 
MeV photons Per Lyman continuum photon -GPL), plus the recently determined \nuk{Fe}{60}$/$\nuk{Al}{26} flux ratio towards the inner 
radiant of the Galaxy, constitute in our opinion the two key experimental data a set of models must satisfy. The reason is that, 
once they are assumed to be equally spatially distributed, their ratio is independent on their spatial distribution. In addition to 
that, also the highly uncertain SFR fortunately does not play any role in these ratios (once a steady state is assumed). The 
experimental value of the $\rm R_{\rm GxL}$ parameter has been determined by \citet{k99b} (even if in different units) to be $\rm 
Log(R_{\rm GxL})=- 10.9~\pm0.3$ while the value of the \nuk{Fe}{60}$/$\nuk{Al}{26} flux ratio (averaged between the RHESSI and SPI 
values) is of the order of $0.14\pm0.03$. As for the first ratio, $\rm R_{\rm GxL}$, its constancy along the galactic longitude 
could hide additional information. Different longitudes probably map combinations of different stellar populations (the smallest 
longitudes probably mapping the region with the highest average metallicity): hence the constancy of $\rm R_{\rm GxL}$ could imply a 
low dependence of this ratio on the initial metallicity. The same argument does not hold for the \nuk{Fe}{60}$/$\nuk{Al}{26} flux 
ratio because in this case we just have one integrated measure towards the central radiant of the Galaxy. 

We have shown that the observed $\rm R_{\rm GxL}$ may be reproduced fairly well for a variety of slopes of the IMF and both sets of 
models, namely the NL00 and the LA89 ones. If we focus on a specific slope, e.g. x=1.8 (a value widely accepted to be reasonable for 
these massive stars) a maximum discrepancy of 0.2 dex is obtained, independently on the adopted $\rm M_ {\rm TOP}$ and mass loss 
rate beyond the WNL phase. The fit even improves for steeper slopes. Within our sets of models, \nuk{Al}{26} is mainly produced by 
stars less massive than, say, 35\msun, the total contribution of the more massive stars (through the wind of the WR stars plus the 
Type Ibc supernovae) never exceeding 30\% of the total. Also this result depends only moderately on the adopted mass loss rate 
beyond the WNL phase and on $\rm M_ {\rm TOP}$.

The theoretical \nuk{Fe}{60} yield, on the contrary, strongly depends on the mass loss rate adopted in the WNE+WCO phases mainly 
because mass loss controls the amount of \nuk{Fe}{60} produced in the He convective shell of the stars more massive than 35\msun 
during the very latest phases that precede their final collapse. Depending on the adopted mass loss rate, the predicted 
\nuk{Fe}{60}$/$\nuk{Al}{26} flux ratio may range between 0.12 and 0.25 for a slope x=1.8. The models that better reproduce the 
observed ratio ($\simeq0.14$) are the LA89 ones since the very strong mass loss rate proposed by \citet{la89} completely dumps out 
the contribution of the He convective shell and also reduces the contribution of the C convective shell because of the much smaller 
CO core mass. The dependence of the \nuk{Fe}{60}$/$\nuk{Al}{26} flux ratio on both the IMF slope and $\rm M_ {\rm TOP}$ is 
negligible in this case because of the overall modest contribution of the stars more massive than 35\msun to the global budget of 
the \nuk{Fe}{60}. The worst fit is obtained for the NL00 models since in this case either a strong contribution to the total yield 
comes from the He convective shell and also the contribution from the C convective shell is larger because the mass size of the CO 
core is definitely much large in this case. A steep IMF plus an $\rm M_ {\rm TOP}$ not larger than, say, 60\msun could anyway 
reconcile also this set of models with the observed flux ratio.

Of course many other unceretainties affect both yields and we have shown in this paper how the \nuk{Al}{26} yield depends, e.g., on 
several cross sections, initial abundances and size of the H-convective core. Also a distortion of the IMF towards the highest 
masses could lead, in practice, to a steepening of the slope. It must also be noted that all the present results have been obtained 
with solar metallicity stars while very probably at least the inner Galaxy has an average metallicity larger than solar. We actually 
do not know how the metallicity influences the yields of the \nuk{Al}{26} and the \nuk{Fe}{60} and hence we leave such an 
exploration to another paper. The dependence of the \nuk{Al}{26} wind component on the initial metallicity has been already 
discussed in a number of papers \citep{pra91,ma93,me97} but unfortunately such a dependence is not representative of the global 
trend because we have shown in this paper that the \nuk{Al}{26} present in the wind always represents a modest contribution to the 
total budget of the \nuk{Al}{26}.

The determination of the total amount of \nuk{Al}{26} present in the Galaxy requires the knowledge of an additional quite uncertain 
parameter, i.e. the SFR. Such a quantity is usually determined by means of either the number of core collapse supernovae per unit 
time or the total Lyman continuum luminosity of the Galaxy. Since it would be desirable to have a SFR linked as strongly as possible 
to the stars that mainly contribute to the Lyman continuum luminosity as well as to the \nuk{Al}{26} and \nuk{Fe}{60} synthesys, the 
adoption of the total Lyman continuum luminosity should be preferred. The core collapse supernovae, in fact, are dominated (in 
number) by the Type II supernovae whose range extends down to masses that do not contribute significantly to neither the Lyman 
continuum luminosity nor the \nuk{Al}{26} nor the \nuk{Fe}{60}. For example, the mass range 11 to 13\msun produces roughly 30\% of 
the total number of Type II supernovae while their contribution to the synthesis of the \nuk{Al}{26} is of the order of 6\%; the 
situation would even worsen if the lower mass limit of the massive stars would reduce to 10 or even 9\msun because these stars would 
strongly affect the frequency of the Type II supernovae without contributing significantly neither to the Lyman continuum luminosity 
nor to the \nuk{Al}{26} or to the \nuk{Fe}{60} budgets. By normalizing to the total Lyman continuum luminosity estimated by 
\citet{ben94} and an IMF slope x=1.8, we predict roughly 2\msun of \nuk{Al}{26} presently in the Galaxy in the NL00 case. The 
adoption of the LA89 models would reduce such an amount to 1.7\msun.  

We have also fitted $\gamma^2$ Velorum, the closest WR star. In particular we predict an abundance of $5.82\times10^{-5}$\msun of 
\nuk{Al}{26} (for a 60\msun), value compatible with the current upper limit of $6.3^{+2.1}_{- 1.4}\times10^{- 5}$\msun.

We did not address the fit to specific OB associations in this paper because it would require an evolutionary synthetic code, as it 
has been correctly addressed by, e.g., \citet{cer00}, that is not available at present.

\section{Acknowledgments} 

It is a pleasure to thank Roland Diehl and Nikos Prantzos for many helpful discussions. A.C. thanks his Institute, the IASF in Rome, 
and in particular the IBIS (INTEGRAL) group for continuous financial support. This research has been partially supported by the 
Ministry of the Education, University and Research (MIUR) through the grant PRIN-2004 (Prot. 2004029938).

\begin{deluxetable}{ccccccccccccccc}
\tablecolumns{15}
\tabletypesize{\scriptsize}
\rotate
\tablewidth{0pt}
\tablecaption{Network reference matrix}
\tablehead{
\colhead{isotope} &
\colhead{(p,$\gamma$)} & \colhead{(p,$\alpha$)} & \colhead{(p,n)} &
\colhead{($\alpha$,$\gamma$)} & \colhead{($\alpha$,p)} & \colhead{($\alpha$,n)} &
\colhead{(n,$\gamma$)} & \colhead{(n,p)} & \colhead{(n,$\alpha$)} &
\colhead{($\gamma$,p)} & \colhead{($\gamma$,$\alpha$)} & \colhead{($\gamma$,n)} &
\colhead{($\beta^+$)} & \colhead{($\beta^-$)} }
\startdata                                                                                                             
N    & BK   &       &       &       &       &       &       &       &       &       &       &       &       &  OD94 \\
H1   & NACR &       &       &       &       &       &       &       &       &       &       &       &  OD94 &       \\
H2   & NACR &       &       &  NACR &       &       &  FKTH &       &       &       &       &       &       &       \\
H3   & CA88 &       &  CA88 &  NACR &       &  CA88 &       &       &       &       &       &       &       &  RATH \\
He3  & FKTH &       &       &  NACR &  NACR &       &  FKTH &  CA88 &       &       &       &       &       &       \\
He4  &      &       &       &       &  NACR &  FKTH &       &       &       &       &       &       &       &       \\
Li6  & NACR &  NACR &       &  CA88 &  NACR &       &  FKTH &       &  CA88 &       &       &       &       &       \\
Li7  &      &  NACR &  CA88 &  NACR &  FKTH &  NACR &  BK   &       &       &       &       &       &       &       \\
Be7  &      &  NACR &       &  NACR &  NACR &       &       &  CA88 &  FKTH &       &       &       &       &       \\
Be9  & NACR &  NACR &       &       &  FKTH &  NACR &  FKTH &       &  FKTH &       &       &       &       &       \\
Be10 & FKTH &  FKTH &       &  FKTH &       &  FKTH &  FKTH &       &       &       &       &       &       &  RATH \\
B10  & NACR &  NACR &       &       &  FKTH &  CA88 &  FKTH &       &  NACR &       &       &       &       &       \\
B11  & NACR &       &       &       &  FKTH &  FKTH &  FKTH &       &  FKTH &       &       &       &       &       \\
C12  & NACR &       &       &  KUNZ &  NACR &  CA88 &  BK   &       &  NACR &       &       &       &       &       \\
C13  & NACR &  FKTH &  NACR &       &  FKTH &  NACR &  BK   &       &  FKTH &       &       &  BK   &       &       \\
C14  & CA88 &  FKTH &  CA88 &  CA88 &       &  FKTH &  BK   &       &  FKTH &       &       &  BK   &       &  RATH \\
N13  & NACR &       &       &       &  CA88 &       &  FKTH &  NACR &  CA88 &  NACR &       &       &  RATH &       \\
N14  & LUNA &  NACR &       &  NACR &  NACR &  NACR &  BK   &  CA88 &  FKTH &  NACR &       &  FKTH &       &       \\
N15  & NACR &  NACR &  NACR &  NACR &  NACR &  CA88 &  BK   &       &  FKTH &  CA88 &       &  BK   &       &       \\
N16  &      &  FKTH &  FKTH &       &  FKTH &  FKTH &       &       &       &       &       &  BK   &       &  RATH \\
O15  &      &  FKTH &       &  CA88 &  FKTH &  RATH &       &  NACR &  CA88 &  LUNA &       &       &  RATH &       \\
O16  & NACR &  CA88 &       &  NACR &  NACR &  RATH &  BK   &  FKTH &  NACR &  NACR &  KUNZ &       &       &       \\
O17  & NACR &  NACR &  FKTH &  CA88 &  FKTH &  NACR &  FKTH &       &  FKTH &       &       &  BK   &       &  OD94 \\
O18  & NACR &  NACR &  FKTH &  NACR &       &  NACR &  BK   &       &  FKTH &       &  CA88 &  FKTH &       &  OD94 \\
O19  &      &  FKTH &  FKTH &       &       &  FKTH &       &       &       &       &       &  BK   &       &  OD94 \\
F17  & FKTH &  CA88 &       &       &  NACR &  RATH &  FKTH &  FKTH &  NACR &  NACR &       &       &  OD94 &       \\
F18  & FKTH &  FKTH &       &       &       &  RATH &  FKTH &  FKTH &  CA88 &  NACR &  NACR &  FKTH &  OD94 &       \\
F19  & NACR &  NACR &       &       &  CA88 &  CA88 &  BK   &  FKTH &  FKTH &  NACR &  NACR &  FKTH &  OD94 &       \\
F20  &      &  FKTH &       &       &       &       &       &       &       &       &       &  BK   &       &  OD94 \\
Ne20 & IL01 &  NACR &       &  NACR &  IL01 &  RATH &  BKRT &       &  NACR &  NACR &  NACR &       &  OD94 &       \\
Ne21 & IL01 &       &  RATH &  CA88 &  RATH &  NACR &  BKRT &       &  NACR &       &  CA88 &  BKRT &       &  OD94 \\
Ne22 & IL01 &  CA88 &  CA88 &  NACR &  RATH &  JAEG &  BKRT &       &  FKTH &       &  NACR &  BKRT &       &  OD94 \\
Ne23 & RATH &       &  RATH &  RATH &       &  RATH &       &       &       &       &       &  BKRT &       &  OD94 \\
Na21 &      &       &       &  RATH &  NACR &       &  RT   &  RATH &  RATH &  IL01 &       &       &  OD94 &       \\
Na22 & IL01 &  RATH &       &  RATH &  RATH &  RATH &  RT   &  CA88 &  CA88 &  IL01 &       &  RT   &  OD94 &       \\
Na23 & IL01 &  IL01 &  NACR &  RATH &  RATH &  NACR &  BKRT &  RATH &       &  IL01 &       &  RT   &  OD94 &  OD94 \\
Na24 & RATH &  RATH &  RATH &  RATH &  RATH &  RATH &       &       &       &  RATH &       &  BKRT &       &  OD94 \\
Mg23 &      &       &       &  RATH &  RATH &       &  RT   &  NACR &  RATH &  IL01 &       &       &  OD94 &       \\
Mg24 & IL01 &  NACR &       &  CA88 &  IL01 &  RATH &  BKRT &  RATH &  NACR &  IL01 &  NACR &  RT   &  OD94 &       \\
Mg25 & IL01 &  RATH &  RATH &  CA88 &  CA88 &  NACR &  BKRT &       &  JAEG &  RATH &  CA88 &  BKRT &       &  OD94 \\
Mg26 & NACR &  RATH &  CA88 &  CA88 &  CA88 &  NACR &  BKRT &       &  RATH &       &  NACR &  BKRT &       &  FFN8 \\
Mg27 & RATH &  RATH &  RATH &  RATH &       &  RATH &       &       &       &       &  RATH &  BKRT &       &  OD94 \\
Al25 &      &       &       &  RATH &  RATH &       &  RT   &  RATH &  RATH &  IL01 &  RATH &       &  OD94 &       \\
Al26 & CA88 &  RATH &       &  RATH &  RATH &  RATH &  BKRT &  CA88 &  NACR &  IL01 &  RATH &  RT   &  FFN8 &       \\
Alg6 & IL01 &  FKTH &       &  FKTH &  FKTH &  FKTH &  FKTH &  CA88 &  NACR &       &       &       &  RATH &       \\
Alm6 & NACR &       &       &       &       &       &       &  CA88 &  NACR &       &       &       &  RATH &       \\
Al27 & IL01 &  IL01 &  RATH &  RATH &  RATH &  NACR &  BKRT &  RATH &  RATH &  IL01 &  RATH &  BKRT &  OD94 &  OD94 \\
Al28 & RATH &  CA88 &  RATH &  RATH &  RATH &  RATH &       &       &       &  RATH &  RATH &  BKRT &       &  OD94 \\
Si27 &      &       &       &  RATH &  RATH &       &  RT   &  RATH &  RATH &  CA88 &  RATH &       &  OD94 &       \\
Si28 & NACR &  RATH &       &  RATH &  FKTH &  RATH &  BKRT &  RATH &  NACR &  IL01 &  CA88 &  RT   &  OD94 &       \\
Si29 & IL01 &  RATH &  RATH &  RATH &  RATH &  RATH &  BKRT &       &  NACR &  RATH &  CA88 &  BKRT &       &  OD94 \\
Si30 & IL01 &  RATH &  RATH &  RATH &  RATH &  RATH &  BKRT &       &  RATH &       &  CA88 &  BKRT &       &  OD94 \\
Si31 & RATH &  RATH &  RATH &  RATH &  RATH &  RATH &  RT   &       &       &       &  RATH &  BKRT &       &  OD94 \\
Si32 & RATH &       &  RATH &  RATH &       &  RATH &       &       &       &       &       &  RT   &       &  OD94 \\
P29  &      &       &       &  RATH &  RATH &       &  RT   &  RATH &  RATH &  IL01 &  RATH &       &  OD94 &       \\
P30  & RATH &  RATH &       &  RATH &  RATH &  RATH &  RT   &  RATH &  NACR &  IL01 &  RATH &  RT   &  OD94 &       \\
P31  & RATH &  FKTH &  RATH &  RATH &  RATH &  RATH &  BKRT &  RATH &  RATH &  IL01 &  RATH &  RT   &  OD94 &  OD94 \\
P32  & RATH &  RATH &  RATH &  RATH &  RATH &  RATH &  RT   &  RATH &       &  RATH &  RATH &  BKRT &  OD94 &  OD94 \\
P33  & RATH &  RATH &  RATH &  RATH &  RATH &  RATH &  RT   &       &       &  RATH &       &  RT   &       &  OD94 \\
P34  & RATH &  RATH &  RATH &  RATH &  RATH &  RATH &       &       &       &       &       &  RT   &       &  OD94 \\
S31  &      &       &       &       &  RATH &       &  RT   &  RATH &  RATH &  RATH &  RATH &       &  OD94 &       \\
S32  & FKTH &  RATH &       &  RATH &  RATH &       &  BKRT &  RATH &  RATH &  RATH &  RATH &  RT   &  OD94 &       \\
S33  & RATH &  RATH &  RATH &  RATH &  RATH &  RATH &  BKRT &  RATH &  RATH &  RATH &  RATH &  BKRT &  OD94 &  OD94 \\
S34  & RATH &  RATH &  RATH &  RATH &  RATH &  RATH &  BKRT &  RATH &  RATH &  RATH &  RATH &  BKRT &  OD94 &  OD94 \\
S35  & RATH &  RATH &  RATH &  RATH &  RATH &  RATH &  RT   &       &  RATH &  RATH &  RATH &  BKRT &       &  OD94 \\
S36  & RATH &  RATH &  RATH &  RATH &       &  RATH &  BKRT &       &       &       &  RATH &  RT   &       &  FFN8 \\
S37  & RATH &  RATH &  RATH &  RATH &       &  RATH &       &       &       &       &       &  BKRT &       &  FFN8 \\
Cl33 &      &       &       &  RATH &  RATH &       &  RT   &  RATH &  RATH &  FKTH &  RATH &       &  OD94 &       \\
Cl34 &      &  RATH &       &  RATH &  RATH &  RATH &  RT   &  RATH &  RATH &  RATH &  RATH &  RT   &  OD94 &       \\
Cl35 & RATH &  RATH &       &  RATH &  RATH &  RATH &  BKRT &  RATH &  RATH &  RATH &  RATH &  RT   &  OD94 &       \\
Cl36 & RATH &  RATH &  RATH &  RATH &  RATH &  RATH &  BKRT &  RATH &  RATH &  RATH &  RATH &  BKRT &  FFN8 &  FFN8 \\
Cl37 & RATH &  RATH &  RATH &  RATH &  RATH &  RATH &  BKRT &  RATH &  RATH &  RATH &  RATH &  BKRT &  FFN8 &  OD94 \\
Cl38 & RATH &  RATH &  RATH &  RATH &  RATH &  RATH &       &       &       &  RATH &  RATH &  BKRT &       &  FFN8 \\
Ar36 & FKTH &  RATH &       &  RATH &  RATH &       &  BKRT &  RATH &  RATH &  RATH &  RATH &       &  FFN8 &       \\
Ar37 & RATH &  RATH &  RATH &  RATH &  RATH &  RATH &  RT   &  RATH &  RATH &  RATH &  RATH &  BKRT &  OD94 &  OD94 \\
Ar38 & RATH &  RATH &  RATH &  RATH &  RATH &  RATH &  BKRT &  RATH &  RATH &  RATH &  RATH &  RT   &  FFN8 &  OD94 \\
Ar39 & RATH &  RATH &  RATH &  RATH &  RATH &  RATH &  BKRT &       &  RATH &  RATH &  RATH &  BKRT &       &  FFN8 \\
Ar40 & RATH &  RATH &  RATH &  RATH &       &  RATH &  BKRT &       &  RATH &       &  RATH &  BKRT &       &  FFN8 \\
Ar41 & RATH &  RATH &  RATH &  RATH &       &  RATH &       &       &       &       &  RATH &  BKRT &       &  FFN8 \\
K37  &      &       &       &  RATH &  RATH &       &  RT   &  RATH &  RATH &  FKTH &  RATH &       &  OD94 &       \\
K38  &      &       &       &  RATH &  RATH &  RATH &  RT   &  RATH &  RATH &  RATH &  RATH &  RT   &  OD94 &       \\
K39  & RATH &  RATH &       &  RATH &  RATH &  RATH &  BKRT &  RATH &  RATH &  RATH &  RATH &  RT   &  FFN8 &       \\
K40  & RATH &  RATH &  RATH &  RATH &  RATH &  RATH &  BKRT &  RATH &  RATH &  RATH &  RATH &  BKRT &  FFN8 &  FFN8 \\
K41  & RATH &  RATH &  RATH &  RATH &  RATH &  RATH &  BKRT &  RATH &  RATH &  RATH &  RATH &  BKRT &  FFN8 &  FFN8 \\
K42  & RATH &  RATH &  RATH &  RATH &  RATH &  RATH &       &       &       &  RATH &  RATH &  BKRT &       &  FFN8 \\
Ca40 & FKTH &  RATH &       &  RATH &  RATH &       &  BKRT &  RATH &  RATH &  RATH &  RATH &       &  FFN8 &       \\
Ca41 & RATH &  RATH &  RATH &  RATH &  RATH &  RATH &  BKRT &  RATH &  RATH &  RATH &  RATH &  BKRT &  FFN8 &  FFN8 \\
Ca42 & RATH &  RATH &  RATH &  RATH &  RATH &  RATH &  BKRT &  RATH &  RATH &  RATH &  RATH &  BKRT &  FFN8 &  FFN8 \\
Ca43 & RATH &  RATH &  RATH &  RATH &  RATH &  RATH &  BKRT &       &  RATH &  RATH &  RATH &  BKRT &       &  FFN8 \\
Ca44 & RATH &  RATH &  RATH &  RATH &  RATH &  RATH &  BKRT &       &  RATH &       &  RATH &  BKRT &       &  FFN8 \\
Ca45 & RATH &  RATH &  RATH &  RATH &  RATH &  RATH &  BKRT &       &       &       &  RATH &  BKRT &       &  FFN8 \\
Ca46 & RATH &       &  RATH &  RATH &  RATH &  RATH &  BKRT &       &       &       &       &  BKRT &       &  LP00 \\
Ca47 & RATH &       &  RATH &  RATH &       &  RATH &  RT   &       &       &       &       &  BKRT &       &  LP00 \\
Ca48 & RATH &       &  RATH &       &       &  RATH &  BKRT &       &       &       &       &  RT   &       &  LP00 \\
Ca49 &      &       &  RATH &       &       &       &       &       &       &       &       &  BKRT &       &  LP00 \\
Sc41 &      &       &       &  RATH &  RATH &       &  RT   &  RATH &  RATH &  FKTH &  RATH &       &  FFN8 &       \\
Sc42 &      &       &       &  RATH &  RATH &  RATH &  RT   &  RATH &  RATH &  RATH &  RATH &  RT   &  FFN8 &       \\
Sc43 & RATH &  RATH &       &  RATH &  RATH &  RATH &  RT   &  RATH &  RATH &  RATH &  RATH &  RT   &  FFN8 &       \\
Sc44 & RATH &  RATH &  RATH &  RATH &  RATH &  RATH &  RT   &  RATH &  RATH &  RATH &  RATH &  RT   &  FFN8 &  FFN8 \\
Sc45 & RATH &  RATH &  RATH &  RATH &  RATH &  RATH &  BKRT &  RATH &  RATH &  RATH &  RATH &  RT   &  FFN8 &  FFN8 \\
Sc46 & RATH &  RATH &  RATH &  RATH &  RATH &  RATH &  RT   &  RATH &       &  RATH &  RATH &  BKRT &  LP00 &  LP00 \\
Sc47 & RATH &  RATH &  RATH &  RATH &  RATH &  RATH &  RT   &  RATH &       &  RATH &       &  RT   &  LP00 &  LP00 \\
Sc48 & RATH &  RATH &  RATH &  RATH &  RATH &  RATH &  RT   &  RATH &       &  RATH &       &  RT   &  LP00 &  LP00 \\
Sc49 & RATH &  RATH &  RATH &       &       &  RATH &       &  RATH &       &  RATH &       &  RT   &  LP00 &  LP00 \\
Ti44 & FKTH &  RATH &       &  RATH &  RATH &       &  RT   &  RATH &  RATH &  RATH &  RATH &       &  FFN8 &       \\
Ti45 & RATH &  RATH &  RATH &  RATH &  RATH &  RATH &  RT   &  RATH &  RATH &  RATH &  RATH &  RT   &  FFN8 &  FFN8 \\
Ti46 & RATH &  RATH &  RATH &  RATH &  RATH &  RATH &  BKRT &  RATH &  RATH &  RATH &  RATH &  RT   &  LP00 &  LP00 \\
Ti47 & RATH &  RATH &  RATH &  RATH &  RATH &  RATH &  BKRT &  RATH &  RATH &  RATH &  RATH &  BKRT &  LP00 &  LP00 \\
Ti48 & RATH &  RATH &  RATH &  RATH &  RATH &  RATH &  BKRT &  RATH &  RATH &  RATH &  RATH &  BKRT &  LP00 &  LP00 \\
Ti49 & RATH &  RATH &  RATH &  RATH &  RATH &  RATH &  BKRT &  RATH &  RATH &  RATH &  RATH &  BKRT &  LP00 &  LP00 \\
Ti50 & RATH &  RATH &  RATH &  RATH &       &  RATH &  BKRT &       &  RATH &  RATH &  RATH &  BKRT &       &  LP00 \\
Ti51 & RATH &  RATH &  RATH &  RATH &       &  RATH &       &       &  RATH &       &  RATH &  BKRT &       &  LP00 \\
V45  &      &       &       &       &  RATH &       &  RT   &  RATH &  RATH &  FKTH &  RATH &       &  FFN8 &       \\
V46  &      &       &       &  RATH &  RATH &       &  RT   &  RATH &  RATH &  RATH &  RATH &  RT   &  LP00 &       \\
V47  & RATH &  RATH &       &  RATH &  RATH &  RATH &  RT   &  RATH &  RATH &  RATH &  RATH &  RT   &  LP00 &       \\
V48  & RATH &  RATH &  RATH &  RATH &  RATH &  RATH &  RT   &  RATH &  RATH &  RATH &  RATH &  RT   &  LP00 &  LP00 \\
V49  & RATH &  RATH &  RATH &  RATH &  RATH &  RATH &  RT   &  RATH &  RATH &  RATH &  RATH &  RT   &  LP00 &  LP00 \\
V50  & RATH &  RATH &  RATH &  RATH &  RATH &  RATH &  BKRT &  RATH &  RATH &  RATH &  RATH &  RT   &  LP00 &  LP00 \\
V51  & RATH &  RATH &  RATH &  RATH &  RATH &  RATH &  BKRT &  RATH &  RATH &  RATH &  RATH &  BKRT &  LP00 &  LP00 \\
V52  & RATH &  RATH &  RATH &  RATH &  RATH &  RATH &       &       &  RATH &  RATH &  RATH &  BKRT &       &  LP00 \\
Cr48 &      &  RATH &       &  RATH &  RATH &       &  RT   &  RATH &  RATH &  RATH &  RATH &       &  LP00 &       \\
Cr49 & RATH &  RATH &       &  RATH &  RATH &  RATH &  RT   &  RATH &  RATH &  RATH &  RATH &  RT   &  LP00 &       \\
Cr50 & RATH &  RATH &  RATH &  RATH &  RATH &  RATH &  BKRT &  RATH &  RATH &  RATH &  RATH &  RT   &  LP00 &  LP00 \\
Cr51 & RATH &  RATH &  RATH &  RATH &  RATH &  RATH &  BKRT &  RATH &  RATH &  RATH &  RATH &  BKRT &  LP00 &  LP00 \\
Cr52 & RATH &  RATH &  RATH &  RATH &  RATH &  RATH &  BKRT &  RATH &  RATH &  RATH &  RATH &  BKRT &  LP00 &  LP00 \\
Cr53 & RATH &  RATH &  RATH &  RATH &  RATH &  RATH &  BKRT &       &  RATH &  RATH &  RATH &  BKRT &       &  LP00 \\
Cr54 & RATH &  RATH &  RATH &  RATH &  RATH &  RATH &  BKRT &       &  RATH &       &  RATH &  BKRT &       &  LP00 \\
Cr55 & RATH &  RATH &  RATH &  RATH &       &  RATH &       &       &       &       &  RATH &  BKRT &       &  LP00 \\
Mn50 &      &       &       &  RATH &  RATH &       &  RT   &  RATH &  RATH &  RATH &  RATH &       &  LP00 &       \\
Mn51 & RATH &  RATH &       &  RATH &  RATH &  RATH &  RT   &  RATH &  RATH &  RATH &  RATH &  RT   &  LP00 &       \\
Mn52 & RATH &  RATH &  RATH &  RATH &  RATH &  RATH &  RT   &  RATH &  RATH &  RATH &  RATH &  RT   &  LP00 &  LP00 \\
Mn53 & RATH &  RATH &  RATH &  RATH &  RATH &  RATH &  RT   &  RATH &  RATH &  RATH &  RATH &  RT   &  LP00 &  LP00 \\
Mn54 & RATH &  RATH &  RATH &  RATH &  RATH &  RATH &  RT   &  RATH &  RATH &  RATH &  RATH &  RT   &  LP00 &  LP00 \\
Mn55 & RATH &  RATH &  RATH &  RATH &  RATH &  RATH &  BKRT &  RATH &  RATH &  RATH &  RATH &  RT   &  LP00 &  LP00 \\
Mn56 & RATH &  RATH &  RATH &  RATH &  RATH &  RATH &  RT   &       &       &  RATH &  RATH &  BKRT &       &  LP00 \\
Mn57 & RATH &  RATH &  RATH &  RATH &  RATH &  RATH &       &       &       &       &       &  RT   &       &  LP00 \\
Fe52 &      &       &       &  RATH &  RATH &       &  RT   &  RATH &  RATH &  RATH &  RATH &       &  LP00 &       \\
Fe53 & RATH &  RATH &       &  RATH &  RATH &  RATH &  RT   &  RATH &  RATH &  RATH &  RATH &  RT   &  LP00 &       \\
Fe54 & RATH &  RATH &  RATH &  RATH &  RATH &  RATH &  BKRT &  RATH &  RATH &  RATH &  RATH &  RT   &  LP00 &  LP00 \\
Fe55 & RATH &  RATH &  RATH &  RATH &  RATH &  RATH &  BKRT &  RATH &  RATH &  RATH &  RATH &  BKRT &  LP00 &  LP00 \\
Fe56 & RATH &  RATH &  RATH &  RATH &  RATH &  RATH &  BKRT &  RATH &  RATH &  RATH &  RATH &  BKRT &  LP00 &  LP00 \\
Fe57 & RATH &  RATH &  RATH &  RATH &  RATH &  RATH &  BKRT &  RATH &  RATH &  RATH &  RATH &  BKRT &  LP00 &  LP00 \\
Fe58 & RATH &  RATH &  RATH &  RATH &  RATH &  RATH &  BKRT &       &  RATH &  RATH &  RATH &  BKRT &       &  LP00 \\
Fe59 & RATH &  RATH &  RATH &  RATH &       &  RATH &  RT   &       &       &       &  RATH &  BKRT &       &  LP00 \\
Fe60 & RATH &  RATH &  RATH &  RATH &       &  RATH &  RT   &       &       &       &       &  RT   &       &  FFN8 \\
Fe61 &      &       &  RATH &  RATH &       &  RATH &       &       &       &       &       &  RT   &       &  LP00 \\
Co54 &      &       &       &  RATH &  RATH &  RATH &  RT   &  RATH &  RATH &  RATH &  RATH &       &  LP00 &       \\
Co55 & RATH &  RATH &       &  RATH &  RATH &  RATH &  RT   &  RATH &  RATH &  RATH &  RATH &  RT   &  LP00 &       \\
Co56 & RATH &  RATH &  RATH &  RATH &  RATH &  RATH &  RT   &  RATH &  RATH &  RATH &  RATH &  RT   &  LP00 &  LP00 \\
Co57 & RATH &  RATH &  RATH &  RATH &  RATH &  RATH &  RT   &  RATH &  RATH &  RATH &  RATH &  RT   &  LP00 &  LP00 \\
Co58 & RATH &  RATH &  RATH &  RATH &  RATH &  RATH &  RT   &  RATH &  RATH &  RATH &  RATH &  RT   &  LP00 &  LP00 \\
Co59 & RATH &  RATH &  RATH &  RATH &  RATH &  RATH &  BKRT &  RATH &  RATH &  RATH &  RATH &  RT   &  LP00 &  LP00 \\
Co60 & RATH &  RATH &  RATH &  RATH &  RATH &  RATH &  RT   &  RATH &  RATH &  RATH &  RATH &  BKRT &  LP00 &  LP00 \\
Co61 & RATH &  RATH &  RATH &  RATH &  RATH &  RATH &       &  RATH &       &  RATH &  RATH &  RT   &  LP00 &  LP00 \\
Ni56 & FKTH &       &       &  RATH &  RATH &       &  RT   &  RATH &  RATH &  RATH &  RATH &       &  LP00 &       \\
Ni57 & RATH &  RATH &  RATH &  RATH &  RATH &  RATH &  RT   &  RATH &  RATH &  RATH &  RATH &  RT   &  LP00 &  FFN8 \\
Ni58 & RATH &  RATH &  RATH &  RATH &  RATH &  RATH &  BKRT &  RATH &  RATH &  RATH &  RATH &  RT   &  LP00 &  LP00 \\
Ni59 & RATH &  RATH &  RATH &  RATH &  RATH &  RATH &  BKRT &  RATH &  RATH &  RATH &  RATH &  BKRT &  LP00 &  LP00 \\
Ni60 & RATH &  RATH &  RATH &  RATH &  RATH &  RATH &  BKRT &  RATH &  RATH &  RATH &  RATH &  BKRT &  LP00 &  LP00 \\
Ni61 & RATH &  RATH &  RATH &  RATH &  RATH &  RATH &  BKRT &  RATH &  RATH &  RATH &  RATH &  BKRT &  LP00 &  LP00 \\
Ni62 & RATH &  RATH &  RATH &  RATH &  RATH &  RATH &  BKRT &       &  RATH &  RATH &  RATH &  BKRT &       &  LP00 \\
Ni63 & RATH &  RATH &  RATH &  RATH &  RATH &  RATH &  BKRT &       &  RATH &       &  RATH &  BKRT &       &  LP00 \\
Ni64 & RATH &  RATH &  RATH &  RATH &       &  RATH &  BKRT &       &  RATH &       &  RATH &  BKRT &       &  LP00 \\
Ni65 & RATH &       &  RATH &  RATH &       &  RATH &       &       &       &       &  RATH &  BKRT &       &  LP00 \\
Cu57 &      &       &       &       &  RATH &       &  RT   &  RATH &  RATH &  FKTH &       &       &  FFN8 &       \\
Cu58 &      &       &       &  RATH &  RATH &       &  RT   &  RATH &  RATH &  RATH &  RATH &  RT   &  LP00 &       \\
Cu59 & RATH &  RATH &       &  RATH &  RATH &  RATH &  RT   &  RATH &  RATH &  RATH &  RATH &  RT   &  LP00 &       \\
Cu60 & RATH &  RATH &  RATH &  RATH &  RATH &  RATH &  RT   &  RATH &  RATH &  RATH &  RATH &  RT   &  LP00 &  LP00 \\
Cu61 & RATH &  RATH &  RATH &  RATH &  RATH &  RATH &  RT   &  RATH &  RATH &  RATH &  RATH &  RT   &  LP00 &  LP00 \\
Cu62 & RATH &  RATH &  RATH &  RATH &  RATH &  RATH &  RT   &  RATH &  RATH &  RATH &  RATH &  RT   &  LP00 &  LP00 \\
Cu63 & RATH &  RATH &  RATH &  RATH &  RATH &  RATH &  BKRT &  RATH &  RATH &  RATH &  RATH &  RT   &  LP00 &  LP00 \\
Cu64 & RATH &  RATH &  RATH &  RATH &  RATH &  RATH &  RT   &  RATH &  RATH &  RATH &  RATH &  BKRT &  LP00 &  LP00 \\
Cu65 & RATH &  RATH &  RATH &  RATH &  RATH &  RATH &  BKRT &  RATH &       &  RATH &  RATH &  RT   &  LP00 &  LP00 \\
Cu66 & RATH &  RATH &  RATH &  RATH &  RATH &  RATH &       &       &       &  RATH &       &  BKRT &       &  RATH \\
Zn60 &      &  RATH &       &  RATH &  RATH &       &  RT   &  RATH &  RATH &  RATH &  RATH &       &  LP00 &       \\
Zn61 & RATH &  RATH &       &  RATH &  RATH &  RATH &  RT   &  RATH &  RATH &  RATH &  RATH &  RT   &  LP00 &       \\
Zn62 & RATH &  RATH &  RATH &  RATH &  RATH &  RATH &  RT   &  RATH &  RATH &  RATH &  RATH &  RT   &  LP00 &  LP00 \\
Zn63 & RATH &  RATH &  RATH &  RATH &  RATH &  RATH &  RT   &  RATH &  RATH &  RATH &  RATH &  RT   &  LP00 &  LP00 \\
Zn64 & RATH &  RATH &  RATH &  RATH &  RATH &  RATH &  BKRT &  RATH &  RATH &  RATH &  RATH &  RT   &  LP00 &  LP00 \\
Zn65 & RATH &  RATH &  RATH &  RATH &  RATH &  RATH &  BKRT &  RATH &  RATH &  RATH &  RATH &  BKRT &  LP00 &  LP00 \\
Zn66 & RATH &  RATH &  RATH &  RATH &  RATH &  RATH &  BKRT &  RATH &  RATH &  RATH &  RATH &  BKRT &       &       \\
Zn67 & RATH &  RATH &  RATH &  RATH &  RATH &  RATH &  BKRT &       &  RATH &  RATH &  RATH &  BKRT &       &       \\
Zn68 & RATH &  RATH &  RATH &  RATH &  RATH &  RATH &  BKRT &       &  RATH &       &  RATH &  BKRT &       &       \\
Zn69 & RATH &  RATH &  RATH &  RATH &  RATH &  RATH &  RT   &       &       &       &  RATH &  BKRT &       &  RATH \\
Zn70 & RATH &       &  RATH &  RATH &       &  RATH &  BKRT &       &       &       &       &  RT   &       &       \\
Zn71 & RATH &       &  RATH &  RATH &       &  RATH &       &       &       &       &       &  BKRT &       &  RATH \\
Ga62 &      &       &       &       &  RATH &       &  RT   &  RATH &  RATH &  RATH &  RATH &       &  LP00 &       \\
Ga63 & RATH &  RATH &       &       &  RATH &       &  RT   &  RATH &  RATH &  RATH &  RATH &  RT   &  LP00 &       \\
Ga64 & RATH &  RATH &  RATH &       &  RATH &       &  RT   &  RATH &  RATH &  RATH &  RATH &  RT   &  LP00 &  LP00 \\
Ga65 & RATH &  RATH &  RATH &       &  RATH &       &  RT   &  RATH &  RATH &  RATH &  RATH &  RT   &  LP00 &  LP00 \\
Ga66 & RATH &  RATH &  RATH &       &  RATH &       &  RT   &  RATH &  RATH &  RATH &  RATH &  RT   &  RATH &       \\
Ga67 & RATH &  RATH &  RATH &  RATH &  RATH &       &  RT   &  RATH &  RATH &  RATH &  RATH &  RT   &  RATH &       \\
Ga68 & RATH &  RATH &  RATH &  RATH &  RATH &  RATH &  RT   &  RATH &  RATH &  RATH &  RATH &  RT   &  RATH &       \\
Ga69 & RATH &  RATH &  RATH &  RATH &  RATH &  RATH &  BKRT &  RATH &  RATH &  RATH &  RATH &  RT   &       &       \\
Ga70 & RATH &  RATH &  RATH &  RATH &  RATH &  RATH &  RT   &  RATH &       &  RATH &  RATH &  BKRT &       &  RATH \\
Ga71 & RATH &  RATH &  RATH &  RATH &  RATH &  RATH &  BKRT &  RATH &       &  RATH &       &  RT   &       &       \\
Ga72 & RATH &  RATH &  RATH &  RATH &  RATH &  RATH &       &       &       &  RATH &       &  BKRT &       &  RATH \\
Ge64 &      &       &       &       &       &       &  RT   &  RATH &  RATH &  RATH &  RATH &       &  LP00 &       \\
Ge65 &      &  RATH &       &       &       &       &  RT   &  RATH &  RATH &  RATH &  RATH &  RT   &  LP00 &       \\
Ge66 &      &  RATH &       &       &       &       &  RT   &  RATH &  RATH &  RATH &  RATH &  RT   &  RATH &       \\
Ge67 &      &  RATH &       &       &       &       &  RT   &  RATH &  RATH &  RATH &  RATH &  RT   &  RATH &       \\
Ge68 &      &  RATH &       &       &  RATH &       &  RT   &  RATH &  RATH &  RATH &  RATH &  RT   &  RATH &       \\
Ge69 &      &  RATH &       &       &  RATH &       &  RT   &  RATH &  RATH &  RATH &  RATH &  RT   &  RATH &       \\
Ge70 & RATH &  RATH &       &  RATH &  RATH &       &  BKRT &  RATH &  RATH &  RATH &  RATH &  RT   &       &       \\
Ge71 & RATH &  RATH &  RATH &  RATH &  RATH &  RATH &  RT   &  RATH &  RATH &  RATH &  RATH &  BKRT &  RATH &       \\
Ge72 & RATH &  RATH &  RATH &  RATH &  RATH &  RATH &  BKRT &  RATH &  RATH &  RATH &  RATH &  RT   &       &       \\
Ge73 & RATH &  RATH &  RATH &  RATH &  RATH &  RATH &  BKRT &       &  RATH &  RATH &  RATH &  BKRT &       &       \\
Ge74 & RATH &  RATH &  RATH &  RATH &  RATH &  RATH &  BKRT &       &  RATH &       &  RATH &  BKRT &       &       \\
Ge75 & RATH &  RATH &  RATH &  RATH &       &  RATH &  RT   &       &       &       &  RATH &  BKRT &       &  RATH \\
Ge76 & RATH &       &  RATH &  RATH &       &  RATH &  BKRT &       &       &       &       &  RT   &       &       \\
Ge77 &      &       &  RATH &  RATH &       &  RATH &       &       &       &       &       &  BKRT &       &  RATH \\
As71 &      &  RATH &       &  RATH &  RATH &       &  RT   &  RATH &  RATH &  RATH &  RATH &       &  RATH &       \\
As72 &      &  RATH &       &  RATH &  RATH &  RATH &  RT   &  RATH &  RATH &  RATH &  RATH &  RT   &  RATH &       \\
As73 & RATH &  RATH &       &  RATH &  RATH &  RATH &  RT   &  RATH &  RATH &  RATH &  RATH &  RT   &  RATH &       \\
As74 & RATH &  RATH &  RATH &  RATH &  RATH &  RATH &  RT   &  RATH &  RATH &  RATH &  RATH &  RT   &  RATH &  RATH \\
As75 & RATH &  RATH &  RATH &  RATH &  RATH &  RATH &  BKRT &  RATH &  RATH &  RATH &  RATH &  RT   &       &       \\
As76 & RATH &  RATH &  RATH &  RATH &  RATH &  RATH &  RT   &  RATH &       &  RATH &  RATH &  BKRT &       &  RATH \\
As77 & RATH &  RATH &  RATH &  RATH &  RATH &  RATH &       &  RATH &       &  RATH &       &  RT   &       &  RATH \\
Se74 & RATH &  RATH &       &  RATH &  RATH &       &  BKRT &  RATH &  RATH &  RATH &  RATH &       &       &       \\
Se75 & RATH &  RATH &  RATH &  RATH &  RATH &  RATH &  RT   &  RATH &  RATH &  RATH &  RATH &  BKRT &  RATH &       \\
Se76 & RATH &  RATH &  RATH &  RATH &  RATH &  RATH &  BKRT &  RATH &  RATH &  RATH &  RATH &  RT   &       &       \\
Se77 & RATH &  RATH &  RATH &  RATH &  RATH &  RATH &  BKRT &  RATH &  RATH &  RATH &  RATH &  BKRT &       &       \\
Se78 & RATH &  RATH &  RATH &  RATH &  RATH &  RATH &  BKRT &       &  RATH &  RATH &  RATH &  BKRT &       &       \\
Se79 & RATH &  RATH &  RATH &  RATH &  RATH &  RATH &  BKRT &       &  RATH &       &  RATH &  BKRT &       &  TAKA \\
Se80 & RATH &  RATH &  RATH &  RATH &  RATH &  RATH &  BKRT &       &  RATH &       &  RATH &  BKRT &       &       \\
Se81 & RATH &       &  RATH &  RATH &       &  RATH &  RT   &       &       &       &  RATH &  BKRT &       &  RATH \\
Se82 & RATH &       &  RATH &  RATH &       &  RATH &  BKRT &       &       &       &       &  RT   &       &       \\
Se83 &      &       &  RATH &  RATH &       &  RATH &       &       &       &       &       &  BKRT &       &  RATH \\
Br75 &      &       &       &  RATH &  RATH &       &  RT   &  RATH &  RATH &  RATH &  RATH &       &  RATH &       \\
Br76 &      &       &       &  RATH &  RATH &  RATH &  RT   &  RATH &  RATH &  RATH &  RATH &  RT   &  RATH &       \\
Br77 & RATH &  RATH &       &  RATH &  RATH &  RATH &  RT   &  RATH &  RATH &  RATH &  RATH &  RT   &  RATH &       \\
Br78 & RATH &  RATH &  RATH &  RATH &  RATH &  RATH &  RT   &  RATH &  RATH &  RATH &  RATH &  RT   &  RATH &       \\
Br79 & RATH &  RATH &  RATH &  RATH &  RATH &  RATH &  BKRT &  RATH &  RATH &  RATH &  RATH &  RT   &       &       \\
Br80 & RATH &  RATH &  RATH &  RATH &  RATH &  RATH &  RT   &  RATH &  RATH &  RATH &  RATH &  BKRT &  RATH &  RATH \\
Br81 & RATH &  RATH &  RATH &  RATH &  RATH &  RATH &  BKRT &  RATH &       &  RATH &  RATH &  RT   &       &       \\
Br82 & RATH &  RATH &  RATH &  RATH &  RATH &  RATH &  RT   &  RATH &       &  RATH &       &  BKRT &       &  RATH \\
Br83 & RATH &  RATH &  RATH &  RATH &  RATH &  RATH &       &  RATH &       &  RATH &       &  RT   &       &  RATH \\
Kr78 & RATH &  RATH &       &       &  RATH &       &  BKRT &  RATH &  RATH &  RATH &  RATH &       &       &       \\
Kr79 & RATH &  RATH &  RATH &       &  RATH &       &  BKRT &  RATH &  RATH &  RATH &  RATH &  BKRT &  RATH &       \\
Kr80 & RATH &  RATH &  RATH &  RATH &  RATH &       &  BKRT &  RATH &  RATH &  RATH &  RATH &  BKRT &       &       \\
Kr81 & RATH &  RATH &  RATH &  RATH &  RATH &  RATH &  BKRT &  RATH &  RATH &  RATH &  RATH &  BKRT &  TAKA &       \\
Kr82 & RATH &  RATH &  RATH &  RATH &  RATH &  RATH &  BKRT &  RATH &  RATH &  RATH &  RATH &  BKRT &       &       \\
Kr83 & RATH &  RATH &  RATH &  RATH &  RATH &  RATH &  BKRT &  RATH &  RATH &  RATH &  RATH &  BKRT &       &       \\
Kr84 & RATH &  RATH &  RATH &  RATH &  RATH &  RATH &  BKRT &       &  RATH &  RATH &  RATH &  BKRT &       &       \\
Kr85 & RATH &  RATH &  RATH &  RATH &  RATH &  RATH &  BKRT &       &  RATH &       &  RATH &  BKRT &       &  TAKA \\
Kr86 & RATH &  RATH &  RATH &  RATH &       &  RATH &  BKRT &       &  RATH &       &  RATH &  BKRT &       &       \\
Kr87 & RATH &       &  RATH &  RATH &       &  RATH &       &       &       &       &  RATH &  BKRT &       &  RATH \\
Rb79 &      &       &       &       &       &       &  RT   &  RATH &  RATH &  RATH &  RATH &       &  RATH &       \\
Rb80 &      &       &       &       &       &       &  RT   &  RATH &  RATH &  RATH &  RATH &  RT   &  RATH &       \\
Rb81 &      &  RATH &       &  RATH &  RATH &       &  RT   &  RATH &  RATH &  RATH &  RATH &  RT   &  RATH &       \\
Rb82 &      &  RATH &       &  RATH &  RATH &  RATH &  RT   &  RATH &  RATH &  RATH &  RATH &  RT   &  RATH &       \\
Rb83 & RATH &  RATH &       &  RATH &  RATH &  RATH &  RT   &  RATH &  RATH &  RATH &  RATH &  RT   &  RATH &       \\
Rb84 & RATH &  RATH &  RATH &  RATH &  RATH &  RATH &  RT   &  RATH &  RATH &  RATH &  RATH &  RT   &  RATH &  RATH \\
Rb85 & RATH &  RATH &  RATH &  RATH &  RATH &  RATH &  BKRT &  RATH &  RATH &  RATH &  RATH &  RT   &       &       \\
Rb86 & RATH &  RATH &  RATH &  RATH &  RATH &  RATH &  BKRT &  RATH &  RATH &  RATH &  RATH &  BKRT &       &  TAKA \\
Rb87 & RATH &  RATH &  RATH &  RATH &  RATH &  RATH &  BKRT &  RATH &       &  RATH &  RATH &  BKRT &       &  RATH \\
Rb88 & RATH &  RATH &  RATH &       &  RATH &  RATH &       &       &       &  RATH &       &  BKRT &       &  RATH \\
Sr84 & RATH &  RATH &       &       &  RATH &       &  BKRT &  RATH &  RATH &  RATH &  RATH &       &       &       \\
Sr85 & RATH &  RATH &  RATH &       &  RATH &       &  RT   &  RATH &  RATH &  RATH &  RATH &  BKRT &  RATH &       \\
Sr86 & RATH &  RATH &  RATH &  RATH &  RATH &       &  BKRT &  RATH &  RATH &  RATH &  RATH &  RT   &       &       \\
Sr87 & RATH &  RATH &  RATH &  RATH &  RATH &  RATH &  BKRT &  RATH &  RATH &  RATH &  RATH &  BKRT &       &       \\
Sr88 & RATH &  RATH &  RATH &  RATH &  RATH &  RATH &  BKRT &  RATH &  RATH &  RATH &  RATH &  BKRT &       &       \\
Sr89 & RATH &  RATH &  RATH &  RATH &       &  RATH &  BKRT &       &  RATH &  RATH &  RATH &  BKRT &       &  RATH \\
Sr90 & RATH &  RATH &  RATH &  RATH &       &  RATH &  RT   &       &  RATH &       &  RATH &  BKRT &       &  RATH \\
Sr91 &      &  RATH &  RATH &  RATH &       &  RATH &       &       &       &       &  RATH &  RT   &       &  RATH \\
Y85  &      &       &       &       &       &       &  RT   &  RATH &  RATH &  RATH &  RATH &       &  RATH &       \\
Y86  &      &       &       &       &       &       &  RT   &  RATH &  RATH &  RATH &  RATH &  RT   &  RATH &       \\
Y87  &      &  RATH &       &  RATH &  RATH &       &  RT   &  RATH &  RATH &  RATH &  RATH &  RT   &  RATH &       \\
Y88  &      &  RATH &       &  RATH &  RATH &  RATH &  RT   &  RATH &  RATH &  RATH &  RATH &  RT   &  RATH &       \\
Y89  & RATH &  RATH &       &  RATH &  RATH &  RATH &  BKRT &  RATH &  RATH &  RATH &  RATH &  RT   &       &       \\
Y90  & RATH &  RATH &  RATH &  RATH &  RATH &  RATH &  RT   &  RATH &  RATH &  RATH &  RATH &  BKRT &       &  RATH \\
Y91  & RATH &  RATH &  RATH &  RATH &  RATH &  RATH &       &  RATH &  RATH &  RATH &  RATH &  RT   &       &  RATH \\
Zr90 & RATH &  RATH &       &  RATH &  RATH &  RATH &  BKRT &  RATH &  RATH &  RATH &  RATH &       &       &       \\
Zr91 & RATH &  RATH &  RATH &  RATH &  RATH &  RATH &  BKRT &  RATH &  RATH &  RATH &  RATH &  BKRT &       &       \\
Zr92 & RATH &  RATH &  RATH &  RATH &  RATH &  RATH &  BKRT &       &  RATH &  RATH &  RATH &  BKRT &       &       \\
Zr93 & RATH &  RATH &  RATH &  RATH &  RATH &  RATH &  BKRT &       &  RATH &       &  RATH &  BKRT &       &  RATH \\
Zr94 & RATH &  RATH &  RATH &  RATH &  RATH &  RATH &  BKRT &       &  RATH &       &  RATH &  BKRT &       &       \\
Zr95 & RATH &       &  RATH &       &       &  RATH &  BKRT &       &       &       &  RATH &  BKRT &       &  TAKA \\
Zr96 & RATH &       &  RATH &       &       &       &  BKRT &       &       &       &       &  BKRT &       &  RATH \\
Zr97 &      &       &  RATH &       &       &       &       &       &       &       &       &  BKRT &       &  RATH \\
Nb91 & RATH &       &       &       &  RATH &       &  RT   &  RATH &  RATH &  RATH &  RATH &       &  RATH &       \\
Nb92 & RATH &       &  RATH &       &  RATH &       &  RT   &  RATH &  RATH &  RATH &  RATH &  RT   &  RATH &  RATH \\
Nb93 & RATH &  RATH &  RATH &       &  RATH &       &  BKRT &  RATH &  RATH &  RATH &  RATH &  RT   &       &       \\
Nb94 & RATH &  RATH &  RATH &       &  RATH &       &  BKRT &  RATH &  RATH &  RATH &  RATH &  BKRT &       &  TAKA \\
Nb95 & RATH &  RATH &  RATH &       &  RATH &       &  RT   &  RATH &       &  RATH &  RATH &  BKRT &       &  RATH \\
Nb96 & RATH &  RATH &  RATH &       &       &       &  RT   &  RATH &       &  RATH &       &  RT   &       &  RATH \\
Nb97 & RATH &  RATH &  RATH &       &       &       &       &  RATH &       &  RATH &       &  RT   &       &  RATH \\
Mo92 &      &       &       &       &       &       &  BKRT &  RATH &       &  RATH &       &       &       &       \\
Mo93 &      &       &       &       &       &       &  RT   &  RATH &  RATH &  RATH &       &  BKRT &  TAKA &       \\
Mo94 &      &  RATH &       &       &       &       &  BKRT &  RATH &  RATH &  RATH &  RATH &  RT   &       &       \\
Mo95 &      &  RATH &       &       &       &       &  BKRT &  RATH &  RATH &  RATH &  RATH &  BKRT &       &       \\
Mo96 &      &  RATH &       &       &       &       &  BKRT &  RATH &  RATH &  RATH &  RATH &  BKRT &       &       \\
Mo97 &      &  RATH &       &       &       &       &  BKRT &  RATH &  RATH &  RATH &  RATH &  BKRT &       &       \\
Mo98 &      &  RATH &       &       &       &       &       &       &  RATH &  RATH &  RATH &  BKRT &       &       \\
\enddata            
\tablecomments{
BK = Bao et al. (2000).
BKRT = Bao et al. (2000) in the energy range 5-100 keV; Rausher \& Thielemann (2000) above this limit
but rescaled to match the experimental values of Bao et al. (2000) at 100 keV.
CA88 = Caughlan \& Fowler (1988).
FKTH = Thielemann et al. (1995), Reaction Rate Library REACLIB in which the experimental values are preferred whenever available.
(for the experimental rates the references are reported in the library).
IL01 = Iliadis et al. (2001).
JAEG = Jaeger et al. (2001), recommended rate.
KUNZ = Kunz et al. (2002), adopted rate.
LUNA = Formicola et al. (2004), LUNA (Laboratory, for Underground Nuclear Astrophysics) collaboration.
NACR = Angulo et al. (1999), NACRE.
RATH = last version of the REACLIB provided by T. Rauscher and F. Thielemann and also available at 
the web site http://quasar.physik.unibas.ch/~tommy.
RT = Rauscher \& Thielemann (2000).
OD94 = Oda et al. (1994).
FFN8 = Fuller, Fowler \& Newman (1982,1985).
LP00 = Langanke \& Martinez Pinedo (2000).
TAKA = Takahashi \& Yokoi (1987)
}
\label{tabele}
\end{deluxetable}

\begin{deluxetable}{lccccccccccccc}
\tablecolumns{14}
\tabletypesize{\scriptsize}
\rotate
\tablewidth{0pt}
\tablecaption{Basic evolutionary properties of the stellar models}
\tablehead{
$\rm M_{\rm ini}$ & $\rm M_{\rm end}$ & $\rm t_{H}$ & $\rm t_{He}$ & $\rm \Delta t_{\rm expl.}$ & $\rm M_{\rm He}$ &
$\rm M_{\rm CO}$ & $\rm M_{\rm bot \& top}(C_{\rm sh})$ & $\rm C_{\rm cen}$ & $\rm t_{\rm O}$ & $\rm t_{\rm WR}$ &
$\rm t_{\rm WNL}$ & $\rm t_{\rm WNE}$ & $\rm t_{\rm WCO}$ \\
\cline{1-14} \\
\colhead{\msun} & \colhead{\msun} & \colhead{\rm Myr} & \colhead{\rm Myr} & \colhead{\rm yr} & \colhead{\msun} & 
\colhead{\msun} & \colhead{\msun} & \colhead{\rm mass fraction} & \colhead{\rm yr} & \colhead{\rm yr} & 
\colhead{\rm yr} & \colhead{\rm yr} & \colhead{\rm yr} 
}
\startdata                                                                                                             
~11  & 10.56 & 20.11 & 1.55 & 7.32E-4 & ~3.473 & ~1.749 & 1.5-1.739  & 0.359 & 0.00E+00 & 0.00E+00 & 0.00E+00 & 0.00E+00 & 0.00E+00 \\
~12  & 11.49 & 17.44 & 1.30 & 6.24E-4 & ~3.912 & ~1.971 & 1.6-1.916  & 0.361 & 0.00E+00 & 0.00E+00 & 0.00E+00 & 0.00E+00 & 0.00E+00 \\
~13  & 12.01 & 15.40 & 1.14 & 4.58E-4 & ~4.363 & ~2.212 & 1.8-2.011  & 0.350 & 0.00E+00 & 0.00E+00 & 0.00E+00 & 0.00E+00 & 0.00E+00 \\
~14  & 12.77 & 13.81 & 1.01 & 3.73E-4 & ~4.816 & ~2.461 & 1.7-2.322  & 0.341 & 2.07E+06 & 0.00E+00 & 0.00E+00 & 0.00E+00 & 0.00E+00 \\
~15  & 13.49 & 12.54 & 0.90 & 3.19E-4 & ~5.293 & ~2.720 & 1.8-2.574  & 0.344 & 5.66E+06 & 0.00E+00 & 0.00E+00 & 0.00E+00 & 0.00E+00 \\
~16  & 14.16 & 11.48 & 0.83 & 2.78E-4 & ~5.765 & ~2.989 & 2.0-2.734  & 0.347 & 6.69E+06 & 0.00E+00 & 0.00E+00 & 0.00E+00 & 0.00E+00 \\
~17  & 14.83 & 10.61 & 0.76 & 2.49E-4 & ~6.209 & ~3.253 & 2.0-3.151  & 0.344 & 7.34E+06 & 0.00E+00 & 0.00E+00 & 0.00E+00 & 0.00E+00 \\
~20  & 16.31 & ~8.68 & 0.64 & 1.91E-4 & ~7.643 & ~4.354 & 1.8-4.354  & 0.321 & 7.08E+06 & 0.00E+00 & 0.00E+00 & 0.00E+00 & 0.00E+00 \\
~25  & 16.35 & ~6.87 & 0.53 & 1.40E-4 & 10.223 & ~6.148 & 2.2-6.148  & 0.287 & 6.02E+06 & 0.00E+00 & 0.00E+00 & 0.00E+00 & 0.00E+00 \\
~30  & 12.92 & ~5.81 & 0.47 & 1.01E-4 & 12.680 & ~8.011 & 1.9-7.154  & 0.263 & 5.19E+06 & 5.05E+04 & 5.05E+04 & 0.00E+00 & 0.00E+00 \\
~35  & 11.94 & ~5.12 & 0.44 & 0.80E-4 & 14.687 & ~8.508 & 2.3-8.203  & 0.251 & 4.62E+06 & 2.68E+05 & 1.16E+05 & 1.52E+05 & 0.00E+00 \\
~40  & 12.52 & ~4.64 & 0.42 & 0.78E-4 & 16.493 & ~8.983 & 2.5-8.600  & 0.247 & 4.16E+06 & 3.41E+05 & 1.10E+05 & 2.04E+05 & 2.64E+04 \\
~60  & 17.08 & ~3.64 & 0.36 & 0.62E-4 & 25.172 & 12.623 & 3.7-11.623 & 0.218 & 3.18E+06 & 3.51E+05 & 6.18E+04 & 1.28E+05 & 1.61E+05 \\
~80  & 22.62 & ~3.18 & 0.33 & 0.57E-4 & 34.706 & 17.408 & 5.0-16.332 & 0.187 & 2.66E+06 & 3.19E+05 & 4.43E+04 & 8.35E+04 & 1.91E+05 \\
120  & 30.83 & ~2.76 & 0.30 & 0.50E-4 & 48.553 & 24.519 & 7.8-22.790 & 0.156 & 2.19E+06 & 4.33E+05 & 1.78E+05 & 5.35E+04 & 2.02E+05 \\
~60t & 19.93 & ~3.93 & 0.34 & 0.64E-4 & 30.034 & 15.815 & 4.4-14.042 & 0.201 & 3.20E+06 & 3.36E+05 & 5.28E+04 & 1.02E+05 & 1.81E+05 \\
120t & 26.83 & ~2.92 & 0.31 & 0.53E-4 & 44.400 & 21.360 & 6.6-19.601 & 0.170 & 2.58E+06 & 5.35E+05 & 2.48E+05 & 5.79E+04 & 2.28E+05 \\
\enddata            
\label{taba}
\end{deluxetable}

\begin{deluxetable}{ccccccccccc}
\tablecolumns{11}
\tabletypesize{\scriptsize}
\rotate
\tablewidth{0pt}
\tablecaption{\nuk{Al}{26} and \nuk{Fe}{60} yields}
\tablehead{
$\rm M_{\rm ini}$ & \nuk{Al}{26} & Wind & $\rm Ne/C_{\rm sh}$ & expl & \nuk{Fe}{60} & $\rm He_{\rm rsh}$ &
$\rm He_{\rm Schwarz.}$ & $\rm He_{\rm Ledoux} $ & $\rm C_{\rm sh}$ & expl \\
\cline{1-11} \\
\colhead{\msun} & \colhead{\msun} & \colhead{\msun} & \colhead{\msun} & \colhead{\msun} &
\colhead{\msun} & \colhead{\msun} & \colhead{\msun} & \colhead{\msun} & \colhead{\msun} & \colhead{\msun} 
}
\startdata                                                                                                             
~11  & 1.60E-5 & 1.16E-11 & 2.20E-6 & 1.38E-5 & 1.71E-6 & 3.0E-7 & ----- &   & 1.4E-6 & 1.0E-8 \\
~12  & 2.11E-5 & 2.91E-11 & 2.30E-6 & 1.71E-5 & 4.33E-6 & 2.0E-6 & ----- &   & 2.3E-6 & ------ \\
~13  & 2.45E-5 & 1.46E-10 & 2.50E-6 & 2.20E-5 & 7.56E-5 & 2.0E-6 & ----- &   & ------ & 7.4E-5 \\
~14  & 1.04E-4 & 4.37E-10 & 7.00E-5 & 3.40E-5 & 5.72E-6 & 1.8E-6 & ----- &   & 3.9E-6 & ------ \\
~15  & 1.32E-4 & 1.17E-09 & 8.00E-5 & 5.20E-5 & 6.28E-6 & 8.0E-7 & ----- &   & 5.4E-6 & 8.0E-8 \\
~16  & 6.80E-5 & 2.74E-09 & 1.00E-5 & 5.80E-5 & 4.39E-6 & 6.0E-7 & ----- &   & 6.0E-7 & 3.2E-6 \\
~17  & 6.87E-5 & 6.76E-09 & 1.50E-5 & 5.37E-5 & 7.96E-6 & 2.0E-7 & ----- &   & 4.0E-6 & 3.8E-6 \\
~20  & 5.43E-5 & 4.32E-08 & 1.80E-5 & 3.63E-5 & 1.56E-5 & ------ & ----- &   & 1.4E-5 & 1.6E-6 \\
~25  & 8.61E-5 & 3.93E-07 & 3.46E-5 & 5.11E-5 & 3.69E-5 & ------ & ----- &   & 3.2E-5 & 4.9E-6 \\
~30  & 9.93E-5 & 2.39E-06 & 2.61E-6 & 9.43E-5 & 1.49E-5 & ------ & ----- &   & 7.0E-6 & 7.9E-6 \\
~35  & 8.38E-5 & 1.14E-05 & 2.06E-5 & 5.18E-5 & 4.03E-5 & ------ & ----- &   & 3.3E-5 & 7.3E-6 \\
~40  & 1.21E-4 & 2.06E-05 & 3.44E-5 & 6.60E-5 & 6.23E-5 & ------ & 1.E-5 & ------ & 4.4E-5 & 8.3E-6 \\
~60  & 2.52E-4 & 6.94E-05 & 5.06E-5 & 1.32E-4 & 2.27E-4 & ------ & 1.E-4 & 6.0E-5 & 8.0E-5 & 4.7E-5 \\
~80  & 4.00E-4 & 1.32E-04 & 8.80E-5 & 1.80E-4 & 7.55E-4 & ------ & 6.E-4 & 1.4E-4 & 1.0E-4 & 5.5E-5 \\
120  & 7.03E-4 & 2.82E-04 & 1.58E-4 & 2.63E-4 & 9.93E-4 & ------ & 8.E-4 & 2.5E-4 & 1.3E-4 & 6.3E-5 \\
~60t & 2.97E-4 & 5.98E-05 & 5.02E-5 & 1.87E-4 & 5.50E-4 & ------ & 4.E-4 & ------ & 1.0E-4 & 5.0E-5 \\
120t & 6.97E-4 & 3.50E-04 & 1.10E-4 & 2.37E-4 & 8.60E-4 & ------ & 7.E-4 & ------ & 1.0E-4 & 6.0E-5 \\
\enddata            
\label{tabb}
\end{deluxetable}

\begin{deluxetable}{lcccc}
\tablecolumns{5}
\tablewidth{0pt}
\tablecaption{Tests on the \nuk{Al}{26} yields}
\tablehead{$\rm test $ & component & 25 & 60 & 120 \\ \cline{1-5} \\ \colhead{} & \colhead{} &\colhead{\msun} & \colhead{\msun} & \colhead{\msun} }
\startdata                                                                                                             
\nuk{Mg}{25}(initial)$\times$2                       & wind & & 1.34(-4) & 5.64(-4)   \\
\nuk{Mg}{25}$\rm (p,\gamma)$\nuk{Al}{26}$\times$10   & wind & & 4.16(-5) & 2.20(-4)   \\
\nuk{Mg}{25}$\rm (p,\gamma)$\nuk{Al}{26}$\times$3    & wind & & 4.60(-5) & 2.33(-4)   \\
\nuk{Mg}{25}$\rm (p,\gamma)$\nuk{Al}{26}$\times$0.33 & wind & & 1.09(-4) & 3.32(-4)   \\
\nuk{Mg}{25}$\rm (p,\gamma)$\nuk{Al}{26}$\times$0.1  & wind & & 7.79(-5) & 2.10(-4)   \\
0.5~$\rm H_p~over.~in~H_{\rm burn.} $ & wind  & & 5.98(-5) & 3.50(-4) \\
0.5~$\rm H_p~over.~in~H_{\rm burn.} $ & total & & 2.97(-4) & 6.97(-4) \\
\nuk{Al}{26}$\rm (n,p)$\nuk{Mg}{26}$\times$2 \& \\ \nuk{Al}{26}$\rm (n,\alpha)$\nuk{Na}{23}$\times$2 & expl. & 3.13(-5) & 7.80(-5) &  \\
\nuk{Mg}{24}$\rm (n,\gamma)$\nuk{Mg}{25}$\times$2 & expl. & 8.00(-5) & 2.11(-4) &  \\
\nuk{Mg}{25}$\rm (p,\gamma)$\nuk{Al}{26}$\times$2 & expl. & 8.20(-5) & 2.09(-4) &  \\
\enddata         
\label{tabc}
\end{deluxetable}

\begin{deluxetable}{ccccccc}
\tablecolumns{7}
\tablewidth{0pt}
\tablecaption{Selected properties of the models computed with the Langer (1989) mass loss rate}
\tablehead{
$\rm M_{\rm ini}$ & $\rm M_{\rm end}$ & $\rm M_{\rm CO}$ & \nuk{Al}{26}$~_{\rm Wind}$ & \nuk{Al}{26}$~_{\rm Cshell+expl.}$ & \nuk{Al}{26}$~_{\rm Total}$ & \nuk{Fe}{60}$~_{\rm Total}$ \\
\cline{1-7} \\
\colhead{\msun} & \colhead{\msun} & \colhead{\msun} & \colhead{\msun} & \colhead{\msun} & \colhead{\msun} & \colhead{\msun} 
}
\startdata                                                                                                             
~40  & 6.88 & 5.25 & 2.17(-5) & 1.053(-4) & 1.27(-4) & 4.54(-5) \\
~60  & 6.01 & 4.54 & 7.51(-5) & 4.490(-5) & 1.20(-4) & 1.52(-5) \\
~80  & 6.10 & 4.61 & 1.42(-4) & 1.120(-4) & 2.54(-4) & 2.17(-5) \\
120  & 6.18 & 4.68 & 3.02(-4) & 4.200(-5) & 3.44(-4) & 1.75(-5) \\
\enddata            
\label{langer}
\end{deluxetable}

\end{document}